\documentclass[11pt]{article}

\usepackage{graphicx}
\usepackage{epsfig,cite}
\usepackage{epstopdf}
\usepackage{amssymb}
\usepackage{amsmath}
\usepackage{dsfont}
\usepackage{url}

\usepackage{url}

\def\smallfrac#1#2{\hbox{${{#1}\over {#2}}$}}
\def\eq#1{Eq.~(\ref{#1})}

\def\half{\hbox{${1\over 2}$}}
\newcommand{\beq}{\begin{equation}}
\newcommand{\eeq}{\end{equation}}
\newcommand{\bea}{\begin{eqnarray}}
\newcommand{\eea}{\end{eqnarray}}
\newcommand{\la}{\left\langle}
\newcommand{\ra}{\right\rangle}

\newcommand{\lp}{\left(}
\newcommand{\rp}{\right)}
\def\{{\left\lbrace}
\def\}{\right\rbrace}

\def\frac#1#2{{{#1}\over {#2}}}
\def\gsim{\mathrel{\rlap{\lower4pt\hbox{\hskip1pt$\sim$}}
    \raise1pt\hbox{$>$}}}         
\def\lsim{\mathrel{\rlap{\lower4pt\hbox{\hskip1pt$\sim$}}
    \raise1pt\hbox{$<$}}}         

\newcommand{\dat}{\mathrm{dat}}

\newcommand{\rep}{\mathrm{rep}}

\newcommand{\new}{\mathrm{new}} 
\newcommand{\old}{\mathrm{old}}

\newcommand{\draft}[1]{}

\def\nn{{\nonumber}}
\def\Aslash{\not{\hbox{\kern-4pt $A$}}}
\def\Eslash{\not{\hbox{\kern-4pt $E$}}}

\begin{document}
\begin{flushright}
Edinburgh 2010/24\\
IFUM-968-FT\\
FR-PHENO-2010-040\\
RWTH TTK-10-55\\
\end{flushright}
\begin{center}
{\Large \bf Reweighting NNPDFs: the W lepton asymmetry}
\vspace{0.8cm}

{\bf  The NNPDF Collaboration:}\\
Richard~D.~Ball$^{1}$, Valerio~Bertone$^2$, Francesco~Cerutti$^3$, \\
 Luigi~Del~Debbio$^1$, Stefano~Forte$^4$, Alberto~Guffanti$^2$, \\
Jos\'e~I.~Latorre$^3$, 
Juan~Rojo$^4$ and Maria~Ubiali$^{5}$.

\vspace{1.cm}
{\it ~$^1$ Tait Institute, University of Edinburgh,\\
JCMB, KB, Mayfield Rd, Edinburgh EH9 3JZ, Scotland\\
~$^2$  Physikalisches Institut, Albert-Ludwigs-Universit\"at Freiburg
\\ Hermann-Herder-Stra\ss e 3, D-79104 Freiburg i. B., Germany  \\
~$^3$ Departament d'Estructura i Constituents de la Mat\`eria, 
Universitat de Barcelona,\\ Diagonal 647, E-08028 Barcelona, Spain\\
~$^4$ Dipartimento di Fisica, Universit\`a di Milano and
INFN, Sezione di Milano,\\ Via Celoria 16, I-20133 Milano, Italy\\
~$^5$ Institut f\"ur Theoretische Teilchenphysik und Kosmologie,\\ RWTH Aachen University, D-52056 Aachen, Germany\\}
\end{center}

\vspace{0.8cm}

\begin{center}
{\bf \large Abstract:}
\end{center}

We present a method for incorporating the information contained in 
new datasets into an 
existing set of parton distribution functions without the need for refitting. 
The method involves reweighting the ensemble of parton densities through the computation of the 
$\chi^2$ to the new dataset. We explain how reweighting may be used to assess the impact of any 
new data or pseudodata on parton densities and thus on their predictions. 
We show that the method works by considering the addition of inclusive jet data to a 
DIS+DY fit, and comparing to the refitted distribution. We then use reweighting to determine the 
impact of recent high statistics lepton asymmetry data from the D0 experiment on the NNPDF2.0 
parton set.
We find that the D0 inclusive muon and electron data are perfectly compatible with the rest of the data included in the 
NNPDF2.0 analysis and impose additional constraints on the large-$x$ $d/u$ ratio. The more exclusive D0 electron datasets are however inconsistent both with the other datasets and among themselves, suggesting 
that here the experimental uncertainties have been underestimated.

\clearpage

\tableofcontents

\clearpage

\section{Introduction}

The determination of parton distribution functions (PDFs) and 
their uncertainties through global fits to datasets taken in deep 
inelastic and hadronic collision experiments, analysed using perturbative 
QCD, is one of the key ingredients in the exploitation of future experiments, 
in particular at LHC. 
Of course such fits can only be as good as the data that goes into them, 
so whenever there is new data or new experiments, the fits have to be redone 
to take the new data into account. This process is cumbersome and time 
consuming, and can only be performed using the same software as 
in the previous fits, and thus by the fitting collaborations themselves.

In this paper we will show how, by using the ensembles of PDFs produced 
by the NNPDF collaboration~\cite{nonsinglet,nnpdf10,nnpdf12,nnpdf20}, 
anyone can determine the effect of new data on the PDFs quickly and 
easily: all that is required is a computation of the $\chi^2$ to the new data 
for each PDF in the ensemble~\cite{Giele:1998gw}. With this information 
one can determine the overall impact of the new data on PDFs, their consistency 
with the older data used in the fit, the effect the new data have on the 
shape and precision of individual PDFs, and thus their effect on 
observables such as benchmark cross-sections or predictions for new 
physics. The same approach can be used just as easily to estimate the 
effects of pseudodata from proposed experiments or machines on PDFs and 
thus on cross-sections.

The technique we use is based on statistical inference. In the NNPDF 
approach~\cite{f2p,nonsinglet,nnpdf10,nnpdf12,nnpdf20} we generate through 
a Monte Carlo procedure an ensemble of $N$ PDF replicas, 
$\mathcal{E} = \{f_k, k=1,\ldots,N\}$, each fitted 
to a data replica generated according to the uncertainties and their 
correlations as measured by the experimental collaborations. 
Each PDF is parametrized by a highly redundant neural network in order to 
avoid parameterization bias which would otherwise spoil the procedure, and 
the stopping point of the fit of each replica is determined using cross-validation 
to prevent over-fitting. 
The final PDF ensemble then forms an accurate representation of the probability 
distribution of PDFs,\footnote{Throughout this paper we will denote 'parton distribution 
function' by PDF, but write out 'probability density function' in full, in order to avoid 
any confusion: both are probability densities, but in very different spaces.} 
conditional on the input data and the particular assumptions (such 
as NLO QCD, a value of $\alpha_s$, etc) used in the fits. 

Given an NNPDF ensemble one can evaluate any quantity or experimental observable 
$\mathcal{O}[f]$ depending on the PDFs (such as the PDF mean, the variance, PDF 
correlations, or indeed the mean, variance etc of any cross-section computed from 
them) by computing $\mathcal{O}[f]$ for each of the replicas, and averaging the results. 
This is because the integral in the space of functions is well approximated by
the average over the ensemble $\mathcal E$, so that the mean value of $\mathcal{O}[f]$ given 
by
\begin{equation}
\la\mathcal{O}\ra=\int \mathcal{O}[f] \, \mathcal{P}(f)\,Df
=\smallfrac{1}{N}\,\sum_{k=1}^{N}\mathcal{O}[f_k]\, .
\label{eq:avg}
\end{equation}
Each of the replicas $f_k$ carries equal weight because they were 
generated through importance sampling: the replicas were fitted to a 
data replica generated according to the probability distribution of 
the experimental data, using a fitting procedure with no bias. 

We can include the effects of a new independent dataset without performing a 
new fit if we instead reweight the old fit according to weights $w_k$, which 
assess the probability that each PDF replica $f_k$ agrees with the new data. 
The reweighted ensemble then forms a representation of the probability distribution 
of PDFs conditional on both the old and the new data. 
The weights are computed straightforwardly by evaluating the $\chi^2$ of the new 
data to each of the replicas. The mean value of the observable $\mathcal{O}[f]$ taking account 
of the new data is then given by the weighted average
\begin{equation}
\la\mathcal{O}\ra_\new = \int \mathcal{O}[f] \, \mathcal{P}_\new(f)\,Df
=\smallfrac{1}{N}\,\sum_{k=1}^{N} w_k \mathcal{O}[f_k]\, .
\label{eq:avgrw}
\end{equation}

The usefulness of this method is clear: it becomes possible to test the impact 
of a new dataset, or indeed the potential impact of MC data generated for 
a new experiment, quickly and simply without the need for a new fit (or 
indeed without considering explicitly any other datasets except the one 
under immediate consideration). This comes at a price: the effective number 
of replicas will be reduced, either because the new data prove to be very constraining 
(good), or because they are inconsistent with the old data (bad). 
We will provide a criterion to distinguish between these two cases. Of course 
if the new data are both consistent and precise, the effective number of 
replicas might be so reduced that a refit becomes necessary. 

One of the advantages of the reweighting method is that it can be used
to assess the impact on the global fit of observables for which no
fast code is available, and thus which cannot be included without 
resorting to $K$--factor approximations. 
One such observable is the Tevatron $W$ lepton charge asymmetry.
Recent measurements\cite{cdfelectron,d0muon,d0electron,d0mnew} of 
this quantity have attracted a
lot of attention, since sizable tension with other data
in the global fit sensitive to the large-$x$ $d$ quark distribution, such as
DIS deuterium structure function data, has been reported~\cite{CT10,MSTWasym}.
It is not clear from these studies whether this tension reflects an 
experimental problem of the recent Tevatron data, or whether the 
problems are with the DIS deuterium 
data, perhaps indicating the need for substantial nuclear corrections. With this
motivation, armed with the statistical power of reweighting, we will 
here study the incorporation of the D0 $W$ lepton charge asymmetry data in 
the NNPDF2.0 fit.

Reweighting is also important from a
conceptual point of view. If more and more data are included
in the fit through reweighting, the resulting PDFs become less and less
dependent on the initial PDF. But PDFs obtained in this way
then by construction satisfy the laws of statistical inference --- for
example, uncertainties will automatically behave upon inclusion of 
new data according to standard statistics.\footnote{Indeed, it was suggested in
Ref.~\cite{Giele:2001mr} that a PDF fit might be performed by including 
all data through reweighting of a first guess based on past experience.}
Hence, checking that the results obtained by reweighting coincide with 
results obtained by simply including the new data in the global fit
provides a highly nontrivial check on the consistency of the NNPDF
global fitting procedure.

The paper is organised as follows. In the next section, we will explain how 
the weights can be computed, and give tests through which one can access 
quantitatively the impact of the new data and their consistency with the 
old data. A detailed proof of these results, with a full discussion 
of the subtleties, is given in section 3: this is important because an earlier 
attempt to implement a reweighting procedure~\cite{Giele:1998gw} used an
expression for the weights which differs from our result 
(a detailed examination of the result of Ref.~\cite{Giele:1998gw} is presented in 
section 3.2).\footnote{A recent study by the LHCb 
collaboration~\cite{DeLorenzi:2010zt} using a reweighting technique 
to assess the impact of low $p_t$ Drell--Yan pairs at the LHC on PDF determinations, 
also appears to use the incorrect formula derived in~\cite{Giele:1998gw}.}
This section may be skipped by readers only interested in applying the new technique. 
In section 4 we show how the method may be used in practice by applying 
it to inclusive 
jet data: since there are existing NNPDF sets with and without this data, this allows 
us to check that reweighting works. Then in section 5 we illustrate the power of 
reweighting by using it to assess the impact of D0 $W$ lepton charge asymmetry 
data on the NNPDF2.0 PDFs. The results are particularly interesting because 
while the inclusive D0 asymmetry data is perfectly compatible with the NNPDF2.0 
set and their inclusion in the global fit results in a moderate 
improvement in the determination of the medium and large-$x$ $d$ quark PDF, the more 
exclusive electron datasets turn out to be inconsistent both with other sets 
in the global analysis and among them.

\section{Reweighting}
\label{sec:reweight}

\subsection{The Weights}
\label{subsec:weights}

We consider the situation where a set
of experimental data have been used to
construct a probability distribution for PDFs,
${\mathcal P}_{\rm old}(f)$. This probability distribution
is delivered as a finite ensemble of PDFs
$\mathcal E = \{f_k, k=1,\ldots,N\}$. Any observable
can be obtained performing averages over this
ensemble as prescribed in \eq{eq:avg},
that is, equally weighting each PDF.

The problem we shall now address is how to update
this probability distribution when new experimental 
data are available. There are two options: either we can construct 
a new probability distribution ${\mathcal P}_{\rm new}(f)$
from scratch by including both old and new data in a new fit, or
we can update the old fit by computing a weight, $w_k$, for each individual
PDF $f_k$ in the ensemble $\mathcal E$ according to the rules
of statistical inference. Then, ${\mathcal P}_{\rm new}(f)$ is
simply understood as an update (or reweighting) of the prior probability
distribution ${\mathcal P}_{\rm old}(f)$.

Both methods incorporate the same information,
the old and the new data, and we will show below that when the weights are 
chosen correctly both techniques are statistically equivalent in the 
sense that when the number of replicas is sufficiently large they 
both give representations of the same probability 
distribution ${\mathcal P}_{\rm new}(f)$. However to calculate the weights 
involves only knowledge of the new data: all the relevant 
information about the old data is already contained in 
${\mathcal P}_{\rm old}(f)$. It is thus substantially easier to implement, 
since no refitting is necessary.

To be specific, we consider a set of $n$ new data that have not been included
in the determination of the initial probability density distribution:
 \begin{equation*}
  y = \left\lbrace y_1,y_2,\cdots,y_n \right\rbrace.
 \end{equation*}
Clearly any instance of such a set of data can be seen as a point $y$
in an $n$-dimensional real space. The experimental uncertainties
are summarised by the $n\times n$ experimental
covariance matrix $\sigma_{ij}$, which includes a term that incorporates 
the overall normalization uncertainty \cite{norm}, but reduces 
to a diagonal matrix
in cases when experimentalists do not provide the correlated systematic
uncertainties. We assume throughout that these new data are statistically 
independent of any of the data included in the original fit.

Using statistical inference we can update the initial probability density
$\mathcal{P}_{\old}(f)$ by taking into account the new data, thereby
obtaining an improved probability density
$\mathcal{P}_{\new}(f)$. To do this we need to know the relative probabilities 
of the new data for different choices of PDF. Since 
the new data are assumed to have Gaussian errors, these probabilities  
will be directly proportional to the probability density of 
the $\chi$ to the new data conditional on $f$:
\begin{equation}
  \mathcal{P}(\chi|f) \propto (\chi^{2}(y,f))^{{1\over 2}(n-1)}
  e^{-\frac{1}{2}\chi^{2}(y,f)},
  \label{eq:csqdist}
\end{equation}
where if $y_i[f]$ is the value predicted for the data
$y_i$ using the PDF $f$,
\begin{equation}
\chi^{2}(y,f) = 
\sum_{i,j=1}^n (y_i-y_i[f])\sigma^{-1}_{ij}(y_j-y_j[f]).
\label{eq:chisq}
\end{equation}

It follows from the statistical independence of the old and new data that 
by the law of multiplication of probabilities:
\begin{equation}
\mathcal{P}_{\rm new}(f)
= \mathcal{N}_{\chi}\mathcal{P}(\chi|f)\;\mathcal{P}_{\rm old}(f),
\label{eq:multiplication}
\end{equation}
where $\mathcal{N}_{\chi}$ is a
normalization factor, independent of $f$. 

Multiplying on both sides by some observable $\mathcal{O}[f]$ and integrating
over the PDFs,
\begin{eqnarray}
\la\mathcal{O}\ra_{\new}&=&\int \mathcal{O}[f] \,
\mathcal{P}_{\rm new}(f)\,Df,\nn\\ &=&\mathcal{N}_\chi\int \mathcal{O}[f]
\,\mathcal{P}(\chi|f)
\mathcal{P}_{\rm old}(f)\,Df,\nn\\ 
&=&\smallfrac{1}{N}\,\sum_{k=1}^{N}
\mathcal{N}_\chi\mathcal{P}(\chi|f_k)
\mathcal{O}[f_k]\, ,
\label{eq:avgnew}
\end{eqnarray}
where in the last line we used \eq{eq:avg}. 

We can thus 
sample the probability density $\mathcal{P}_{\rm new}(f)$ using the
$N$ replicas $f_k$, but reweighted: in place of \eq{eq:avg} we now have 
\begin{equation}
\la\mathcal{O}\ra_{\new}=\smallfrac{1}{N}\sum_{k=1}^{N} w_k \, \mathcal{O}[f_k]\,
,
\label{eq:reweighting}
\end{equation}
where 
\begin{equation}
w_k = \mathcal{N}_\chi\mathcal{P}(\chi|f_k)
= 
\mathcal{N}^\prime_{\chi}(\chi^{2}_k)^{{1\over 2}(n-1)} 
e^{-\frac{1}{2}\chi^{2}_k},
\label{eq:weightsN}
\end{equation}
and $\chi^2_k\equiv \chi^2(y,f_k)$ is evaluated using \eq{eq:chisq}. 
The normalization factor $\mathcal{N}^\prime_\chi$ is fixed by normalizing the
new probability density $\mathcal{P}_{\rm new}(f)$: taking the operator 
$\mathcal{O}[f]$ to be the unit operator, $\la 1\ra_{\new}=1$, so from 
\eq{eq:reweighting} this fixes 
$\sum_{k=1}^{N} w_k=N$, and thus using \eq{eq:weightsN}
\begin{equation}
w_k = 
\frac{(\chi^{2}_k)^{{1\over 2}(n-1)} 
e^{-\frac{1}{2}\chi^{2}_k}}
{\smallfrac{1}{N}\sum_{k=1}^{N}(\chi^{2}_k)^{{1\over 2}(n-1)}
e^{-\frac{1}{2}\chi^{2}_k}}.
\label{eq:weights}
\end{equation}
The weights $w_k$, when divided by $N$, are then simply the 
probabilities of the replicas $f_k$, given the $\chi^2$ to the new data. 

Note that our formula \eq{eq:weights} for the weights is different from the
one derived in Ref.~\cite{Giele:1998gw} whenever the number of new data 
points is greater than one. The reason 
for this is that the use of Bayes theorem for multidimensional probability 
densities is subtle, since without care one may fall foul of 
the Borel-Kolmogorov paradox (see for example Ref.~\cite{jaynes}). A careful 
proof of the weights
\eq{eq:weights} using the elementary rules of statistical inference 
is given in sec.~\ref{sec:bayesianreweighting}: the subtle error in 
the argument used in Ref.~\cite{Giele:1998gw} is explained in 
sec.~\ref{sec:gielekosower}.

\subsection{Measuring Information Loss and Consistency}

The original ensemble of replicas $\mathcal E = \{f_k, k=1,\ldots,N\}$ 
is constructed through importance sampling of the probability density 
$\mathcal{P}_{\rm old}(f)$, and thus each 
replica has equal weight. It is maximally efficient, in the sense 
this is the best representation of the underlying density 
$\mathcal{P}_{\rm old}(f)$ for a given number of replicas $N$: the only 
way to improve it is by increasing $N$. After reweighting, this will 
no longer be the 
case, since in fact the weights give the relative importance of the different 
replicas, and the replicas with very small weights will become almost 
irrelevant in ensemble averages. The reweighted replicas will thus no 
longer be as efficient as the old: for a given $N$, the accuracy of the 
representation of the underlying distribution $\mathcal{P}_\new(f)$ will 
be less than it would be in a new fit. 

We can quantify this loss of efficiency by using the Shannon 
entropy to compute the effective number of replicas left after reweighting:
\begin{equation}
N_{\rm eff}\equiv \exp \{\smallfrac{1}{N}\sum_{k=1}^N w_k \ln (N/w_k)\}.
\label{eq:effective}
\end{equation}
Clearly $0<N_{\rm eff}<N$: the reweighted fit has the same accuracy as a 
refit with $N_{\rm eff}$ replicas. Thus if $N_{\rm eff}$ becomes too low, the 
reweighting procedure will no longer be reliable, either because 
the new data contain a lot of information on the PDFs, necessitating a 
full refitting with more replicas, or because
the new data are inconsistent with the old. 

These two cases can be distinguished by examining the $\chi^2$ profile 
of the new data: if
in the reweighted fit there are very few replicas with a $\chi^2$ 
per data point of order unity, the errors in the new dataset 
have probably been underestimated. This profile may be easily evaluated:
\begin{equation}
\mathcal{P}(\chi^2)= \smallfrac{1}{N}\sum_{k} w_k,
\label{eq:evidence}
\end{equation}
where the sum is over all replicas $k$ such that 
$\chi^{2}_k\in [\chi^2,\chi^2+d\chi^2]$.

Alternatively, we consider inconsistent data as data whose errors have been 
underestimated. If we rescale the uncertainties of the data by a factor 
$\alpha$, we can use inverse probability to calculate the probability density 
for the rescaling parameter $\alpha$: 
\begin{equation}
\mathcal{P}(\alpha)\propto \smallfrac{1}{\alpha}\sum_{k=1}^N w_k(\alpha).
\label{eq:rescaling}
\end{equation}
Here $w_k(\alpha)$ are the weights \eq{eq:weightsN} evaluated by replacing 
$\chi^2_k$ with $\chi^2_k/\alpha^2$ (and are thus proportional to 
the probability of $f_k$ given the new data with rescaled errors): 
averaging them in the reweighted fit thus gives the probability density 
for $\alpha$. If this probability density peaks close to one the new 
data are consistent, while if it peaks far above one, then it is 
likely that the errors in the data have been underestimated.

If the new data are reasonably consistent, we can assess whether they 
should be included in a new fit by calculating the $\chi^2$ distribution 
of the dataset that would be used in the new fit (i.e. all the old data 
plus the new data), using the reweighting procedure as in \eq{eq:evidence}. 
Comparison to the old $\chi^2$ distribution then tells us whether the new 
data would improve the fit: if so the peak should shift a little towards one, 
with a slight narrowing due to the increase in the total number of data points.
This may be quantified by comparing the area under the curves in a 
given range.

\section{Statistical Inference}
\label{sec:Bayes}

\subsection{Proof of the Weight Formula}
\label{sec:bayesianreweighting}

Here we give a careful derivation of the rules for reweighting 
presented in the previous section. Some readers
may consider skipping this section and simply use
the practical prescription as given in \eq{eq:weights}.
Our argument is based on the standard use of statistical inference. However 
some of the details are subtle, since we need to use probability densities 
in multi-dimensional spaces, and thus need to take care with limits.

By the probability $P(f)$ for the PDF $f$ what we actually mean 
is the probability $P(f|K)$, where $K$ denotes all the data used in the 
determination and their associated errors, the values of parameters such 
as $\alpha_s$ and heavy quark masses used in the computation of the data 
expected from the PDF, and finally also the theoretical framework used (for 
example NLO QCD). If we then wish to extend the dataset by including new 
data $y$, the new probability $P_\new(f)$ is then $P(f|yK)$: besides $K$ 
we now also assume the new data $y$. 

The new probability is then determined from the old probability
using the sampling distribution $P(y|fK)$ and multiplicative rule for 
probabilities (often known as Bayes theorem):
\begin{equation}
P(AB|C)=P(A|BC)P(B|C) = P(B|AC)P(A|C),
\label{eq:Bayes}
\end{equation}
where $P(A|C)$ is the probabilities of $A$ given $C$, etc, and $AB$ denotes 
$A$ and $B$. Naively applying this result in the present case we have
\begin{equation}
P(f|yK)P(y|K) = P(y|fK)P(f|K),
\label{eq:Bayesmult}
\end{equation}
whence (replacing $P(f|K)$ with $\mathcal{P}(f|K)Df$, $P(f|yK)$ 
with $\mathcal{P}(f|yK)Df$)
\begin{equation}
\mathcal{P}(f|yK)= P(y|fK)\mathcal{P}(f|K)/P(y|K).
\label{eq:Bayespdf}
\end{equation}
Note that $P(y|K)$ does not depend on the PDF $f$, and can thus be 
determined simply by insisting that $\mathcal{P}(f|yK)$ is properly 
normalized: we then find 
\begin{equation}
P(y|K)= \int P(y|fK)\mathcal{P}(f|K)Df,
\label{eq:marginalization}
\end{equation}
so
\begin{equation}
\mathcal{P}(f|yK)= P(y|fK)\mathcal{P}(f|K)\Big/\int P(y|fK)\mathcal{P}(f|K)Df,
\label{eq:Bayesdiscrete}
\end{equation}
where everything on the right hand side is now known.

This argument would work without problems if the data $y$ could only take 
discrete values. The difficulty in the present case is that our data are 
continuous, so rather than the probability $P(y|fK)$ we have to work with 
a multi-dimensional probability density $\mathcal{P}(y|fK)d^ny$, in a 
limit in which the volume element $d^ny$ goes to zero. Of course in this 
limit the probabilities $P(y|fK)$ and $P(y|K)$ also go to zero, and we find 
a ratio of two zeros in 
\eq{eq:Bayespdf}. The conditional probability $P(f|yK)$ is then only well 
defined if we specify carefully the way in which the limit is to be taken: 
probabilities conditional on sets of measure zero are ambiguous. Failure 
to specify the limiting process can result in contradictions (the 
Borel-Kolmogorov paradox~\cite{jaynes}).

Consider then the probability density for the data $y$. Assuming that 
the new experiments are not correlated with any
of the experiments used in the determination of the initial
probability density, the probability density of $y$ is then
given by \eq{eq:marginalization}:
\begin{equation}
  \mathcal{P}(y|K) = \int \mathcal{P}(y| fK)
  \mathcal{P}(f|K)\,Df = \smallfrac{1}{N}\sum_{k=1}^{N}
  \mathcal{P}(y|f_kK)
  \label{eq:dP},
\end{equation}
where in the second step we used \eq{eq:avg}. The density $\mathcal{P}(y|f K)$ 
gives the probability that the new data lie in an infinitesimal volume $d^n y$ 
centred at $y$ in the space of possible data given a particular choice of PDF 
$f$: it is often called the sampling distribution or (when considered as a 
function of $f$) the likelihood function. Assuming that the uncertainties 
in the data are purely Gaussian, 
\begin{equation}
  \mathcal{P}(y|fK)d^ny =(2\pi)^{-n/2}(\det \sigma_{ij})^{-1/2}
  e^{-\frac{1}{2}\chi^{2}(y,f)}d^ny,
  \label{eq:condP}
\end{equation}
where $\chi^{2}(y,f)$ is calculated using \eq{eq:chisq} (and of course using the assumptions $K$ in the computation of the predictions $y_i[f]$). The volume element $d^ny$ is independent of $f$: without a specific prediction, all data are assumed equally likely.

Since to compute $\mathcal{P}(y|fK)$ it is sufficient to compute $\chi^{2}(y,f)$, it is sufficient for our purposes to consider the probability density for the $\chi^2$ to the new dataset:
\begin{equation}
  \mathcal{P}(\chi|fK)d\chi =
  2^{1-n/2}(\Gamma(n/2))^{-1}(\chi(y,f))^{n-1}
  e^{-\frac{1}{2}\chi^{2}(y,f)}d\chi,
  \label{eq:condPcsq}
\end{equation}
where $\chi(y,f)\equiv(\chi^2(y,f))^{1/2}$. This distribution 
may be readily derived from \eq{eq:condP} by
diagonalising the covariance matrix and rescaling the data to a set
$\{Y_i\}$ of independent Gaussian variables each with unit
variance. Then $d^ny = (\det \sigma_{ij})^{1/2}d^nY$, and $\chi^2 =
\sum_{i=1}^n Y_i^2$.  Choosing $n$-dimensional spherical co-ordinates
in the space of data (with $\chi$ as the radial co-ordinate,
and thus $y=y[f]$ as the origin), we may write $d^nY =
A_n\chi^{n-1}d\chi d^{n-1}\Omega$, where $d^{n-1}\Omega$
is the measure on the sphere and $A_n = 2\pi^{n/2}(\Gamma(n/2))^{-1}$
is the area of the unit sphere in $n$-dimensions. The probability
\eq{eq:condP} may thus be written
\begin{eqnarray}
  \mathcal{P}(y|fK)d^ny &=&(2\pi)^{-n/2}
  e^{-\frac{1}{2}\chi^{2}(y,f)}d^nY \nn \\ 
  &=& 2^{1-n/2}(\Gamma(n/2))^{-1}(\chi(y,f))^{n-1}
  e^{-\frac{1}{2}\chi^2(y,f)} d\chi d^{n-1}\Omega\,,
  \label{eq:condPcsqderiv}
\end{eqnarray}
in agreement with \eq{eq:condPcsq} provided
\begin{equation}
  \mathcal{P}(y|fK)d^ny 
  = \mathcal{P}(\chi|fK)d\chi d^{n-1}\Omega.
  \label{eq:ytochirep}
\end{equation}
Again the probability density $\mathcal{P}(\chi|K)$ for the $\chi$
of the new dataset is obtained by averaging over replicas:
\begin{equation}
  \mathcal{P}(\chi|K) = \int \mathcal{P}(\chi|fK )
  \mathcal{P}(f|K)\,Df 
  = \smallfrac{1}{N}\sum_{k=1}^{N} \mathcal{P}(\chi|f_kK);
\label{eq:dPcsq}
\end{equation}
so combining \eq{eq:dP}, \eq{eq:ytochirep}, and \eq{eq:dPcsq} 
\begin{equation}
  \mathcal{P}(y|K)d^ny 
  = \smallfrac{1}{N}\sum_{k=1}^{N} \mathcal{P}(\chi|f_kK)d\chi d^{n-1}\Omega 
  = \mathcal{P}(\chi|K)d\chi d^{n-1}\Omega, 
  \label{eq:ytochi}
\end{equation}
since both the volume factor $d^{n-1}\Omega$ and the interval
$d\chi$ are independent of the choice of replica, and may thus be
taken out of the sum: this follows directly from the assumption that the 
measure $d^ny$ in \eq{eq:condP} is independent of $f$. 

The advantage of using $\mathcal{P}(\chi|fK)$ instead of $\mathcal{P}(y|fK)$ when evaluating \eq{eq:Bayespdf} is that $\mathcal{P}(\chi|fK)$ is only a one dimensional density, so taking the limit in which the volume element goes to zero is straightforward and unambiguous.
We may write \eq{eq:Bayespdf} as
\begin{equation}
\mathcal{P}(f|\chi K)Df\;\mathcal{P}(\chi |K)d\chi 
= \mathcal{P}(\chi|fK)d\chi \;\mathcal{P}(f|K)Df.
\label{eq:Bayescsq}
\end{equation}
The marginalization \eq{eq:dPcsq} follows directly on
integration over $f$, since if $\mathcal{P}(f|\chi)$ is correctly
normalized, $\int \mathcal{P}(f|\chi K)Df = 1$.  Now, cancelling the
$d\chi$ from either side of \eq{eq:Bayescsq} (since this is just a
pre-assigned interval),
\begin{equation}
 \mathcal{P}(f|\chi K)Df =
\frac{\mathcal{P}(\chi |fK)}{\mathcal{P}(\chi |K)}\mathcal{P}(f|K)Df.
\label{eq:newden}
\end{equation}
Multiplying on both sides by some observable $\mathcal{O}[f]$ and integrating
over the PDFs,
\begin{eqnarray}
\la\mathcal{O}\ra_{\new}&=&\int \mathcal{O}[f] \,
\mathcal{P}(f|\chi K)\,Df,\nn\\ &=&\int \mathcal{O}[f]
\,\frac{\mathcal{P}(\chi|fK)}{\mathcal{P}(\chi|K)}
\mathcal{P}(f|K)\,Df,\nn\\ 
&=&\frac{1}{N}\,\sum_{k=1}^{N}
\frac{\mathcal{P}(\chi|f_kK)}{\mathcal{P}(\chi|K)}
\mathcal{O}[f_k]\, ,
\label{eq:avgnewx}
\end{eqnarray}
where in the last line we used \eq{eq:avg}. This corresponds to the reweighting
\eq{eq:reweighting}
with weights
\begin{equation}
w_k = \frac{\mathcal{P}(\chi|f_kK)}{\mathcal{P}(\chi|K)}.
\label{eq:weightscsq}
\end{equation}
Combining \eq{eq:weightscsq} with
\eq{eq:condPcsq} and \eq{eq:dPcsq}, we obtain \eq{eq:weights}.

Note that a further application of Bayes' theorem to \eq{eq:weightscsq} gives 
the alternative form
\begin{equation}
w_k = \frac{P(f_k|\chi K)}{P(f_k|K)}= N P(f_k|\chi K),
\label{eq:weightsprob}
\end{equation}
since because the replicas are uniformly distributed, $P(f_k|K)=1/N$. 
Thus $w_k/N$ is the probability of replica $f_k$ given the 
$\chi$ to the new data.

\subsection{The Naive Prescription}
\label{sec:gielekosower}

It is instructive to also derive the weights working directly with the 
probability density $\mathcal{P}(y|fK)$: using Bayes' theorem we may
write instead of \eq{eq:Bayescsq}
\begin{equation}
  \mathcal{P}(f|yK)Df\;\mathcal{P}(y|K)d^ny 
  = \mathcal{P}(y|fK)d^ny\;\mathcal{P}(f|K)Df.
  \label{eq:Bayesy}
\end{equation}
Again, the marginalization \eq{eq:dP} follows directly from
the requirement that $\mathcal{P}(f|yK)$ be normalized, i.e. that 
$\int \mathcal{P}(f|yK)Df = 1$.

Naively cancelling the volume factor
$d^ny$ from either side, and pursuing the same argument as before
yields:
\begin{eqnarray}
\la\mathcal{O}\ra_{\new}^G&=&\int \mathcal{O}[f] \, \mathcal{P}(f|yK)\,Df,\nn\\ 
&=&\int
\mathcal{O}[f] \,\frac{\mathcal{P}(y|fK)}{\mathcal{P}(y|K)}
\mathcal{P}(f|K)\,Df,\nn\\ 
&=&\frac{1}{N}\,\sum_{k=1}^{N}\frac{\mathcal{P}(y|f_kK)}{\mathcal{P}(y|K)}
\mathcal{O}[f_k]\, .
\label{eq:avgnewy}
\end{eqnarray}
This would then lead to the conclusion of 
Giele-Keller~\cite{Giele:1998gw} that the weights are proportional to
$\mathcal{P}(y|f_kK)/\mathcal{P}(y|K)$, and thus (using \eq{eq:condP})
are given by
\begin{equation}
w^G_k = \frac{e^{-\frac{1}{2}\chi^{2}_k}}
{\smallfrac{1}{N}\sum_{k=1}^N e^{-\frac{1}{2}\chi^{2}_k}}.
\label{eq:weightsgauss}
\end{equation}
When $n>1$ this result is clearly different from our previous result
\eq{eq:weightscsq}. We see explicitly that the 
densities $\mathcal{P}(f|\chi K)$ and
$\mathcal{P}(f|yK)$ are not the same, despite the fact that when the
data $y$ take a given value, $\chi$ takes a corresponding value.

It is also clear that the Gaussian weights \eq{eq:weightsgauss} must 
be incorrect: in the limit where the number of new (and consistent) 
data $n$ becomes very large,
$\chi^{2}_k$ peaks around $n$, and only the very few replicas in the 
tail of the distribution ($\chi^{2}_k\ll n$ is very unlikely) will 
survive. By contrast with the correct weights \eq{eq:weightscsq}, 
replicas with $\chi^{2}_k\sim n$ will dominate the fit, replicas 
with very small or very large $\chi^2$ being suppressed.

The reason for the difference between the results \eq{eq:weightscsq} and 
\eq{eq:weightsgauss} is the Borel-Kolmogorov paradox~\cite{jaynes}:
when dealing with multi-dimensional probability densities care must be 
taken with limits, since a conditional 
probability on a set of measure zero is not well defined. Here
the limits used to derive $\mathcal{P}(f|\chi K)$ and
$\mathcal{P}(f|yK)$ are different, and thus so are the results. The correct 
result can only be obtained by taking the appropriate limit.

The probability density $\mathcal{P}(f|\chi K)$ is defined as the 
probability density for $f$ given that
the $\chi$ lies in the finite interval $[\chi,\chi+d\chi]$, in
the limit $d\chi\to 0$. In this case the conditioning variable spans
a one--dimensional manifold, and therefore there is no freedom in the
choice of the limiting procedure. The definition of 
$\mathcal{P}(f|\chi K)$ is unique, and thus
the argument which leads to \eq{eq:weightscsq} unambiguous. 
However the probability density $\mathcal{P}(f|yK)$ is defined as 
the probability density for $f$ given that $y$ lies in some 
volume $V_n$, in the limit $V_n\to 0$. In a multi-dimensional space 
such as this, the conditional
probability density $\mathcal{P}(f|yK)$ is ambiguous, since it depends
on the way the volume element $V_n$ is chosen, and then taken to zero. 
Different definitions
correspond to different physical settings. In the argument which led
us to \eq{eq:weightsgauss}, we implicitly assumed that $V_n$ was the
compact volume $d^ny$ centred on $y$, so that as $V_n\to 0$, the point
$y$ was uniquely selected. However there are many points in the 
space of data which have the same
$\chi^2$, and thus the same effect on $f$. We must include all
these points with equal weight when determining the conditional
probability density of $f$ given $y$, so we need to sum over all the
compact volumes $d^ny$ that build the $(n-1)$-dimensional level surfaces of
$\chi(y,f)$ through the point $y$. Thus $V_n$ is a thin shell with
thickness $d\chi$, and hence its total volume is proportional to $A_n
d\chi$. The limit $V_n\to 0$ is then taken by letting $d\chi\to
0$. We should thus write \eq{eq:Bayesy} as (using \eq{eq:ytochirep} and
\eq{eq:ytochi})
\begin{equation}
\mathcal{P}(f|yK) Df\;\mathcal{P}(\chi|K) A_n d\chi
= \mathcal{P}(\chi|fK) A_n d\chi \;\mathcal{P}(f|K)Df.
\label{eq:Bayesys}
\end{equation}
Cancelling the volume factor $A_n d\chi$, since this is independent
of $f$, this definition is the same as \eq{eq:newden}, and thus
yields the correct weights \eq{eq:weights} in the limit $V_n\to 0$.

\subsection{Derivation of the Consistency Tests}

Finally we consider the derivation of the two diagnostic results \eq{eq:evidence} and \eq{eq:rescaling}. The first is simply the `evidence' \eq{eq:dPcsq}, evaluated by binning in $\chi^2$. The second is more involved: using Bayes Theorem
\begin{equation}
\mathcal{P}(\alpha|\chi f_k K)d\alpha \mathcal{P}(\chi|f_k K)d\chi
= \mathcal{P}(\chi|\alpha f_k K)d\chi 
\mathcal{P}(\alpha|f_k K)d\alpha.\label{eq:alfBayes}
\end{equation}
Now the likelihood $\mathcal{P}(\chi|\alpha f_k K)$ may be evaluated using the usual formula \eq{eq:condPcsq}, and by noting that 
the effect of $\alpha$ is to rescale $\chi^2\to \chi^2/\alpha^2$: we accordingly denote the result by $w_k(\alpha)$. The prior density $\mathcal{P}(\alpha|f_k, K)$ we assume is uniform in $\ln\alpha$, since $\alpha$ is a scale parameter (this ensures that the results are invariant under $\alpha\to 1/\alpha$). We thus find 
\begin{equation}
\mathcal{P}(\alpha|\chi, f_k, K) = \frac{w_k(\alpha)}{\alpha\int d(\ln\alpha')w_k(\alpha')},
\label{eq:weightscsqalf}
\end{equation}
where the overall normalization has been fixed by integrating over $\alpha$. Then as usual
\begin{eqnarray}
\mathcal{P}(\alpha|\chi, K) 
&=& \int Df\, \mathcal{P}(\alpha|\chi, f, K)\mathcal{P}(f|\chi, K)\nn\\
&=& \smallfrac{1}{N}\sum_{k=1}^N\frac{w_k(\alpha)}
{\alpha\int d(\ln\alpha')w_k(\alpha')}.
\label{eq:alfBayesx}
\end{eqnarray}
It is easy to show by a change of variable that the integrals in the denominator are the same for all $k$, whence we find \eq{eq:rescaling}.

\section{Validation: Inclusive Jets}

As a demonstration of the effectiveness of our reweighting procedure, 
we first apply it 
to a dataset that has already been included and studied in the NNPDF2.0 
analysis~\cite{nnpdf20}. We thus start with the fit obtained including
only the DIS and Drell--Yan data, call this NNPDF2.0(DIS+DYP), and then add 
the inclusive jet data from Tevatron Run II~\cite{Abulencia:2007ez,D0:2008hua}, 
which were included in the NNPDF2.0 analysis, through reweighting. 
The resulting reweighted fit can then be compared directly with 
the NNPDF2.0 fit, which includes the same DIS, Drell--Yan and  
Tevatron inclusive jet data. Given the consistency of 
the inclusive jet data with the DIS and Drell--Yan data demonstrated in 
Ref.~\cite{nnpdf20}, we expect the reweighted and refitted 
distributions to give results that are equivalent up to statistical 
fluctuations.

Note that from this section on we will slightly change the notation
to make it more similar to that of previous NNPDF studies: $N_{\rm rep}$
will denote the number of replicas in the sample and $N_{\rm dat}$ the
number of data points in the set which is added through reweighting.

To obtain the reweighted PDFs, all that has to be done is to compute 
the $\chi^{2}_k$ of replica $k$ to the inclusive jet data, using \eq{eq:chisq}, 
for each of the $N_{\rm rep}=1000$ replicas of the NNPDF2.0(DIS+DYP) parton set. 
For the inclusive jet data the total number of data points is 
$N_{\rm dat}=186$, and the covariance matrices are as given by the CDF and 
D0 collaborations, the normalization uncertainty are incorporated using the 
$t_0$-method, as discussed in Ref.\cite{norm,nnpdf20}.
The weight associated with each replica in then computed according to 
\eq{eq:weights}: specifically we evaluate
\begin{equation}
  e_{k}\equiv \smallfrac{1}{2}\big((N_{\rm dat} - 1)\log\chi^{2}_k-\chi^{2}_k\big) 
\label{eq:exponent}
\end{equation}
hence if $\la e_k\ra \equiv \smallfrac{1}{N_{\rm rep}}\sum_{k=1}^{N_{\rm rep}}
 e_k$, the weights are given by 
\begin{equation}
  w_k = \mathcal{N}\exp[e_k-\la e_k\ra],\qquad 
  \mathcal{N} = N_{\rm rep}\big/\sum_{k=1}^{N_{\rm rep}} \exp[e_k-\la e_k\ra].
  \label{eq:weightscomp}
\end{equation}
The subtraction of $\la e_k\ra$ in the exponent is introduced
to avoid numerical problems. We set to zero all weights for which 
$\exp[e_k-\la e_k\ra] < 10^{-12}$.

\begin{figure}[ht]
  \begin{center}
    \includegraphics[width=0.45\textwidth]{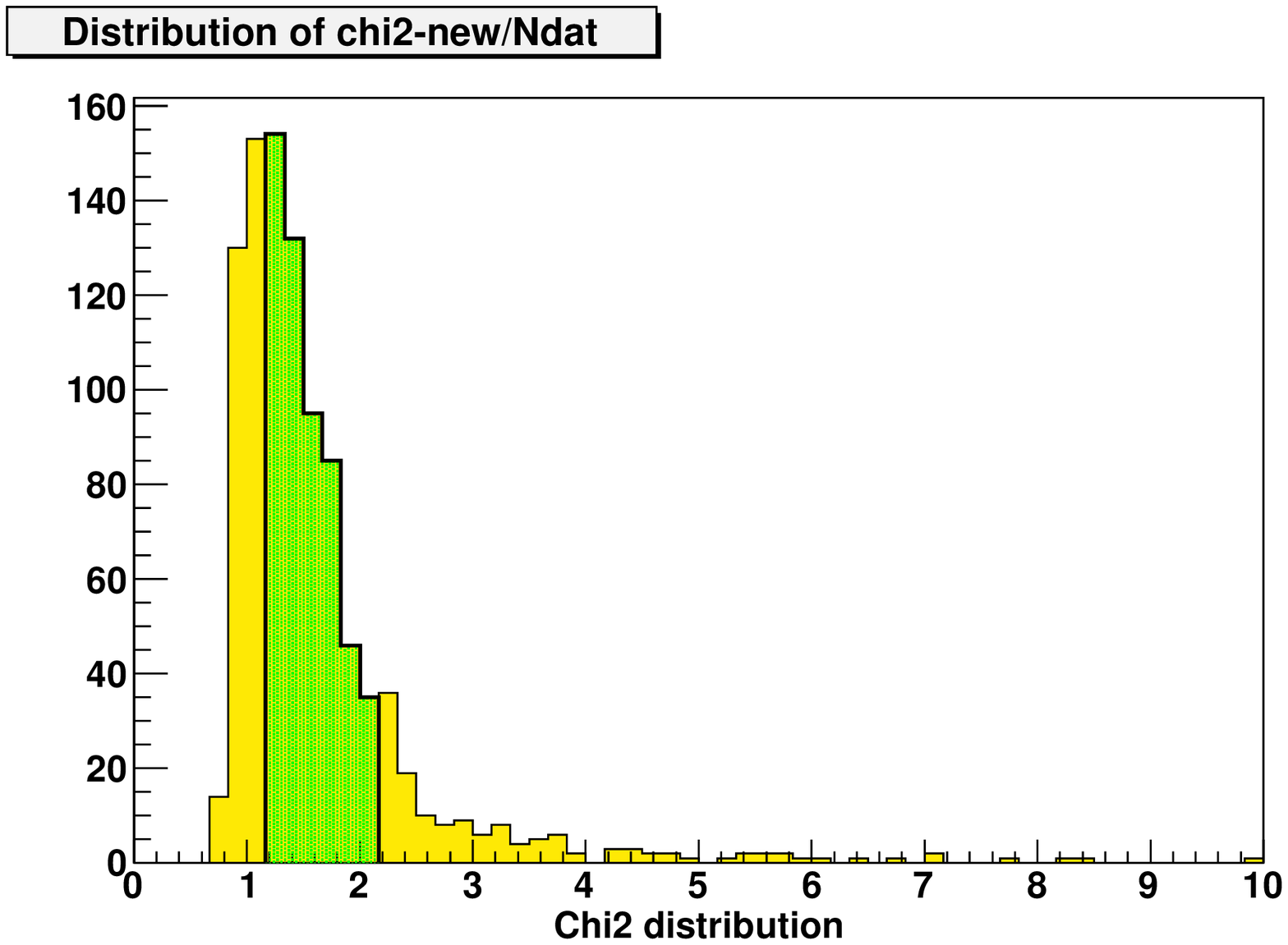}
    \includegraphics[width=0.45\textwidth]{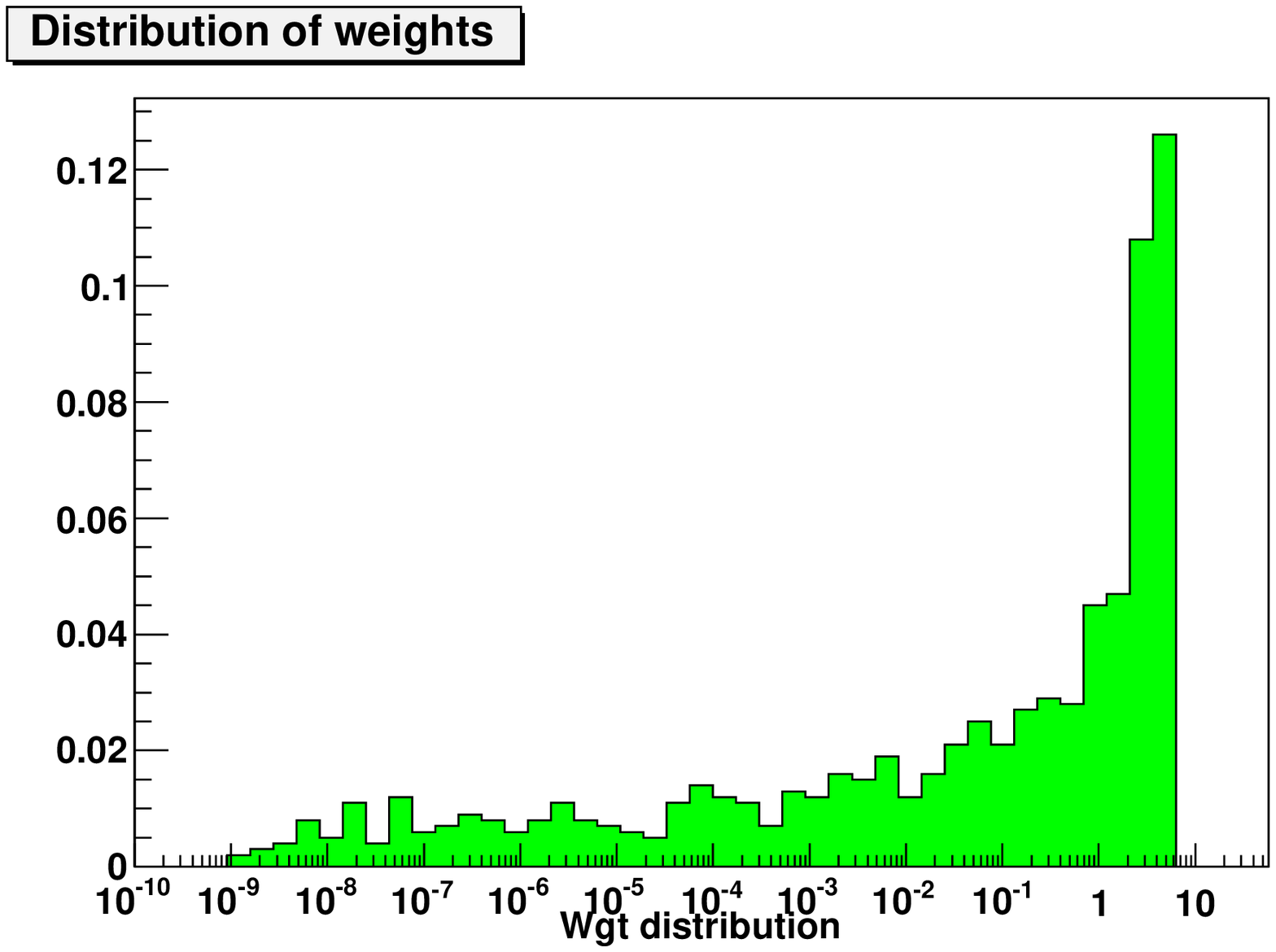}
    \includegraphics[width=0.45\textwidth]{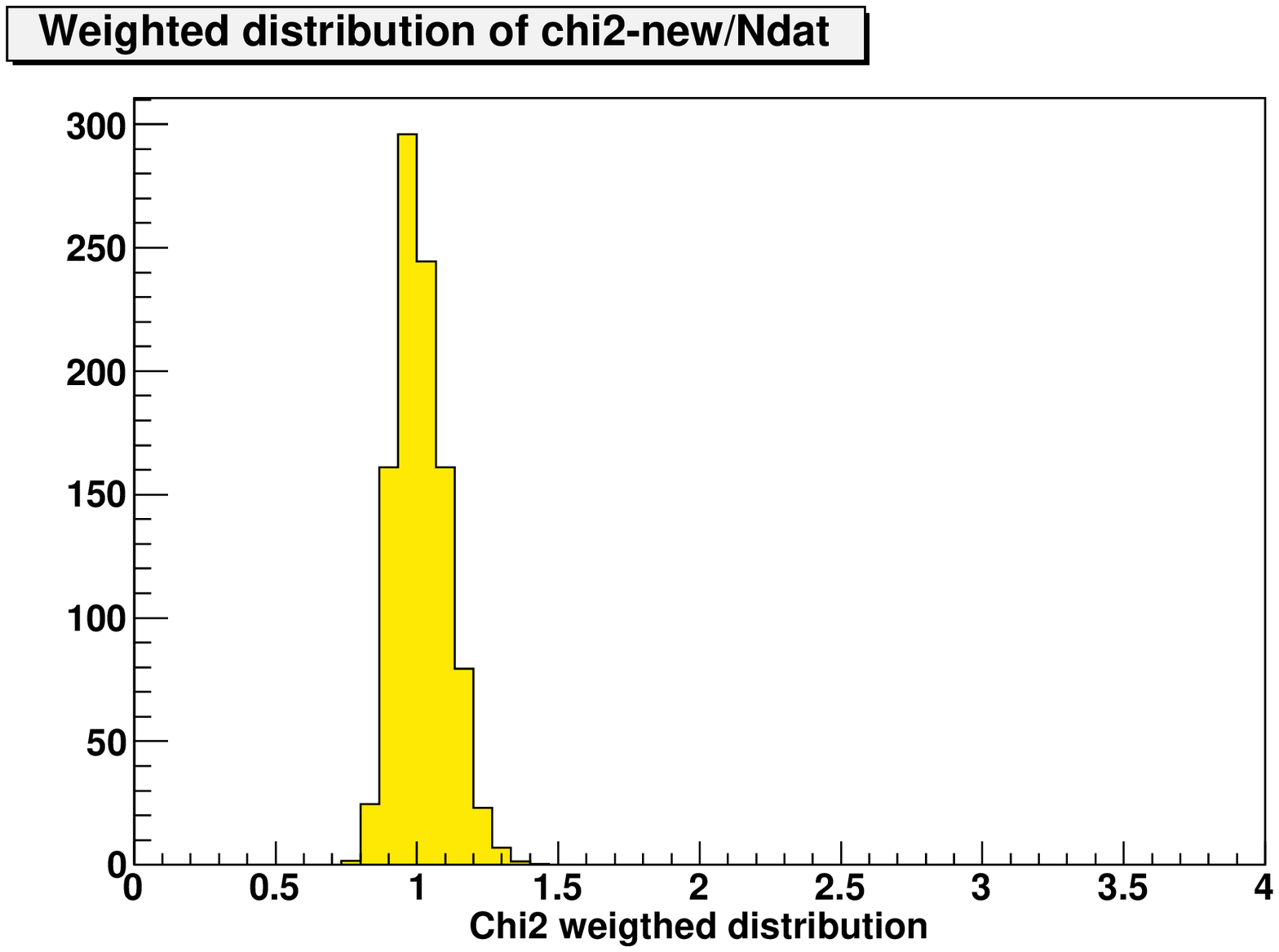}
    \includegraphics[width=0.45\textwidth]{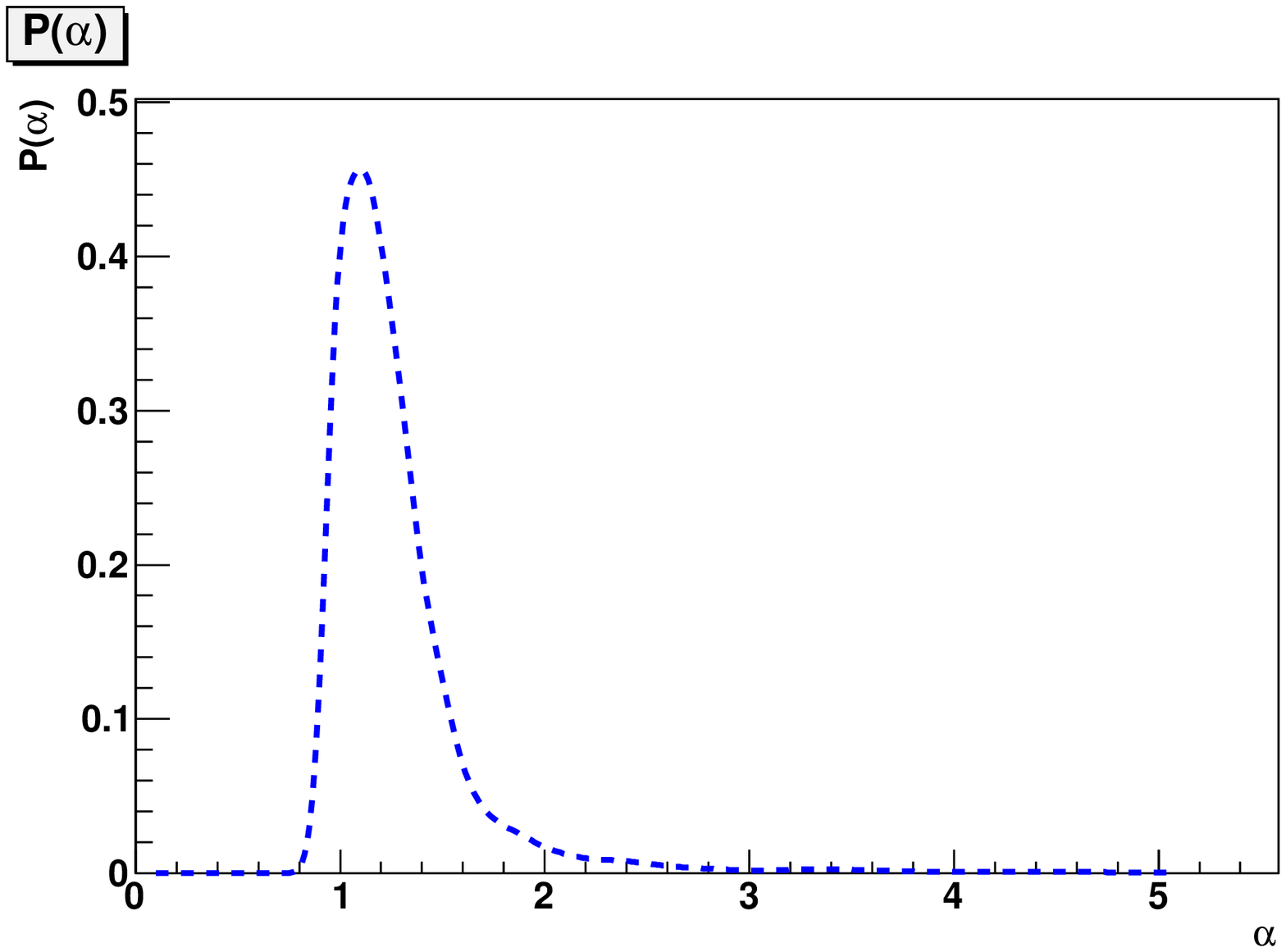}
  \end{center}
  \caption{\small Upper plots: distribution of $\chi^2_k/N_{\rm dat}$ 
    (the $\chi^2$ per data point) and the weights $w_k$ in the 
    reweighting of the NNPDF20(DIS+DYP) set using the inclusive jet data. 
    Lower plots: Distribution of the reweighted $\chi^2$ distribution of the 
    inclusive jet data, and the probability distribution 
    $\mathcal{P}(\alpha)$ of the error rescaling parameter $\alpha$. }
  \label{fig:chi2-jets} \label{fig:consistency-jets}
\end{figure}

\begin{figure}[ht]
  \begin{center}
    \includegraphics[width=0.9\textwidth]{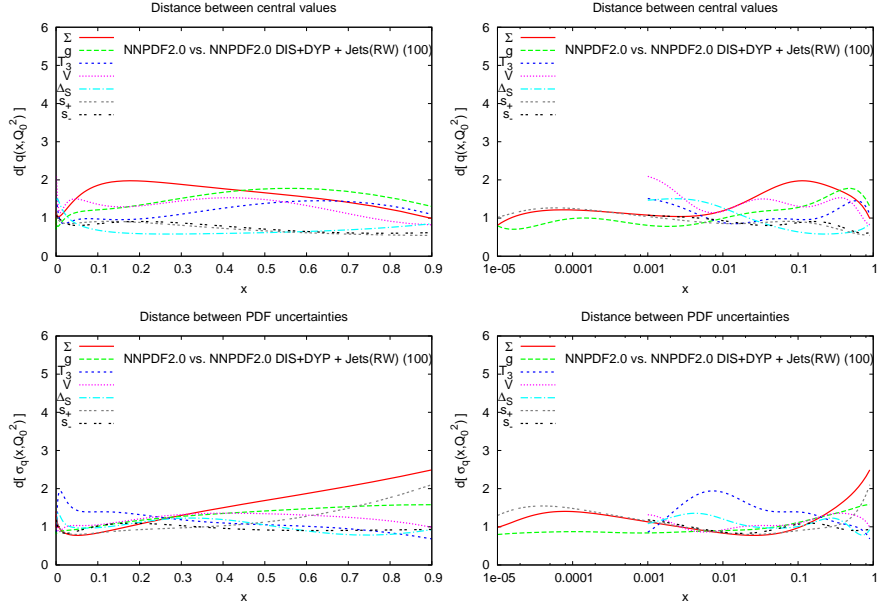}
  \end{center}
  \caption{\small Distances between PDFs (above) and uncertainties (below) 
    for the NNPDF2.0 set and a set obtained adding the Tevatron inclusive jet 
    production data to the NNPDF2.0(DIS+DY) fit using the reweighting technique. 
    The distances have been computed between sets of $N_{\rm rep}=100$ replicas.}
  \label{fig:distances-jets}
\end{figure}
  
  \begin{figure}[b!]
    \begin{center}
      \includegraphics[width=0.45\textwidth]{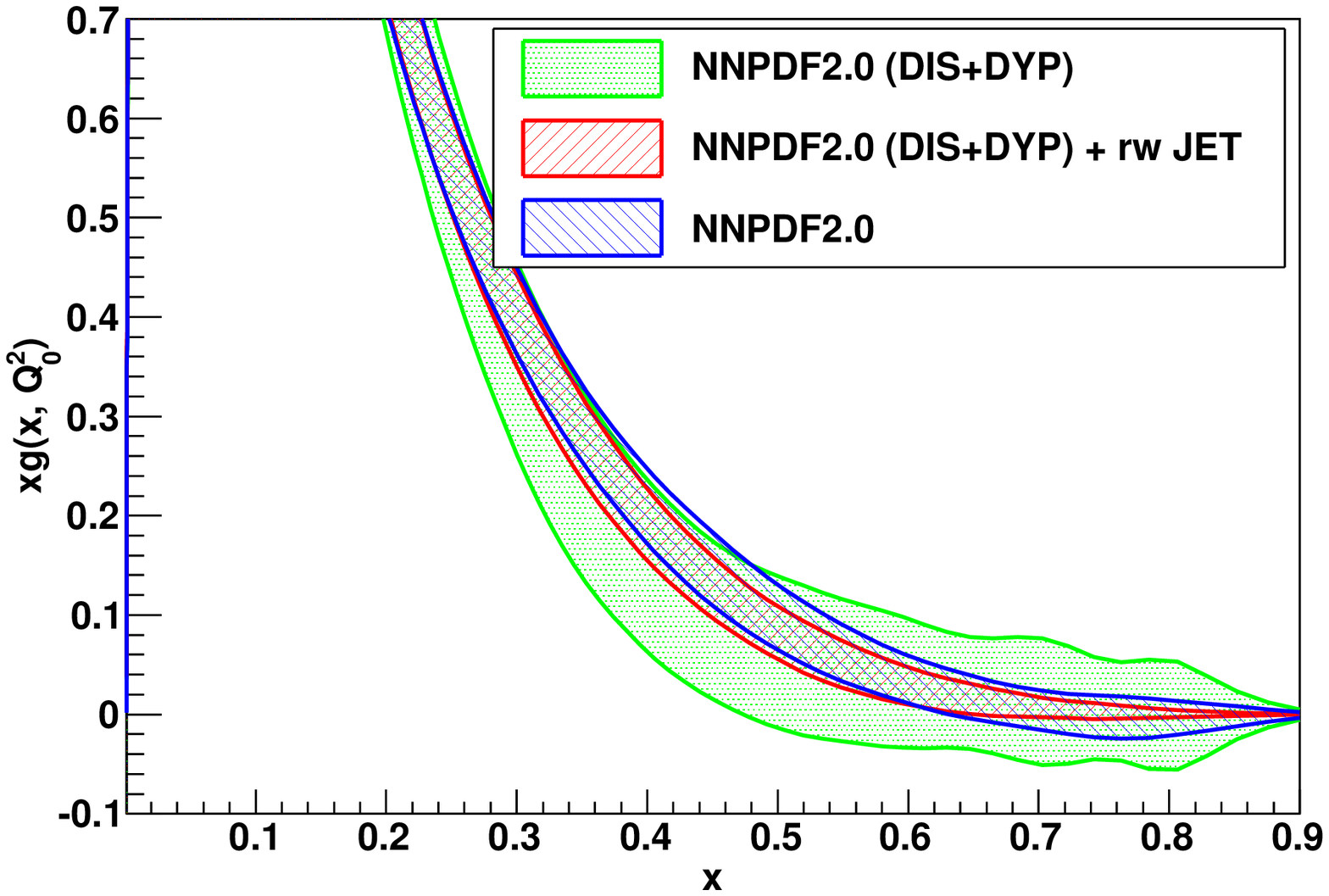}
      \includegraphics[width=0.45\textwidth]{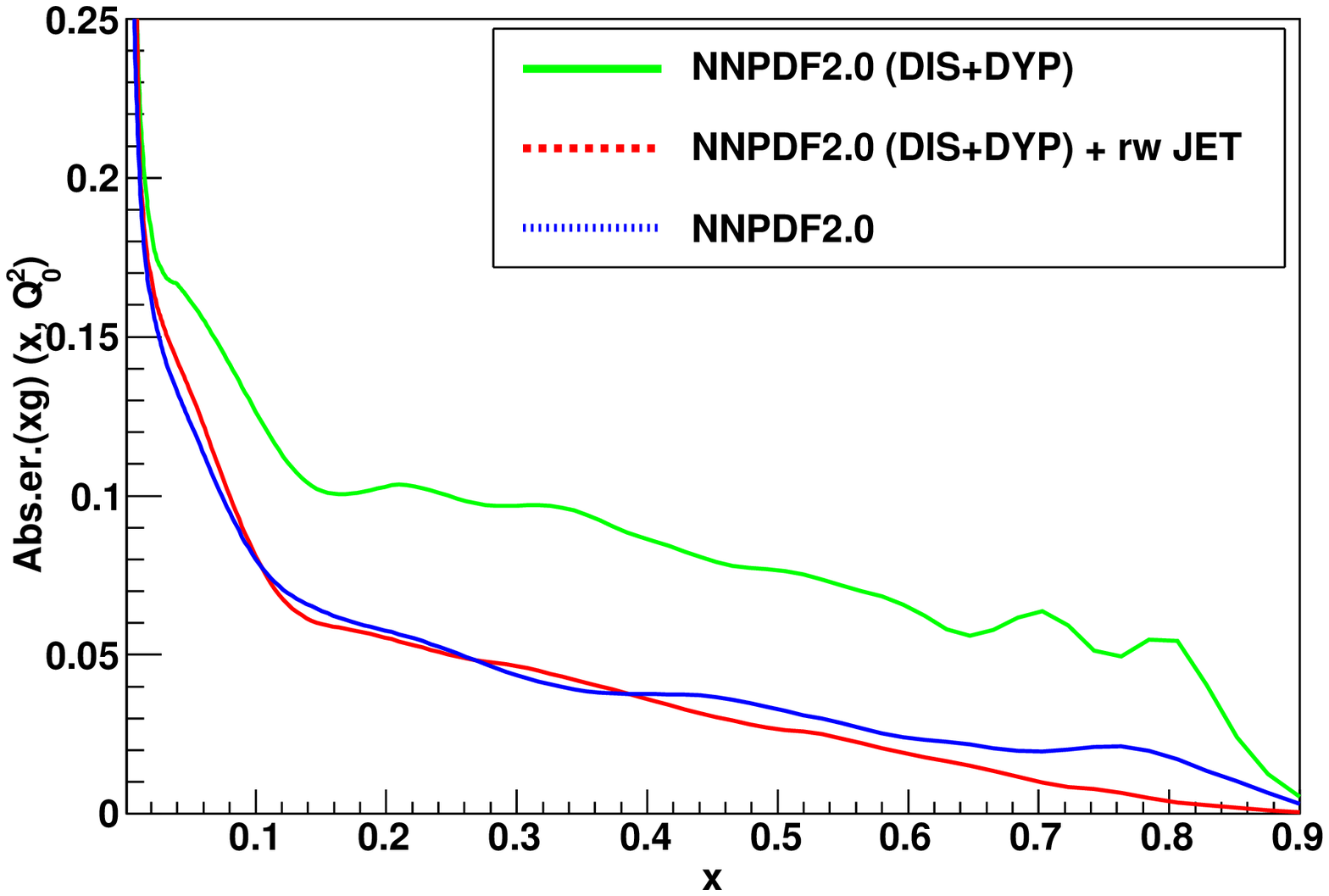}
    \end{center}
    \vskip-0.5cm
    \caption{\small The gluon distribution (left) and its uncertainty (right) of 
      the NNPDF2.0(DIS+DY) fit before and after reweighting with the inclusive jet data 
      compared to the refitted gluon from NNPDF2.0 on a linear scale.}
    \label{fig:pdf-jets}
\end{figure}

The $\chi^2_k$ distributions for the jet data before and after reweighting,
the $P(\alpha)$ estimator and the distribution of weights are shown in 
Fig.~\ref{fig:chi2-jets}. 
We notice that before reweighting the distribution of $\chi^2$ per data point
is peaked close to one, but with a long tail extending to higher values of 
$\chi^2$. This has to be expected since the inclusive jet data 
are not included in the NNPDF2.0(DIS+DYP) set. However $82\%$ of the replicas 
have $0.5<\chi^2_k<2$, confirming that the 
inclusive jet data are likely to be consistent with the other data in the fit 
and their inclusion in the fit will have only a moderate impact. 
Indeed a significant fraction of the weights are of order 
one, with however a long tail of small weights for replicas which will be 
effectively eliminated once the inclusive jet data are included.

To make these statements more quantitative, we can now evaluate the 
number of effective replicas, determined through the Shannon entropy according 
to \eq{eq:effective}: the effective number of replicas after reweighting 
using the jet data is $N_{\rm eff}=332$, i.e. around a third of the replicas are left. 

To examine the consistency of the inclusive jet data with the DIS 
and Drell-Yan data used in NNPDF20(DIS+DYP), we show in 
Fig.~\ref{fig:consistency-jets} the reweighted $\chi^2$ distribution 
computed according to \eq{eq:evidence}. Clearly the replicas which gave a
poor fit to the jet data have now been removed, and the result is
a distribution of $\chi^2$ peaked at one, which shows that the jet data are 
consistent with the DIS and Drell-Yan data. 
This conclusion is reinforced by  the probability distribution $\mathcal{P}(\alpha)$, 
plotted in Fig.~\ref{fig:chi2-jets}: the most probable value for $\alpha$ is close 
to one, showing that the overall size of the experimental errors of these data have 
been well estimated by CDF and D0.

In order to determine quantitatively if indeed the refitted and reweighted
PDF sets represent statistically identical distributions, we can
compute the distances between central values and uncertainties of different
PDF combinations, as defined in Ref.~\cite{nnpdf20} with the
required modifications to account for the individual weights
of each replica.\footnote{The expressions for the distances
for reweighted PDF sets are collected in Appendix~\ref{sec:distances}.}
In Fig.~\ref{fig:distances-jets} we plot the distances between PDFs' central values 
and uncertainties for the reweighted set and the (refitted) NNPDF2.0 set. Note 
that distances have been computed between sets of $N_{\rm rep}=100$ replicas. 
The distance is normalised such that distances $d\sim 1$ correspond to 
statistically identical distributions.
We see that to a very good approximation the refitted and the reweighted sets are 
statistically equivalent, both for central values and uncertainties. The very largest 
distances, $d\sim 2$, corresponding a difference of about one seventh of a 
standard deviation of the measured quantity.

Given that as shown in Fig.~\ref{fig:distances-jets} the refitted
and reweighted sets are statistically equivalent, we know from~\cite{nnpdf20}
that inclusive jet data constrain only the large--$x$ gluon, leaving virtually unchanged all other distributions. 
The reweighted gluon distribution and its uncertainty are shown in 
Fig.~\ref{fig:pdf-jets}, compared with the original distribution, 
the NNPDF20(DIS+DYP) fit,  and with the full NNPDF2.0 fit. On the left hand side
we plot the gluon distribution with its uncertainty band and on the right hand side
the absolute value of the uncertainty. The reweighted and refitted 
distributions  are indeed shown to be equivalent within errors. 
In particular the error of the medium and large--$x$ gluon is sensibly 
reduced by the inclusion of the Tevatron inclusive jet data while, the 
other PDFs are essentially unchanged in
both the refitted and the reweighted sets.

This statistical equivalence is an important check on the consistency 
of the NNPDF fitting methodology and the reweighting method
presented here. In particular, it shows that an NNPDF parton fit (at
least in the case examined here) behaves in a way which is consistent
with the laws of statistical inference: since reweighting is simply 
an application of probability theory, and since reweighting and refitting 
can be used interchangeably, the results obtained from the
global fits indeed behave as probability distributions.

\section{Application: the W lepton asymmetry}

Now that we have explicitly verified that reweighting works, we can use it to 
assess the impact on PDF determination of data which were not included in the 
NNPDF2.0 fit. 
In this section we consider the Tevatron D0 $W$ lepton charge asymmetry high 
luminosity data from Run II~\cite{d0electron,d0muon}. 
This data have attracted a lot of interest recently because of their
potential inconsistency with other datasets which are traditionally 
included in the global fit like the deuterium DIS data~\cite{CT10,MSTWasym}.

\subsection{Motivation}

In proton--antiproton scattering, $W^{\pm}$ bosons are mainly produced by the 
annihilation of a $u(d)$ quark in the proton with the $\bar{d}$($\bar{u}$) in 
the anti--proton. An asymmetry in the $W^+$ and $W^-$ rapidity distributions 
is the result of a difference between the $u$ and $d$ distributions in the proton. 
Therefore, the information on the W charge asymmetry~\cite{cdfboson} provides a 
further constraint on the $u$ and $d$ PDFs. However, due to the unknown longitudinal
momentum of the neutrino, the vector boson rapidity is difficult to determine 
directly.
What is typically measured~\cite{cdfelectron,d0muon,d0electron,d0mnew} is 
instead the lepton charge asymmetry. The vector boson rapidity may then 
be deduced in terms of the pseudo-rapidity of the charged lepton and 
its transverse energy $E^l_T$. Moreover, if the transverse energy $E^l_T$ of the 
outgoing lepton is relatively small, the leading sea contribution 
$\bar{u}-\bar{d}$ is enhanced relative to the valence--valence contributions, 
so the lepton charge asymmetry also probes the separation into valence and sea 
quarks. For this reason in some experimental analysis~\cite{d0electron,d0mnew}, 
the lepton asymmetry is measured in different bins of $E^l_T$.

\begin{figure}[ht]
  \begin{center}
    \includegraphics[width=0.45\textwidth]{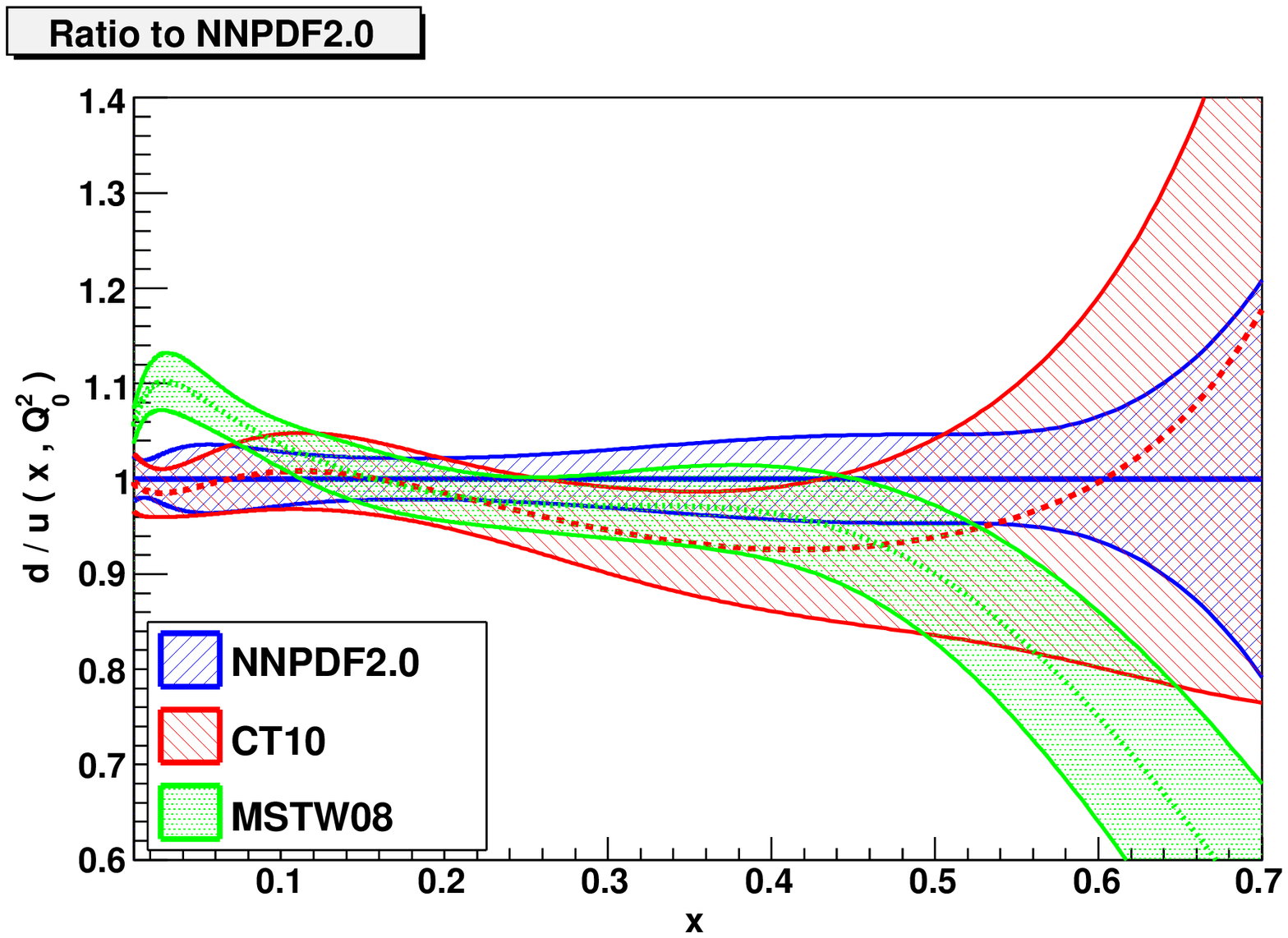}
    \includegraphics[width=0.45\textwidth]{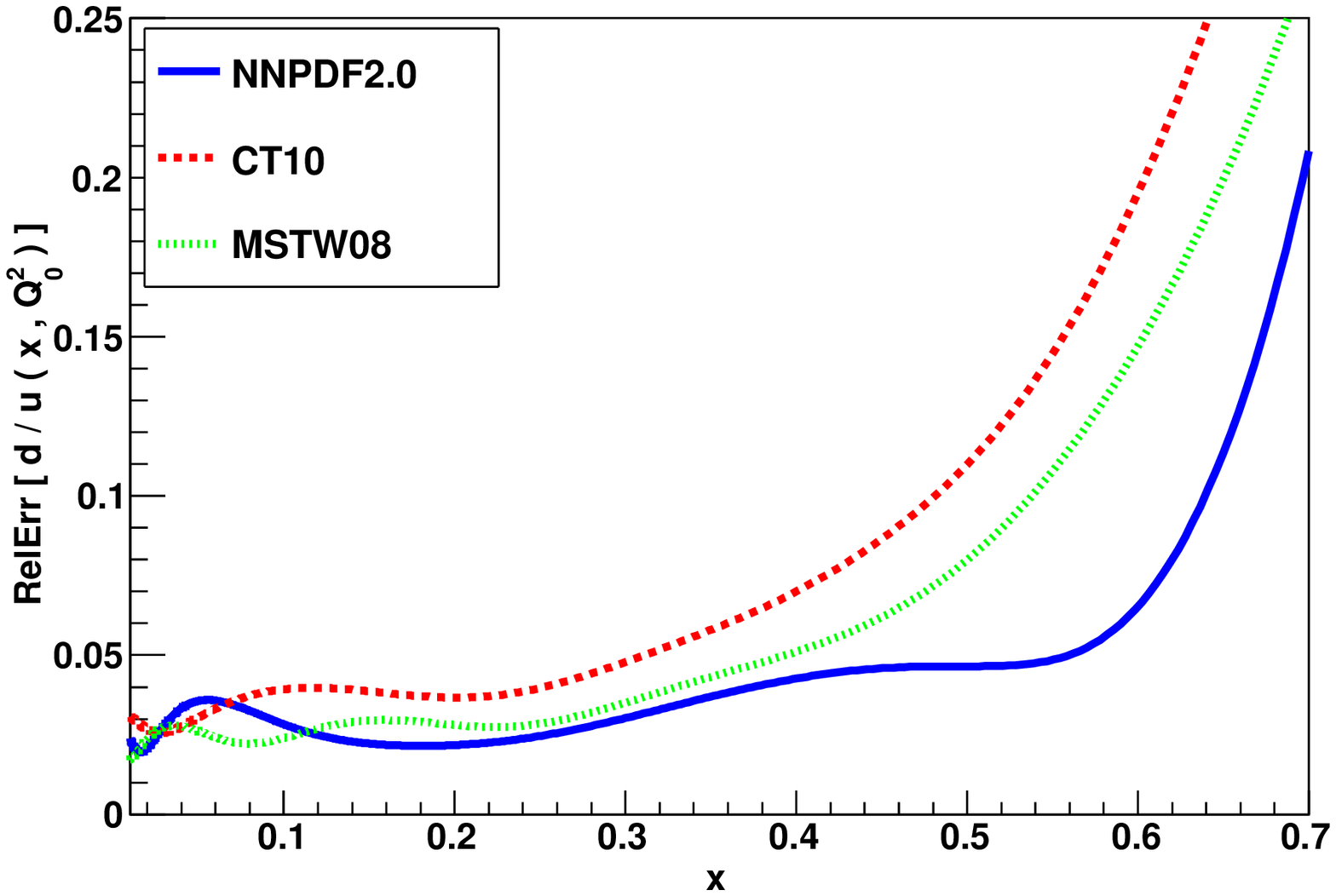}
  \end{center}
  \caption{\small The $d/u$ ratio at large $x$ computed at $Q_0^2=2$ GeV$^2$ 
    from the NNPDF2.0, MSTW08 and CT10 sets. We show the results for the ratio 
    normalized to NNPDF2.0 (left plot) and the relative PDF uncertainties in 
    each case (right plot). All uncertainties are 1$\sigma$.
    \label{fig:d-over-u-init}   }
\end{figure}

Historically, the Tevatron $W$ lepton charge asymmetry data have been
used in global fits together with the deuterium DIS data from BCDMS and NMC
to constrain the ratio of $d$ to $u$ quarks at large--$x$.
One advantage with respect DIS data is that theoretical uncertainties linked to
the deuterium target, like nuclear effects, are not present for the 
lepton asymmetries, where only proton PDFs are involved.
In Fig.~\ref{fig:d-over-u-init} we show the $d/u$ ratio computed using 
different recent PDF sets: NNPDF2.0, CT10 and MSTW08, together
with the relative uncertainties. 
It is clear that PDF uncertainties are sizable for this combination at large-$x$,
thus additional precision measurements of the $W$ asymmetry are useful to reduce 
PDF errors in this region. We notice that in the kinematic region probed by the 
Tevatron measurements ($0.1 \lsim x \lsim 0.7$) the predictions from the three 
sets are in reasonable agreement within the respective uncertainties.

In the NNPDF2.0 analysis only the CDF $W$ boson direct asymmetry data of 
Ref.~\cite{cdfboson} are included. This observable is implemented in the fitting
code at next--to--leading order, without reverting to a K--factor approximation,
using the FastKernel method described in~\cite{nnpdf20}. 
The $W$ lepton asymmetry measurements, on the other hand, were not included 
in the analysis due to the lack of a fast implementation. However, the recent 
development of the APPLGRID~\cite{Carli:2010rw} interface is likely to facilitate 
the future inclusion of these data directly in our fits. 
Thanks to the reweighting technique presented earlier in this paper, we 
can now study the 
impact of the lepton asymmetry data consistently in NLO QCD. 

\begin{figure}[t]
  \begin{center}
    \includegraphics[width=0.45\textwidth]{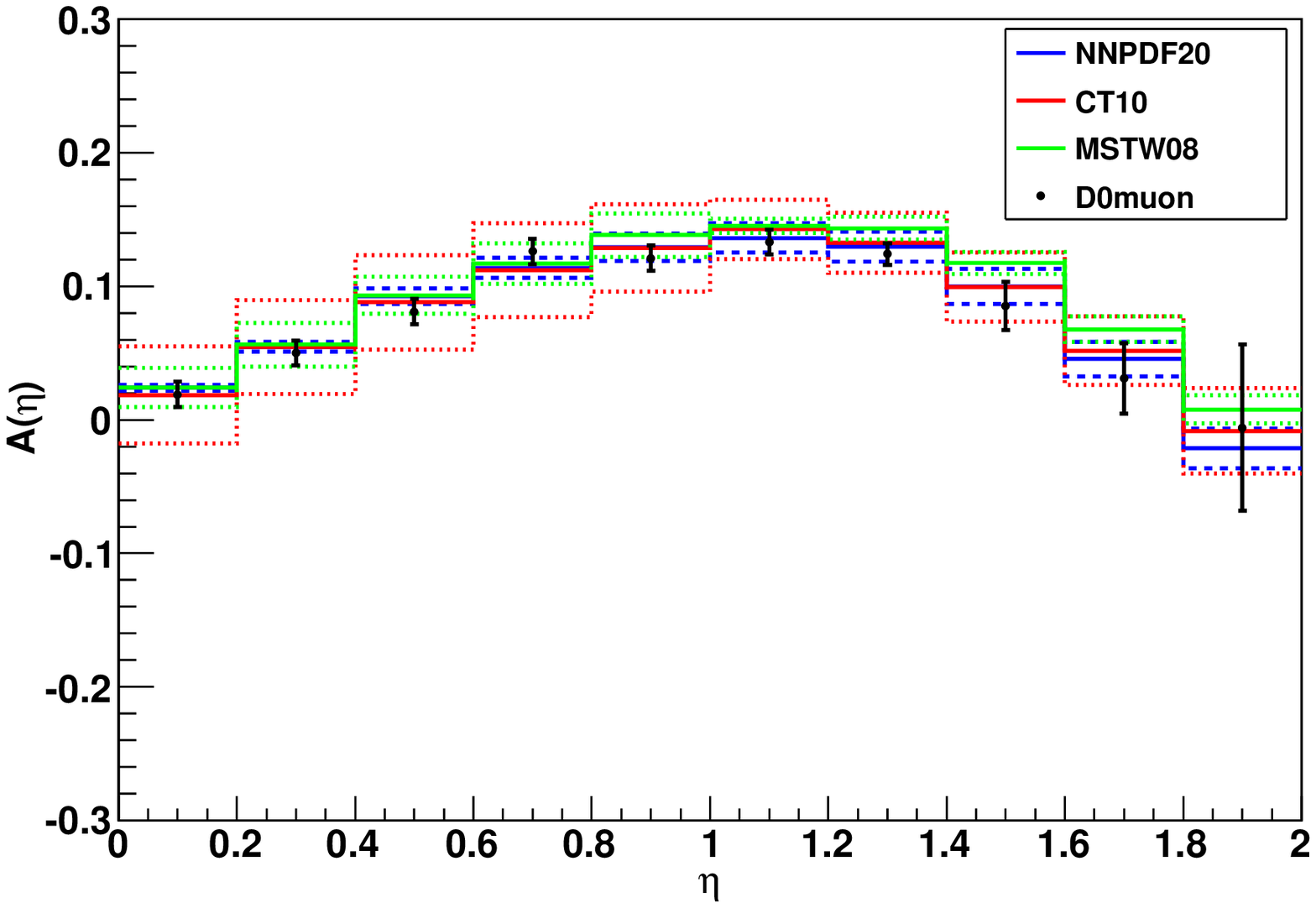}
    \includegraphics[width=0.45\textwidth]{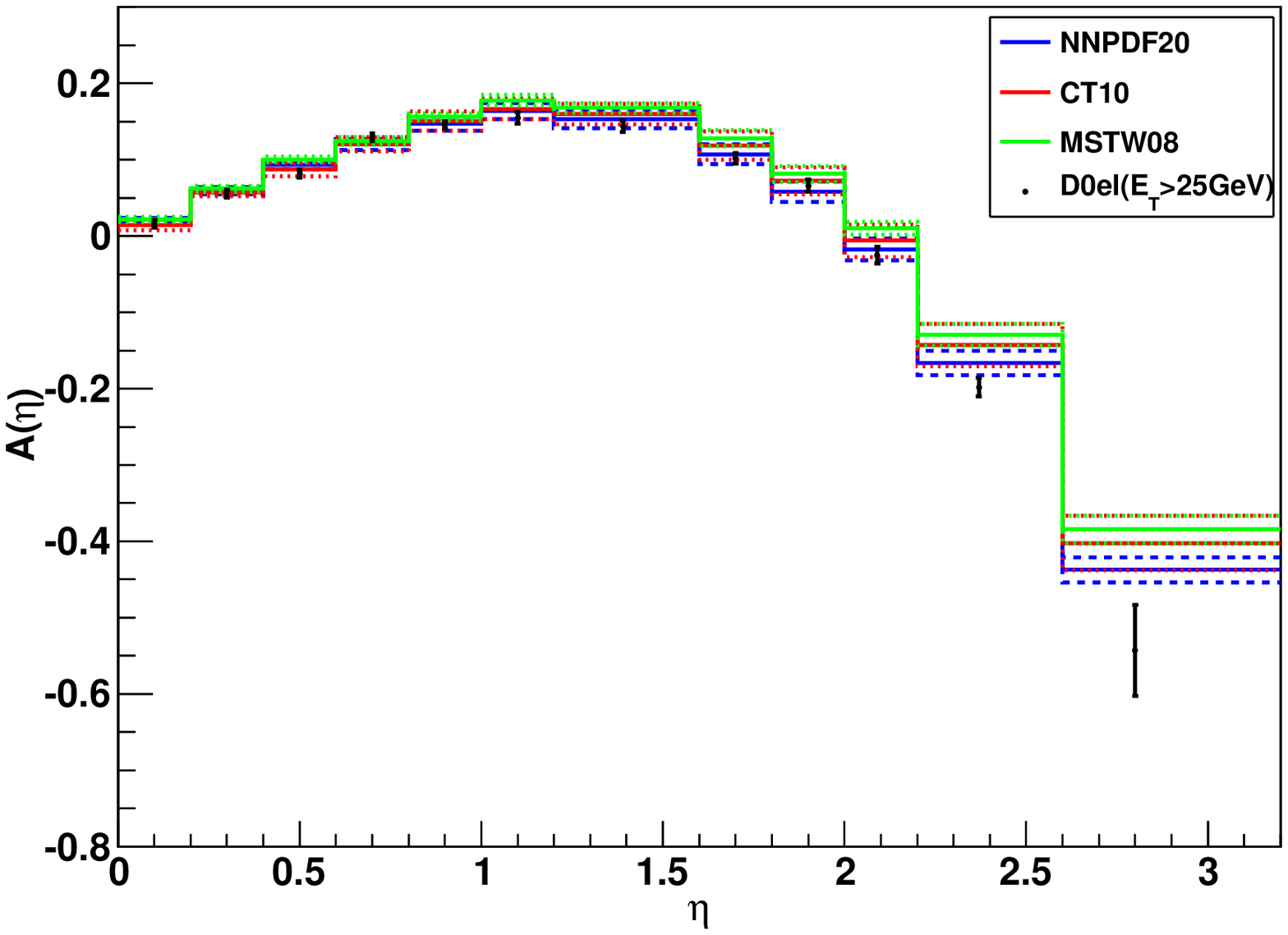}
    \includegraphics[width=0.45\textwidth]{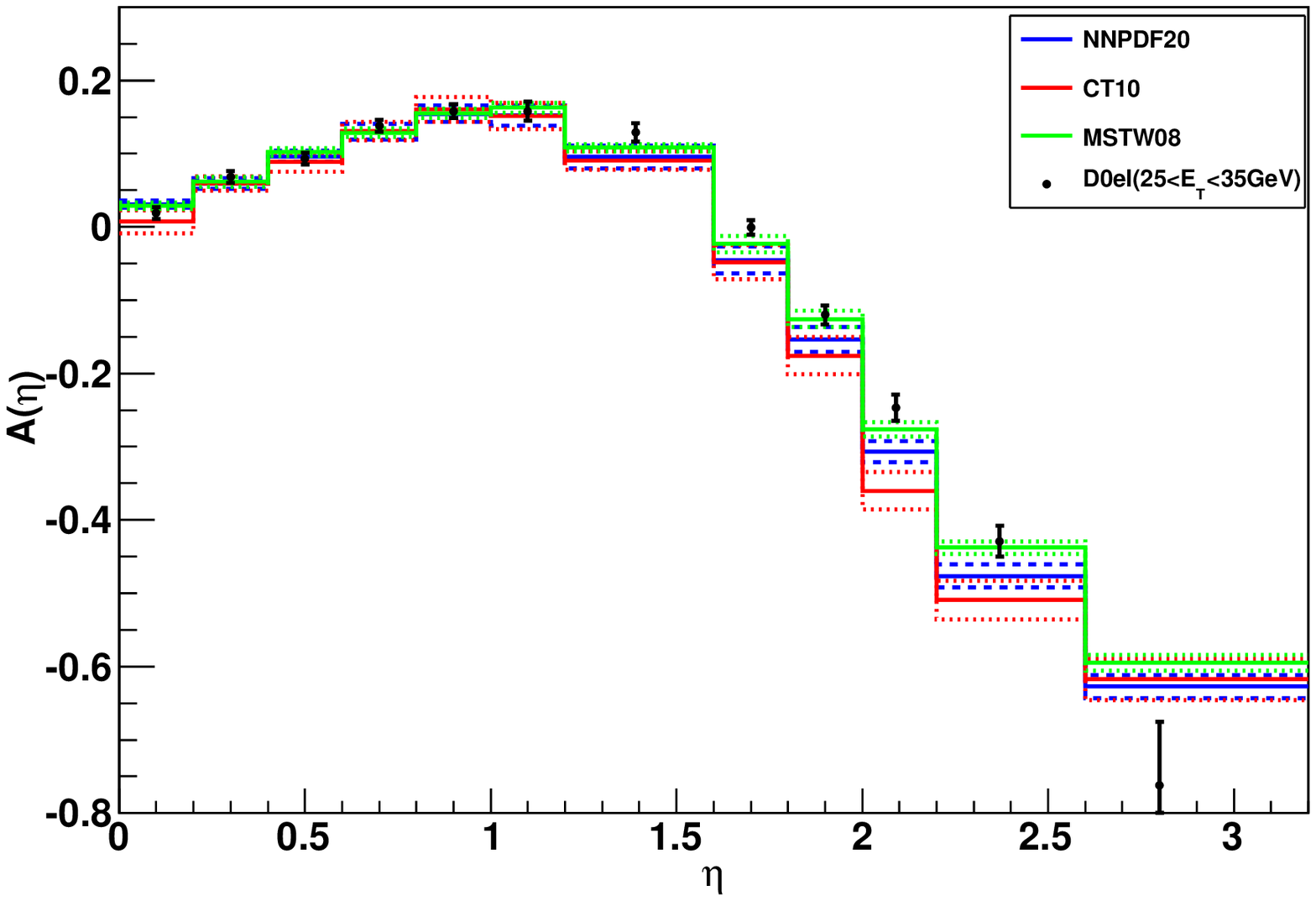}
    \includegraphics[width=0.45\textwidth]{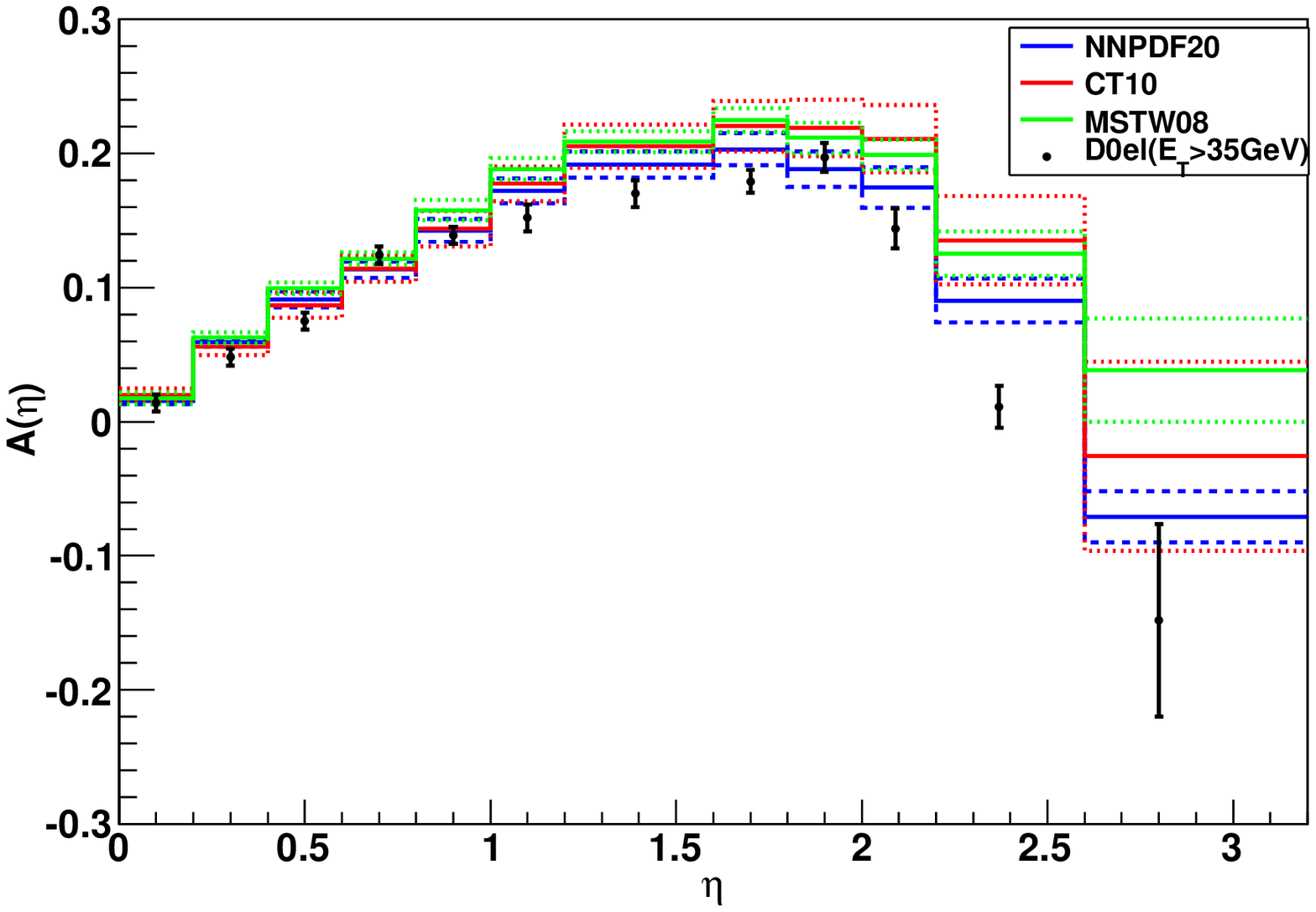}
  \end{center}
  \caption{\small Predictions for the D0 $W$ lepton charge asymmetry obtained 
    with the DYNNLO code at next-to-leading order, using the NNPDF2.0~\cite{nnpdf20}, 
    CT10~\cite{CT10} and MSTW08~\cite{MSTW08} parton sets. 
    We show results for the muon charge asymmetry (top left),  
    and the electron charge asymmetry in the inclusive 
    bin, $E^e_T>25$ GeV, binA(top right), 
    and then in less inclusive bins, 25 GeV $<E_T^e<35$ GeV, bin B (bottom left), 
    and $E^e_T>35$ GeV, 
    binC (bottom right). 
    \label{fig:predictionmu} }
\end{figure}

Here we will consider the electron and muon asymmetry measurements performed 
by the D0 collaboration at Run II of the Tevatron and published in 
Refs.~\cite{d0electron,d0muon}. The more recent D0 muon analysis of 
Ref.~\cite{d0mnew} has not been included since the data are still 
preliminary. The datasets included in our analysis are the same as those 
included in the dedicated CT10W analysis~\cite{CT10}.
The lepton asymmetry measurements from CDF~\cite{cdfelectron} are not considered 
here since the direct CDF $W$ asymmetry data is already included in the 
NNPDF2.0 fit.

Let us discuss in more detail the lepton asymmetry data that we consider.
In Ref.~\cite{d0muon} a measurement of the muon charge asymmetry based on
0.3 fb$^{-1}$ of data is presented. 
The asymmetry measurement is binned in ten bins in the muon pseudo-rapidity 
in the range $|\eta_{\mu}|<2$, and cuts are imposed on the transverse energy and 
mass of the muon: $E^\mu_T>20$ GeV and $M_T > 40$ GeV. 
In Ref.~\cite{d0electron} a similar measurement of the electron charge 
asymmetry is presented based on 0.75 fb$^{-1}$ of data. 
The asymmetry is binned in twelve bins in the electron pseudo-rapidity 
in the range $|\eta_e|<3.2$, and  cuts are imposed on the missing energy and 
transverse mass: $\Eslash>25$ GeV and $M_T > 50$ GeV. 
Three sets of measurements are then given, which have different cuts in 
the transverse energy of the electron: an 'inclusive' bin which has $E^e_T>25$ 
GeV (which we refer to here as bin A), and two less inclusive bins with 
more restrictive cuts on the transverse energy, $25~{\rm GeV} <E^e_T<35~{\rm GeV}$ 
(bin B) and $E^e_T>35$ GeV (bin C). Note that bins B and C together cover the 
same kinematic range as the more inclusive bin A.
 
To analyse these data using the reweighting technique we use the DYNNLO 
code~\cite{dynnlo} to compute the theoretical predictions for the lepton 
asymmetries at NLO, using NNPDF2.0 as input parton densities. 
This code is a parton level Monte Carlo program designed to compute 
exclusive hadronic processes up to NNLO, and it enables the user to 
implement the same cuts used in the experimental analyses.

Before considering reweighting, let us first compare in Fig.~\ref{fig:predictionmu} 
the predictions obtained with DYNNLO and different PDF sets at NLO, for the
various D0 datasets.
It is perhaps surprising that, even though none of these data are included 
into the NNPDF2.0 fit, the prediction obtained from the NNPDF2.0 is 
in general closer to the experimental data than the predictions obtained with 
the other parton sets: the reason can be traced back to the somewhat larger $d/u$ 
ratio in the range  $0.2 \lsim x \lsim 0.6$ (Fig.~\ref{fig:d-over-u-init}) for the
NNPDF2.0 set. The exception is bin B, for which MSTW08 provides the best description.

\begin{table}[t]
\begin{center}
\scriptsize
\begin{tabular}{|l|c|c|c|c|}
\hline
Set & $N_{\dat}$ & NNPDF2.0 &  MSTW08 & CT10 \\ 
\hline
\hline
D0 $\mu$   ($E_T^\mu>20$ GeV) ~\cite{d0muon}    & 10 & 0.62  & 1.51 & 0.70 \\ 
\hline
D0 electron, $E_T^e>25$ GeV (bin A) ~\cite{d0electron} & 12 & 2.12 & 9.20 & 4.07 \\ 
D0 electron, $25$ GeV$<E_T^e<35$ GeV (bin B)~\cite{d0electron}& 12 & 4.75 & 1.66 & 9.48 \\ 
D0 electron, $E_T^e>35$ GeV (bin C)    ~\cite{d0electron}     & 12 & 5.06 & 13.4 & 11.7 \\
\hline
\end{tabular} 
\end{center}
\caption{\small The D0 $W$ lepton charge asymmetry datasets that are included in 
  the present analysis, together with the $\chi^2$ per data point obtained 
  from the NLO 
  predictions of various PDFs sets. The electron data of Ref.\cite{d0electron} 
  is divided into three
  bins that we denote by bin A, bin B and bin C. 
\label{tab:chi2comptab}}
\end{table}

The quality of the comparison of various PDF sets with the asymmetry data 
can be quantified by evaluating the $\chi^2$ to each data set. 
For all the Tevatron Run II D0 lepton asymmetry only the statistical and uncorrelated
systematic errors are quoted. 
The covariance matrix is therefore diagonal and its elements are given by the sum 
in quadrature of the statistical and the uncorrelated systematic errors. There is 
no normalization uncertainty since the asymmetry is a ratio of cross-sections. 

The value of the $\chi^2$ per data point and the number of data points 
for each set considered in the present analysis are shown in 
Table~\ref{tab:chi2comptab}.
The results confirm the studies performed in Ref.~\cite{Catani:2010en}. 
In particular, the less inclusive data (bins B and C) are rather poorly 
described by all the current
PDF fits, with the exception of bin B which MSTW08 describes reasonably 
well (though at the cost of a very bad fit to bins A and C).
Note however that 
Ref.~\cite{CT10} uses the RESBOS program~\cite{Balazs:1997xd} to compute the 
predictions for the $W$ lepton asymmetry. RESBOS computes on top of the NLO higher 
order corrections from $p_T$ resummation. The differences between NLO and RESBOS 
are maximal in the kinematics of the electron bin B data. This differences might 
explain, at least in part, the values of the $\chi^2$ for CT10
obtained in Table~\ref{tab:chi2comptab} compared to those given in 
Ref.~\cite{CT10}.

We now consider the effect of including the Run II D0 muon and electron asymmetry 
data in the NNPDF2.0 analysis using reweighting. 
We will consider each dataset in turn, concentrating first on the inclusive sets (muon and electron bin A),  
and turning later to the less inclusive data sets (bins B and C). 
For each case we will proceed as follows: first we provide the distribution 
of $\chi^2_k$ before and after reweighting, the probability distribution 
$P\lp \alpha_s\rp$ and the distribution of weights. We then compare the 
reweighted PDFs to experimental data. 
Finally we compute the distances between the original and the reweighted sets, 
and compare the corresponding PDFs where they differ substantially from the 
original ones.

Unless otherwise stated, PDFs and their uncertainties will be plotted 
at the scale $Q^2=Q_0^2=2$ GeV$^2$. 
The $N_{\rm rep}=1000$ NNPDF2.0 set is used throughout, with the
exception of the computation of distances, where we instead use sets of 
100 replicas. 


\subsection{Inclusive data}

\begin{figure}[t]
  \begin{center}
    \includegraphics[width=0.45\textwidth]{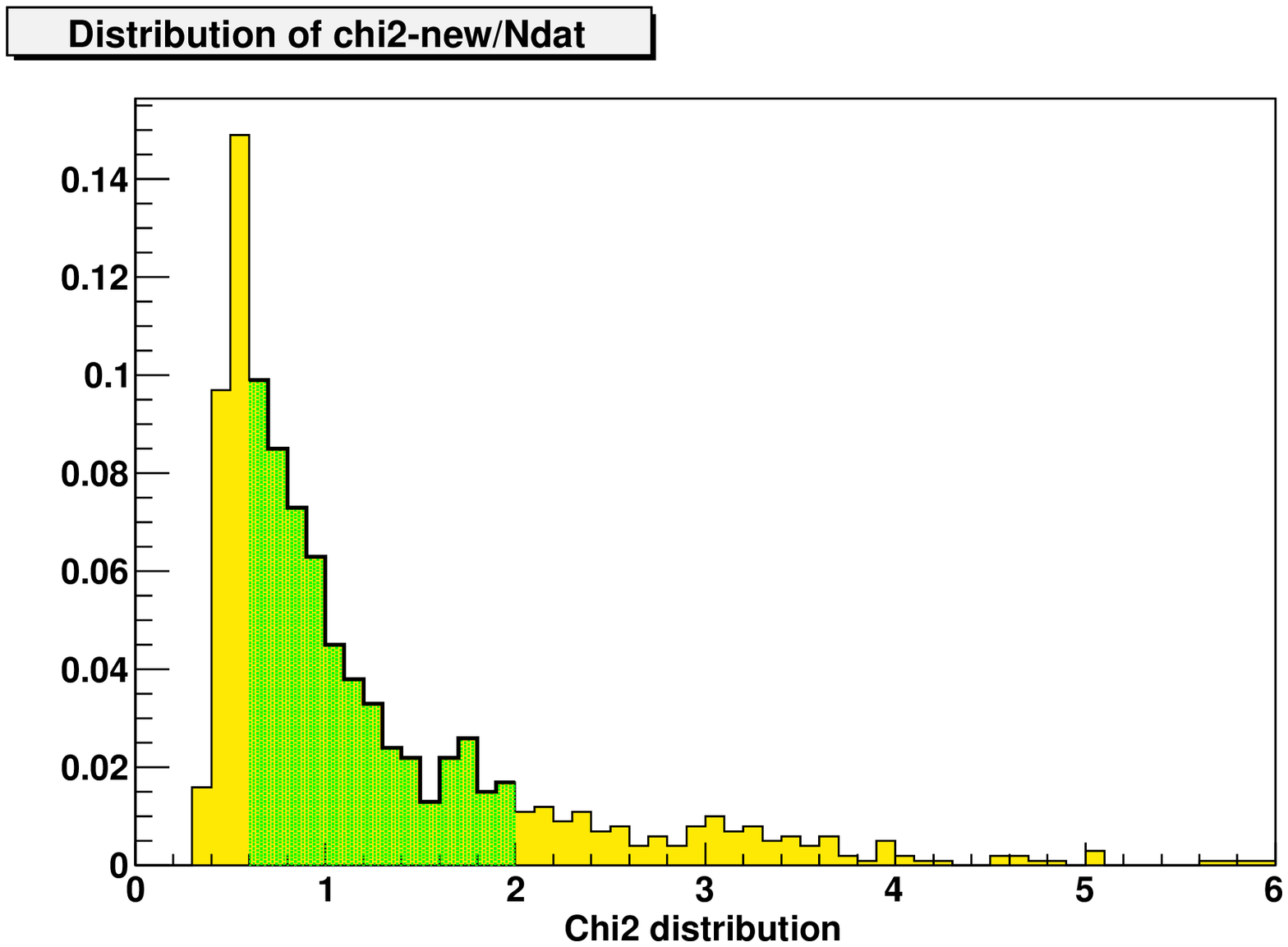}
    \includegraphics[width=0.45\textwidth]{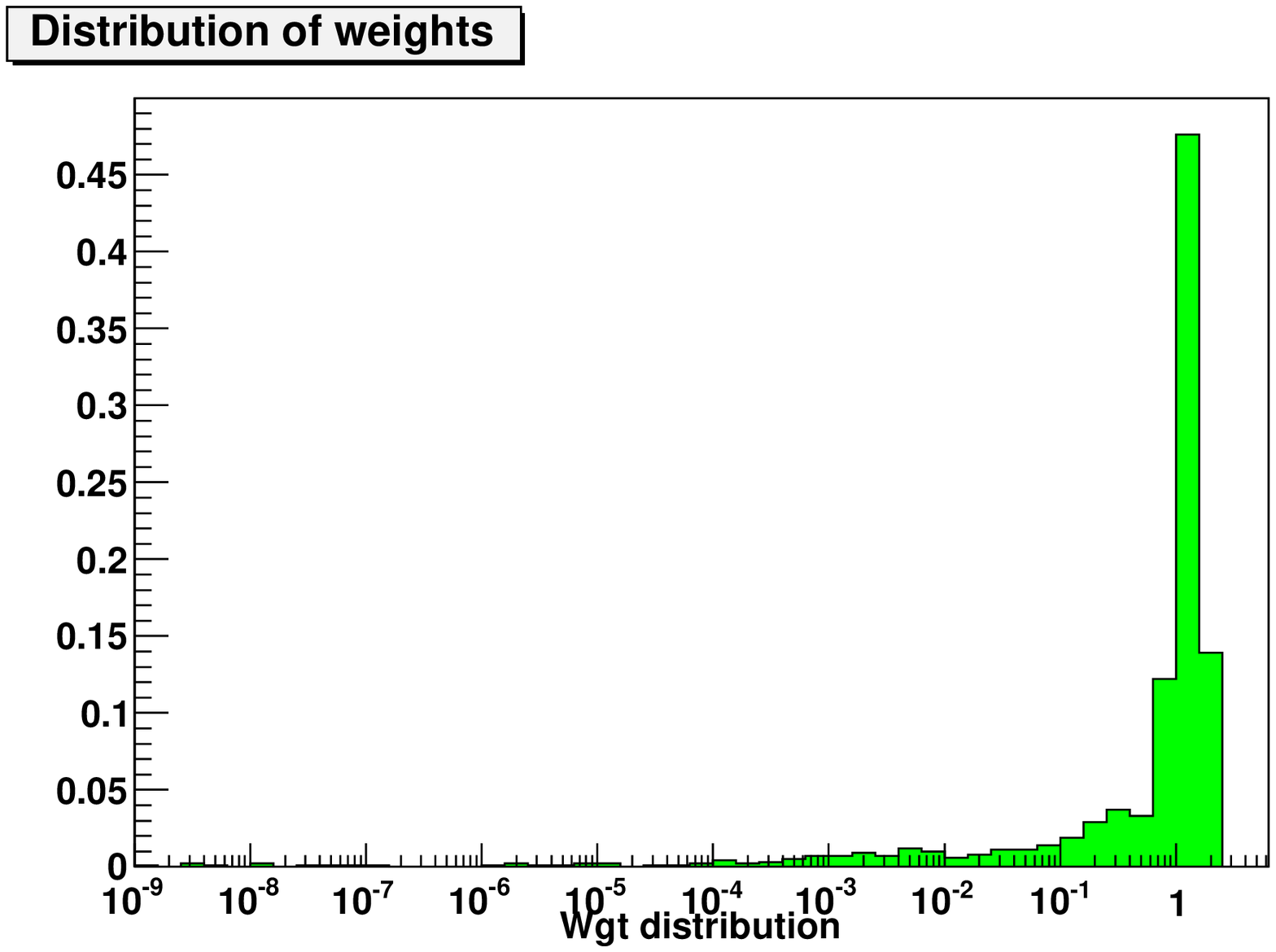}\\
    \includegraphics[width=0.45\textwidth]{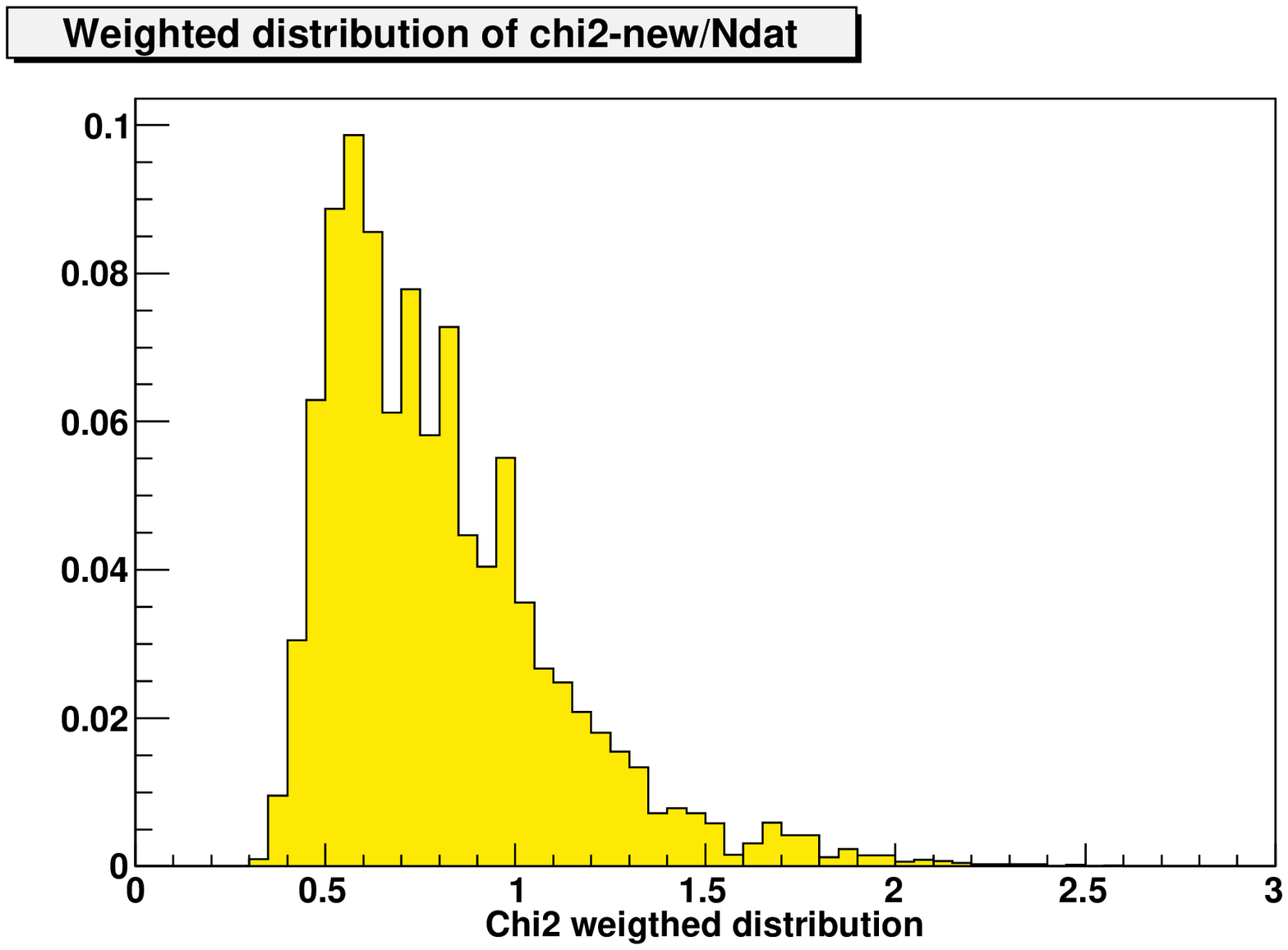}
    \includegraphics[width=0.45\textwidth]{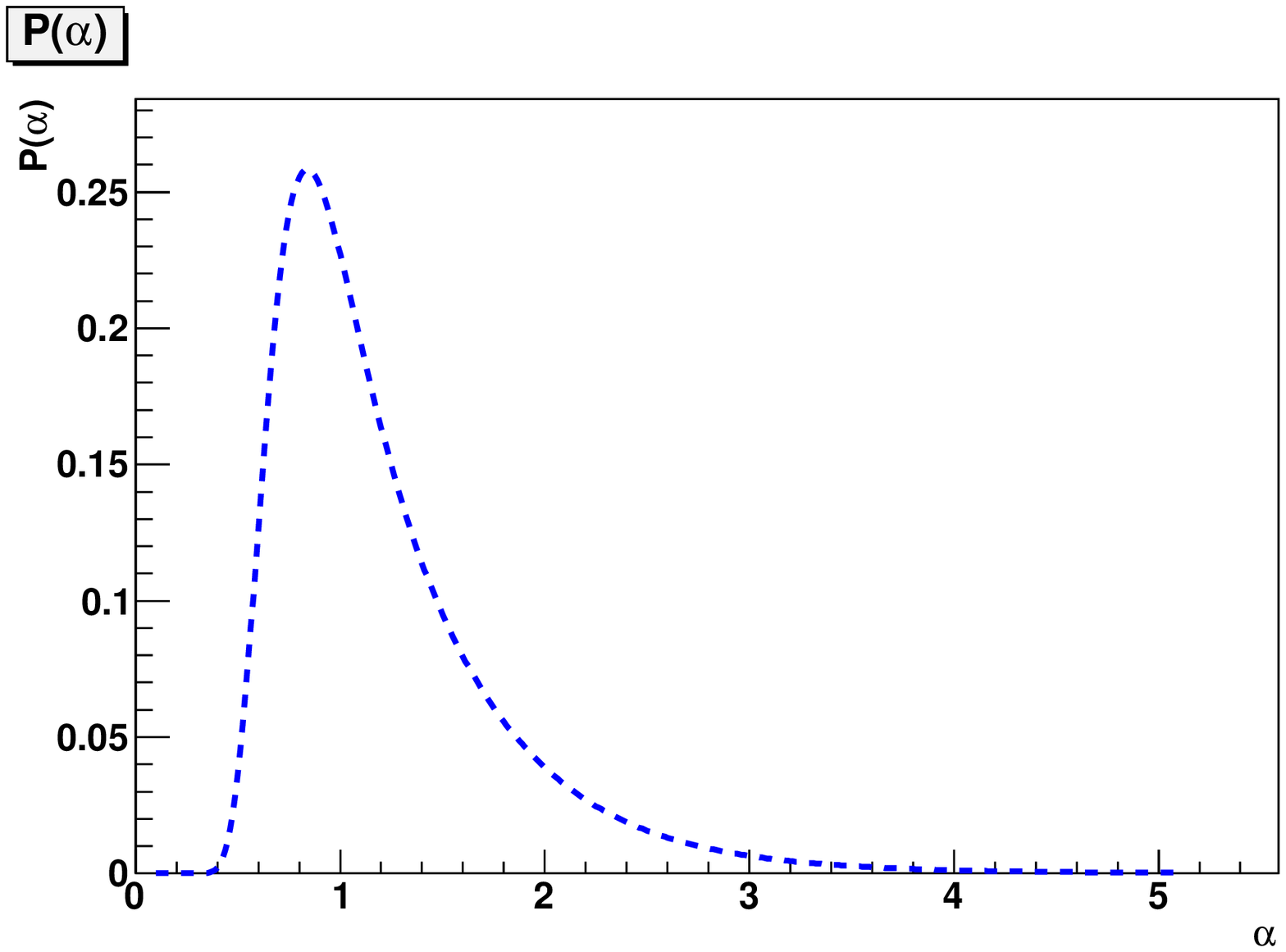}
  \end{center}
  \caption{\small Distribution of the $\chi^2_k$ and the weights $w_k$, the 
    reweighted $\chi^2$-distribution and the probability distribution 
    ${\mathcal P}(\alpha)$ in the reweighting of the NNPDF2.0 PDF set
    using the D0 muon asymmetry data~\cite{d0muon}. \label{fig:chi2-muon} }
\end{figure}

Let us first consider the inclusion of the D0 muon charge asymmetry data~\cite{d0muon}.
The distribution of the $\chi^{2}_k$ and corresponding weights $w_k$ for these 
data is shown in the upper plots of Fig.~\ref{fig:chi2-muon}.
Since the $\chi^2$-distribution is peaked close to one, the weights are also mostly 
of order unity. The reweighted $\chi^2$, and probability density for the rescaling 
parameter $\alpha$ are shown in the lower plots: they peak a little below one, 
suggesting that the errors on these data are actually likely to have been
overestimated by D0. 
After reweighting the $\chi^2$ per data point drops from $0.62$ to $0.51$, and 
the number of effective replicas is $N_{\rm eff}=795$.

\begin{figure}[ht]
 \begin{center}
   \includegraphics[width=0.45\textwidth]{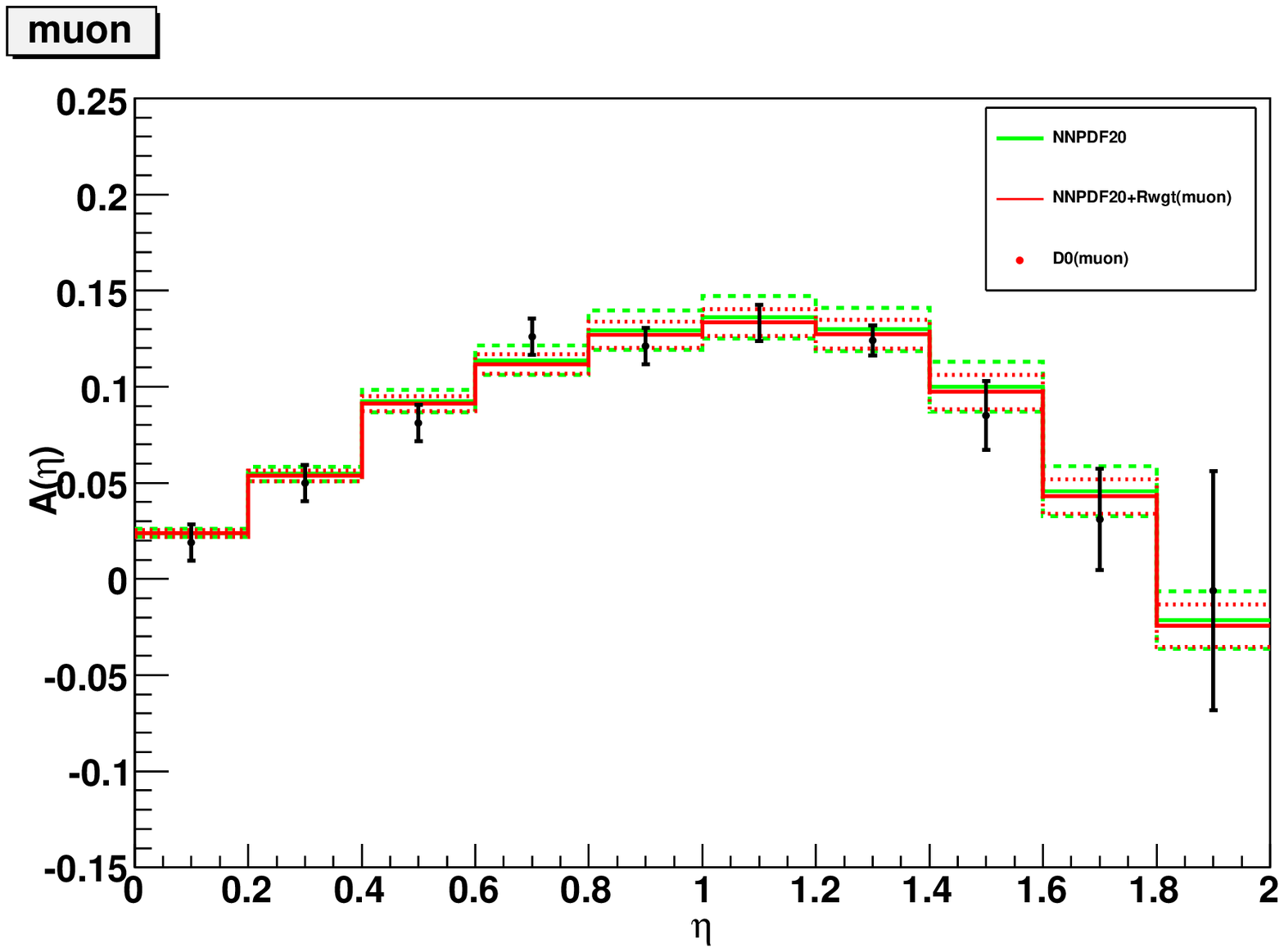}
   \includegraphics[width=0.45\textwidth]{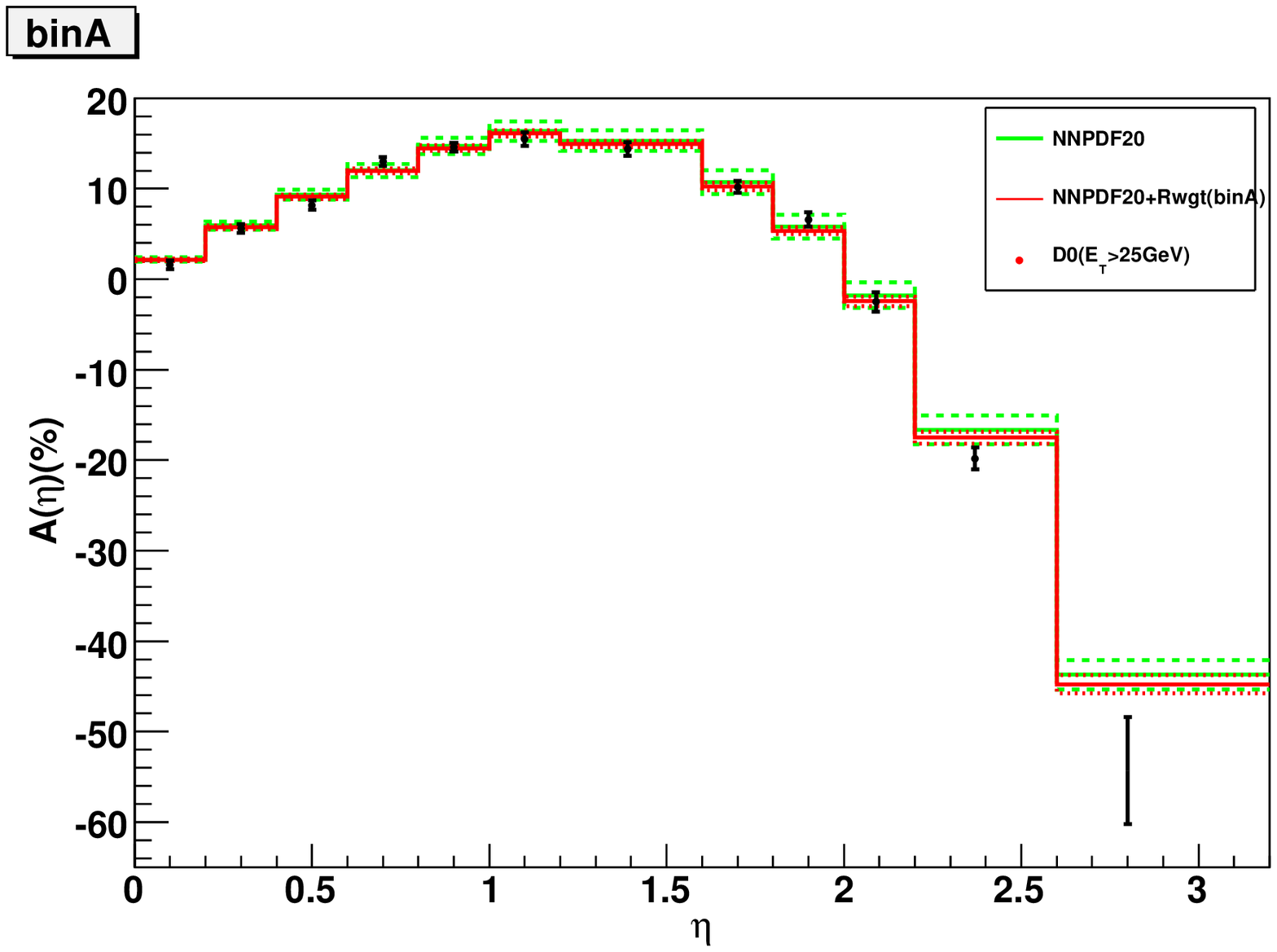}
 \end{center}
 \caption{\small Left: $W$ muon charge asymmetry computed for the NNPDF2.0 PDFs
   before and after the reweighting of these data into the parton analysis.
   Right: $W$ electron charge asymmetry (inclusive bin) computed with the NNPDF2.0 
   PDFs before and after the reweighting of these data in the parton analysis. 
   \label{fig:obs-incl}}
\end{figure}
\begin{figure}[h!]
  \begin{center}
    \includegraphics[width=0.45\textwidth]{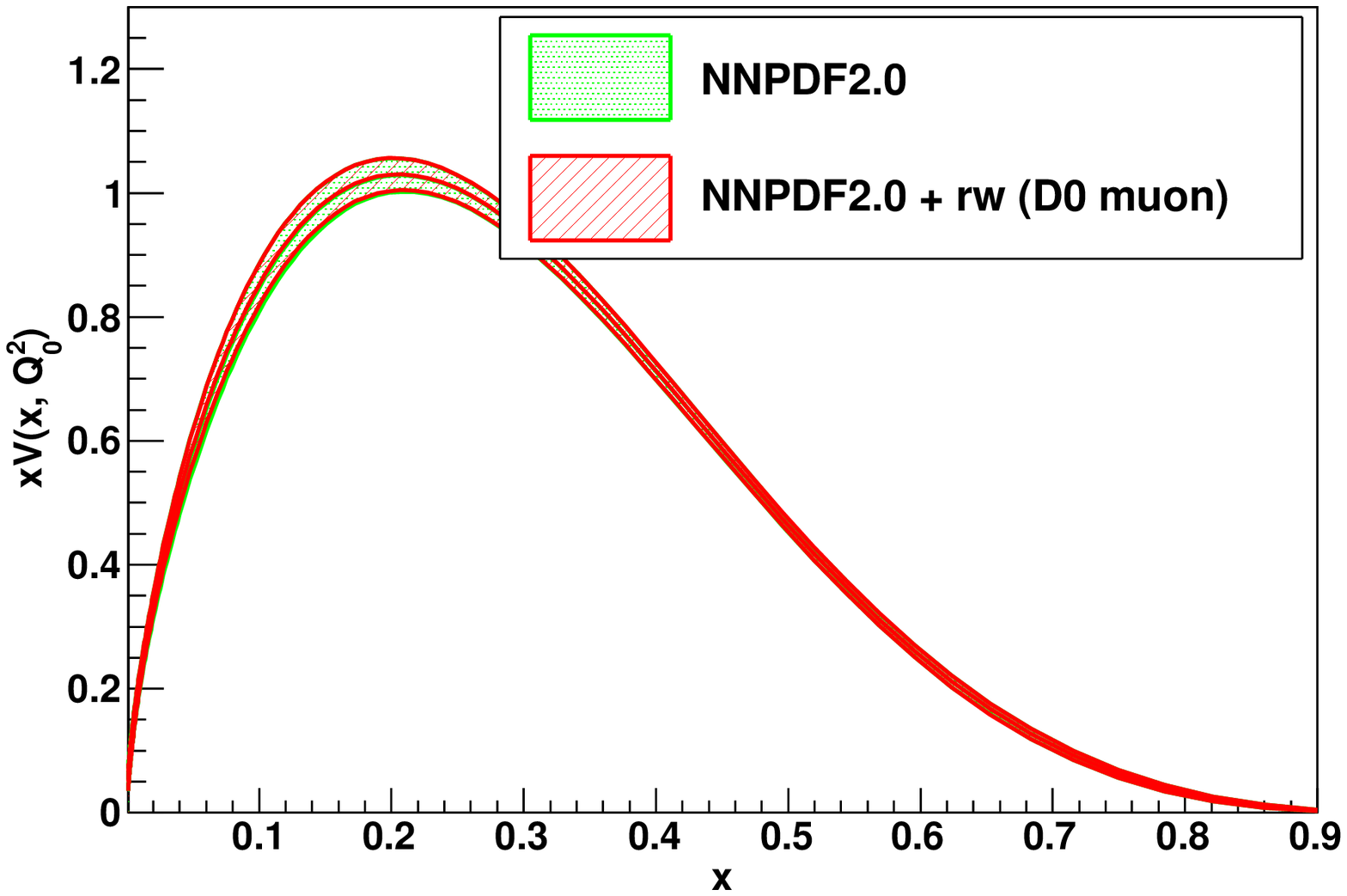}
    \includegraphics[width=0.45\textwidth]{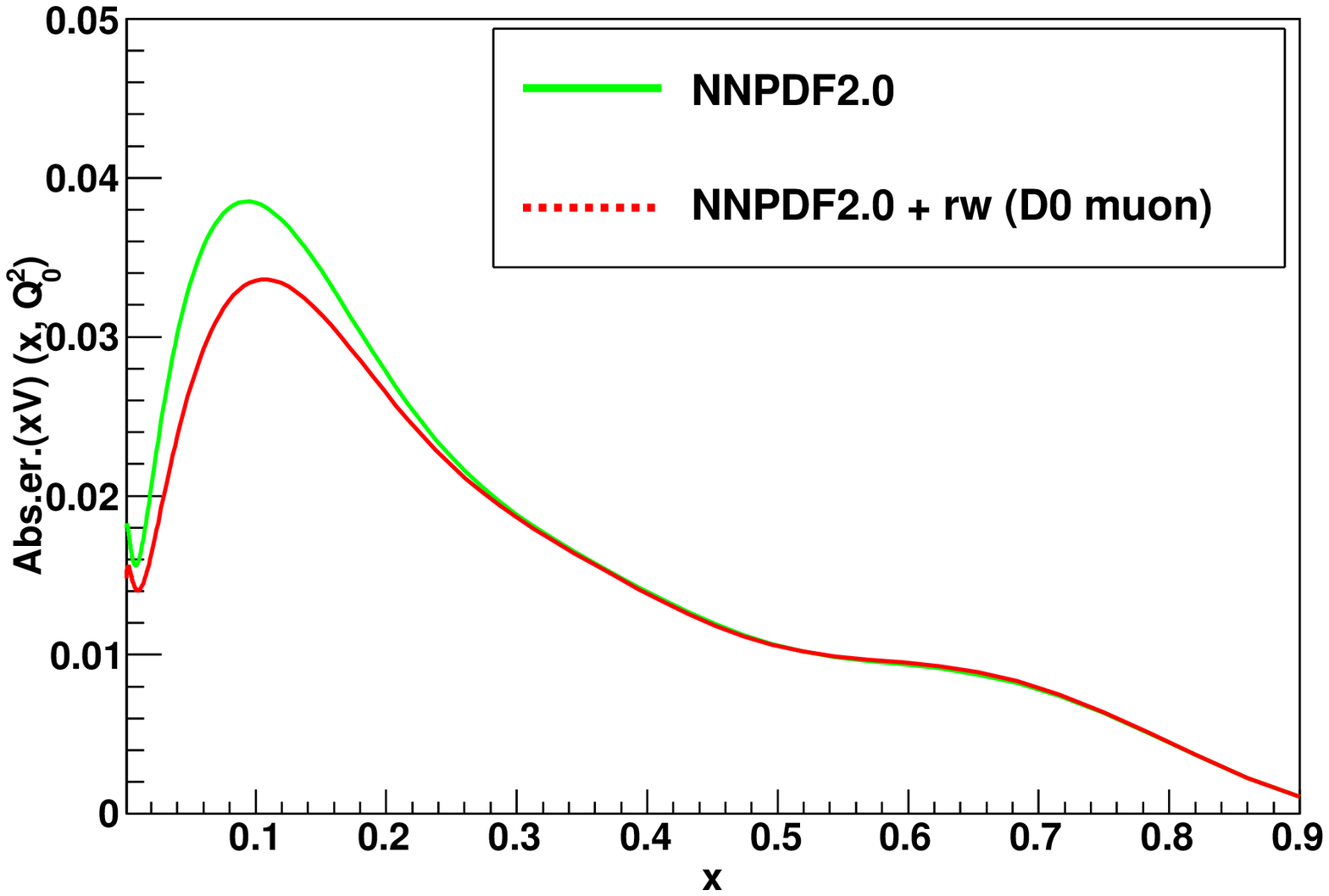}
  \end{center}
  \caption{\small Total valence PDF for the NNPDF2.0 and NNPDF2.0 + D0 muon data PDF sets.
    \label{fig:valence-muon}}
  \vskip 1.0cm
  \begin{center}
    \includegraphics[width=0.9\textwidth]{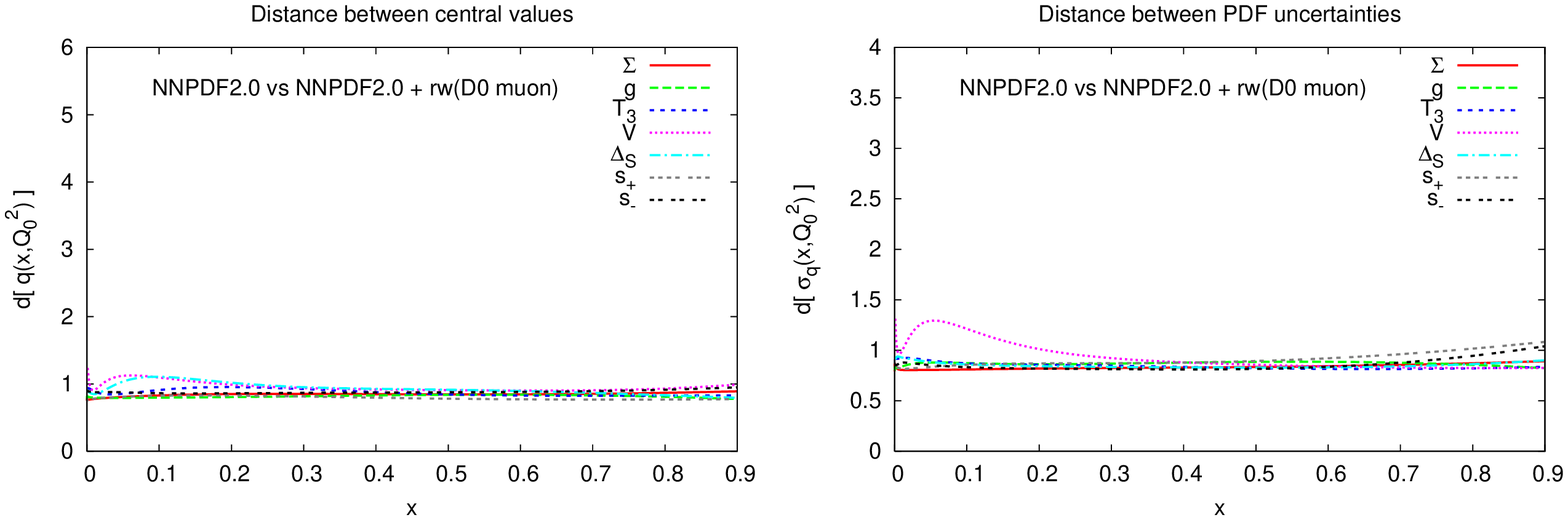}\\
  \end{center}
  \caption{\small Distances between NNPDF2.0 and NNPDF2.0 + D0 W lepton asymmetry 
    measurements from the muon dataset. The NNPDF2.0 set with $N_{\rep}=100$ has 
    been used in the computation of the distances.
    \label{fig:distances-muon}}
\end{figure}

On the left in Fig.~\ref{fig:obs-incl} we show the muon charge asymmetry before and 
after the reweighting. Indeed the predictions get closer to the data, once the 
PDFs are reweighted. We have also examined the effect on the shape of the PDFs, 
but the effects are negligible apart from a slight reduction in the uncertainty 
of the total valence distribution, shown in Fig.~\ref{fig:valence-muon}. 
This is confirmed by the distance analysis, Fig.~\ref{fig:distances-muon}, that 
shows that central values and uncertainties for all PDFs are essentially unchanged, 
with the exception of the total valence PDF where the inclusion of muon data
has a moderate effect.

\begin{table}[t]
  \begin{center}
    {\footnotesize
      \begin{tabular}{|c|c|c|c|c|c|c|}
        \hline
        Set & $\chi^2_{\rm 2.0}$ & $\chi^2_{\rm 2.0+\mu}$ & $\chi^2_{\rm 2.0+binA}$ 
        & $\chi^2_{\rm 2.0+binA+\mu}$  & $\chi^2_{\rm 2.0+binB}$ 
        & $\chi^2_{\rm 2.0+binC}$  \\ 
        \hline
        NMC-pd          &    0.99  &   0.98 &  0.98 & 0.98  & 0.97 & {\bf 1.13}\\
        NMC             &    1.72  &   1.72 &  1.69 & 1.70  & 1.72 & 1.72\\
        SLACp           &    1.55  &   1.55 &  1.53 & 1.54  & 1.50 & {\bf 1.63}\\
        SLACd           &    1.12  &   1.12 &  1.07 & 1.09  & 1.05 & {\bf 1.24}\\
        BCDMSp          &    1.35  &   1.35 &  1.33 & 1.34  & {\bf 1.41} & 1.35\\
        BCDMSd          &    1.16  &   1.16 &  1.16 & 1.16  & {\bf 1.24} & 1.14\\
        HERA1-NCep      &    1.35  &   1.35 &  1.34 & 1.34  & 1.33 & 1.35\\
        HERA1-NCem      &    0.86  &   0.86 &  0.86 & 0.86  & 0.86 & 0.86\\
        HERA1-CCep      &    0.96  &   0.96 &  0.94 & 0.94  & 1.02 & 0.92\\
        HERA1-CCem      &    0.56  &   0.56 &  0.56 & 0.56  & 0.56 & 0.57\\
        CHORUSnu        &    1.08  &   1.08 &  1.08 & 1.08  & 1.11 & 1.10\\
        CHORUSnb        &    0.86  &   0.86 &  0.86 & 0.86  & 0.87 & 0.90\\
        FLH108          &    1.50  &   1.50 &  1.50 & 1.50  & 1.47 & 1.50\\
        NTVnuDMN        &    0.69  &   0.66 &  0.67 & 0.65  & 0.82 & 0.60\\
        NTVnbDMN        &    0.70  &   0.70 &  0.69 & 0.69  & 0.72 & 0.81\\
        Z06NC           &    1.24  &   1.24 &  1.24 & 1.24  & 1.23 & 1.26\\
        Z06CC           &    1.19  &   1.19 &  1.19 & 1.19  & 1.15 & 1.21\\
        DYE605          &    0.86  &   0.86 &  0.84 & 0.85  & 0.87 & 0.85\\
        DYE886p         &    1.31  &   1.32 &  1.29 & 1.30  & 1.28 & {\bf 1.36}\\
        DYE886r         &    0.83  &   0.79 &  {\bf 0.67} & { 0.71}  & {\bf 1.08} & 0.72\\
        CDFWASY         &    1.88  &   1.88 &  {\bf 1.78} & 1.82  & {\bf 2.05} & {\bf 1.60}\\
        CDFZRAP         &    1.74  &   1.77 &  1.75 & 1.77  & {\bf 1.37} & {\bf 1.97}\\
        D0ZRAP          &    0.59  &   0.59 &  0.59 & 0.59  & 0.60 & 0.61\\
        CDFR2KT         &    1.02  &   1.02 &  0.95 & 0.97  & {\bf 1.21} & 0.93 \\
        D0R2CON         &    0.86  &   0.86 &  0.84 & 0.84  & {\bf 0.91} & 0.84\\
        \hline
        TOTAL           &    1.14   &  1.14 &  1.13 & 1.13 & {\bf 1.16} & {\bf 1.16}\\
        \hline
      \end{tabular}
    }
  \end{center}
  \caption{\small $\chi^2$ per data point  of all the experiments included in 
    the NNPDF2.0 fit evaluated before and after reweighting with the various 
    lepton asymmetry data sets. Note that here we use the $t_0$ covariance matrix 
    in the evaluation of the $\chi^2$: the numbers are thus slightly different 
    from those shown in Ref.~\cite{nnpdf20}. The cases in which the $\chi^2$ 
    varies significatively as compared to the reference are highlighted in 
    boldface.
    \label{tab:chi2-20y}}
\end{table}

To study the compatibility of these data with the data included in 
the NNPDF2.0 analysis, in Tab.~\ref{tab:chi2-20y} we show the $\chi^2$ of each of 
the datasets included in the NNPDF2.0 analysis evaluated with the original 
NNPDF2.0 PDFs and then with these PDFs reweighted by the inclusion of 
the D0 muon asymmetry data. If anything, there is a slight improvement in 
the description of most of the datasets.
To summarize, the D0 muon asymmetry data~\cite{d0muon} are perfectly 
consistent with NNPDF2.0, but are not sufficiently precise to add much information 
to the PDFs.
It will be interesting to assess the impact of the higher statistics
D0 Run II muon data set~\cite{d0mnew} once the analysis is completed.

\begin{figure}[ht]
  \begin{center}
    \includegraphics[width=0.45\textwidth]{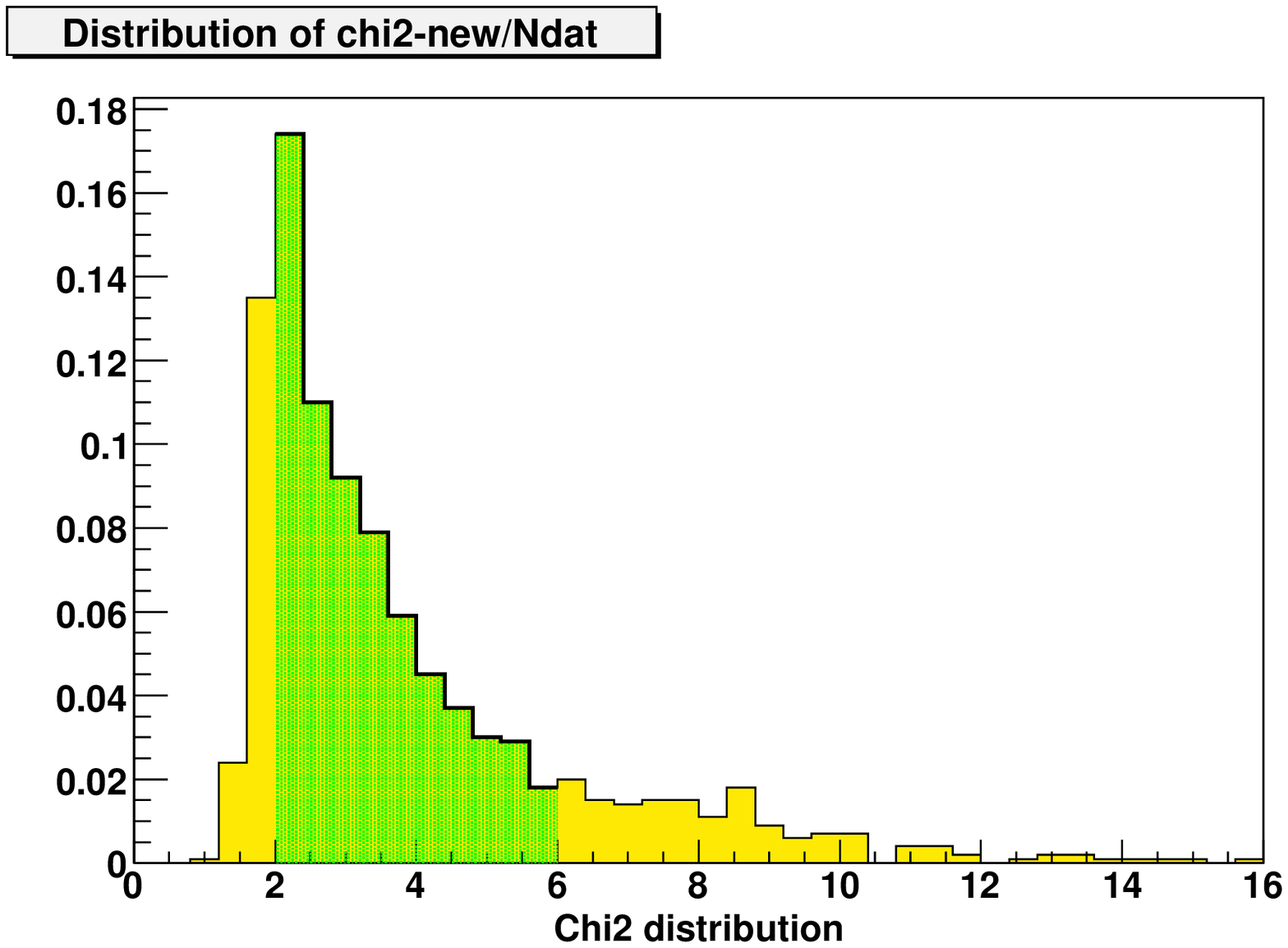}
    \includegraphics[width=0.45\textwidth]{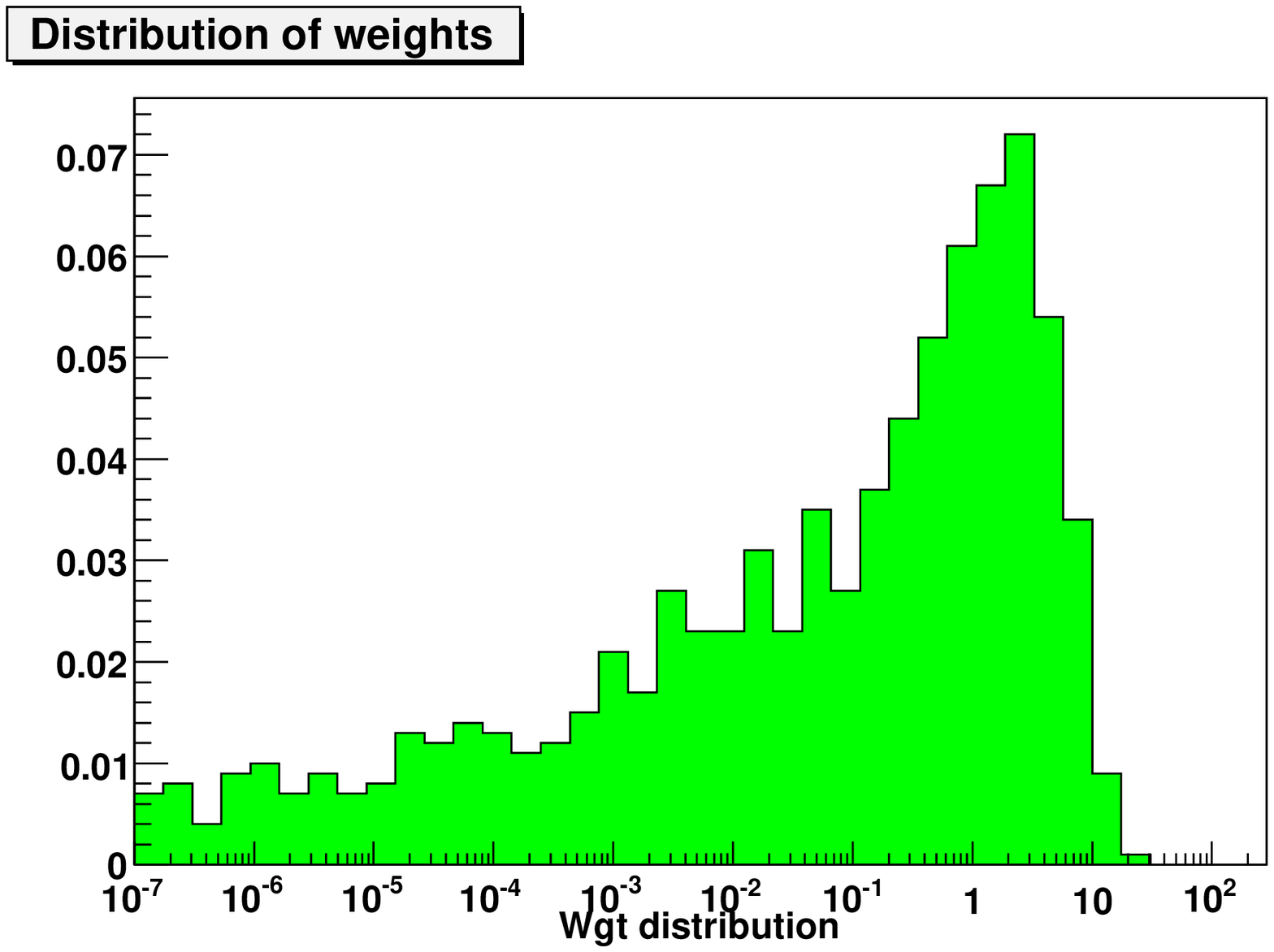}\\
    \includegraphics[width=0.45\textwidth]{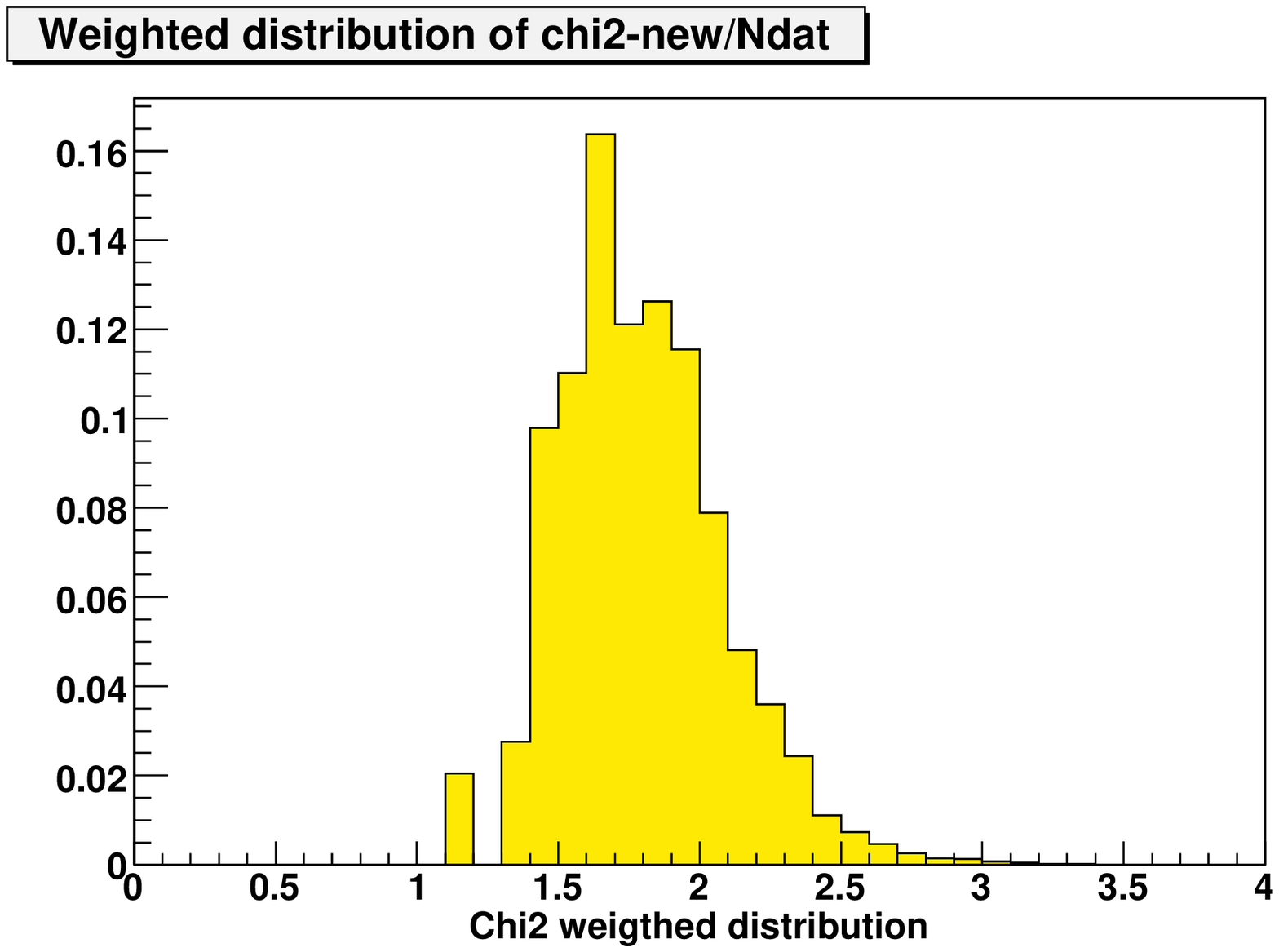}
    \includegraphics[width=0.45\textwidth]{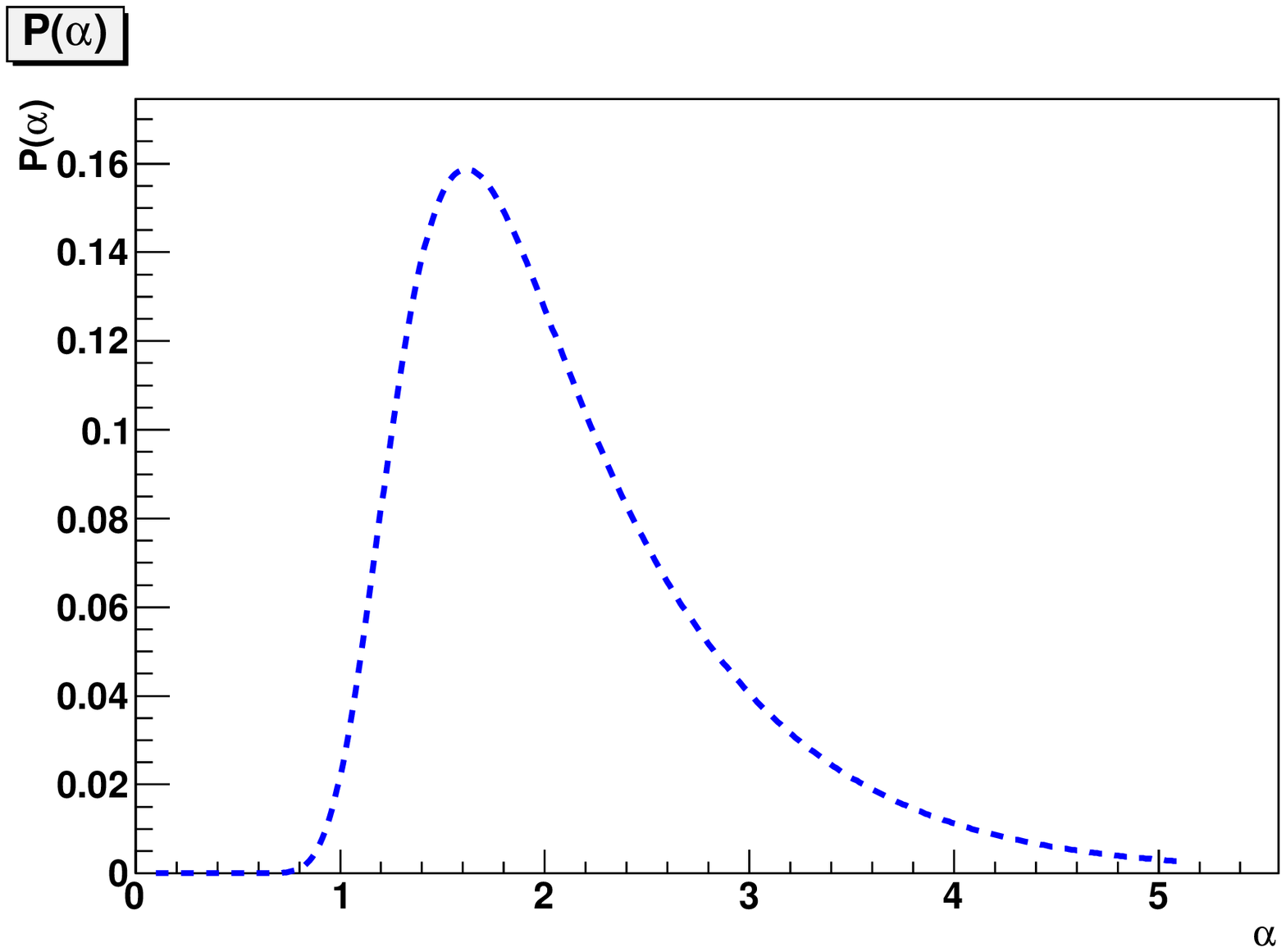}
  \end{center}
  \caption{\small Distribution of the $\chi^2_k$ and the weights $w_k$, the 
    reweighted $\chi^2$-distribution and the probability distribution 
    ${\mathcal P}(\alpha)$ in the reweighting of the NNPDF2.0 PDF set
    using the D0 electron asymmetry data (bin A)~\cite{d0electron}.
    \label{fig:chi2-binA}}
\end{figure}

Next we consider the inclusive D0 electron data (bin A) with $E^e_T > 25$ GeV. 
The results are shown in Fig.~\ref{fig:chi2-binA}. 
Once included in the fit through reweighting the $\chi^2$ for this set drops from 
2.12 to 1.55.  
While the distribution of the unweighted $\chi^2_k$ is peaked above two and 
has a long tail to higher values, after reweighting the peak is shifted 
much closer to one.
This is  achieved through a substantial reduction in the effective number of 
replicas: after reweighting $N_{\rm eff} = 262$. However, while before reweighting 
only $16\%$ of replicas lie in the region $\half<\chi^2<2$, after reweighting 
this figure rises to $78\%$. 
This behaviour is confirmed by the plot of $\mathcal{P}(\alpha)$: the data indicate 
that the most probable value of $\alpha$ is around $1.6$, indicating that 
experimental errors on these data are underestimated. 
Taken together, this shows that these data might have a significant effect on 
constraining the PDFs, while still being broadly consistent with all the other 
data included in the fit.

\begin{figure}[t]
  \begin{center}
    \includegraphics[width=0.45\textwidth]{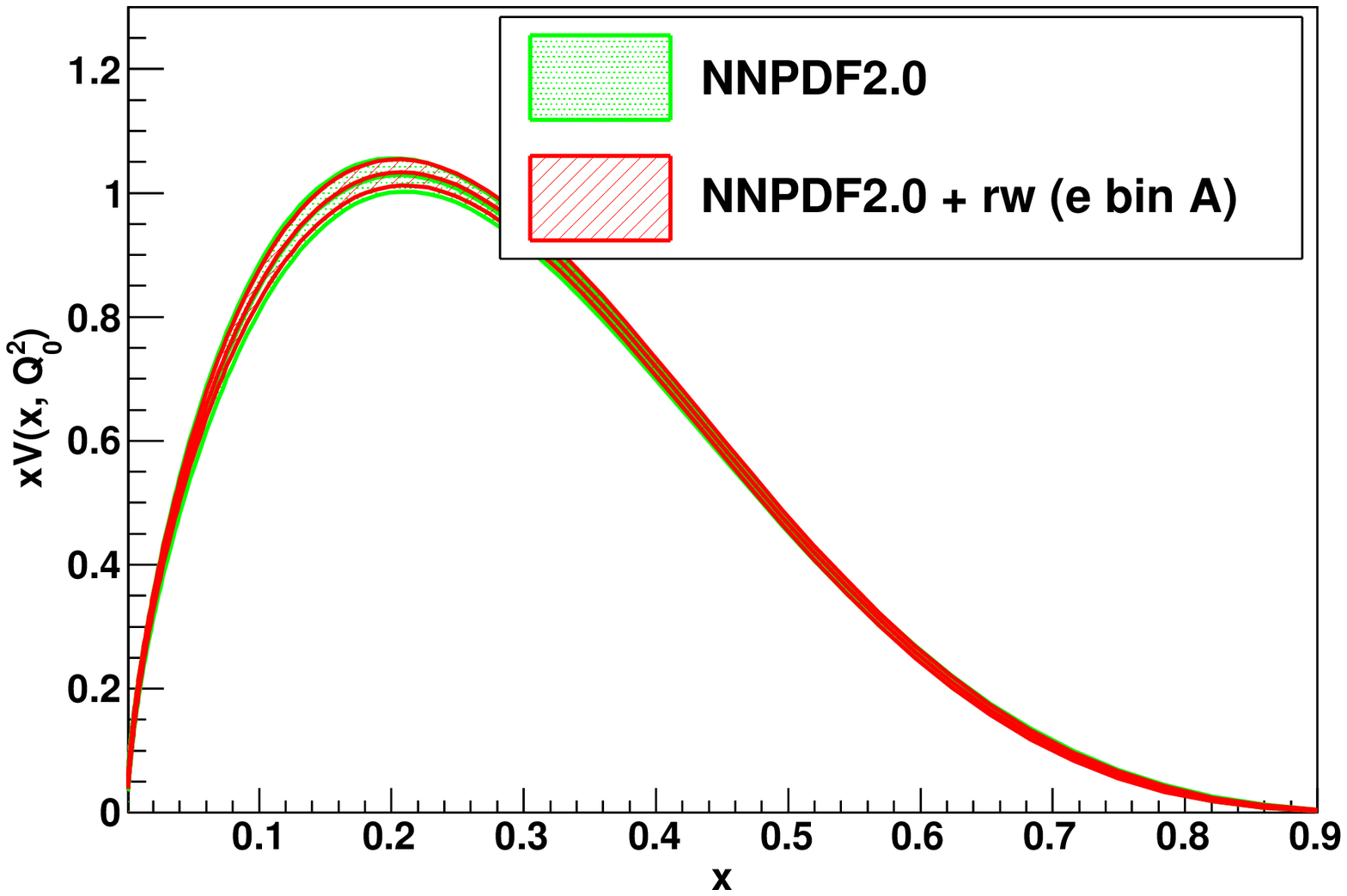}
    \includegraphics[width=0.45\textwidth]{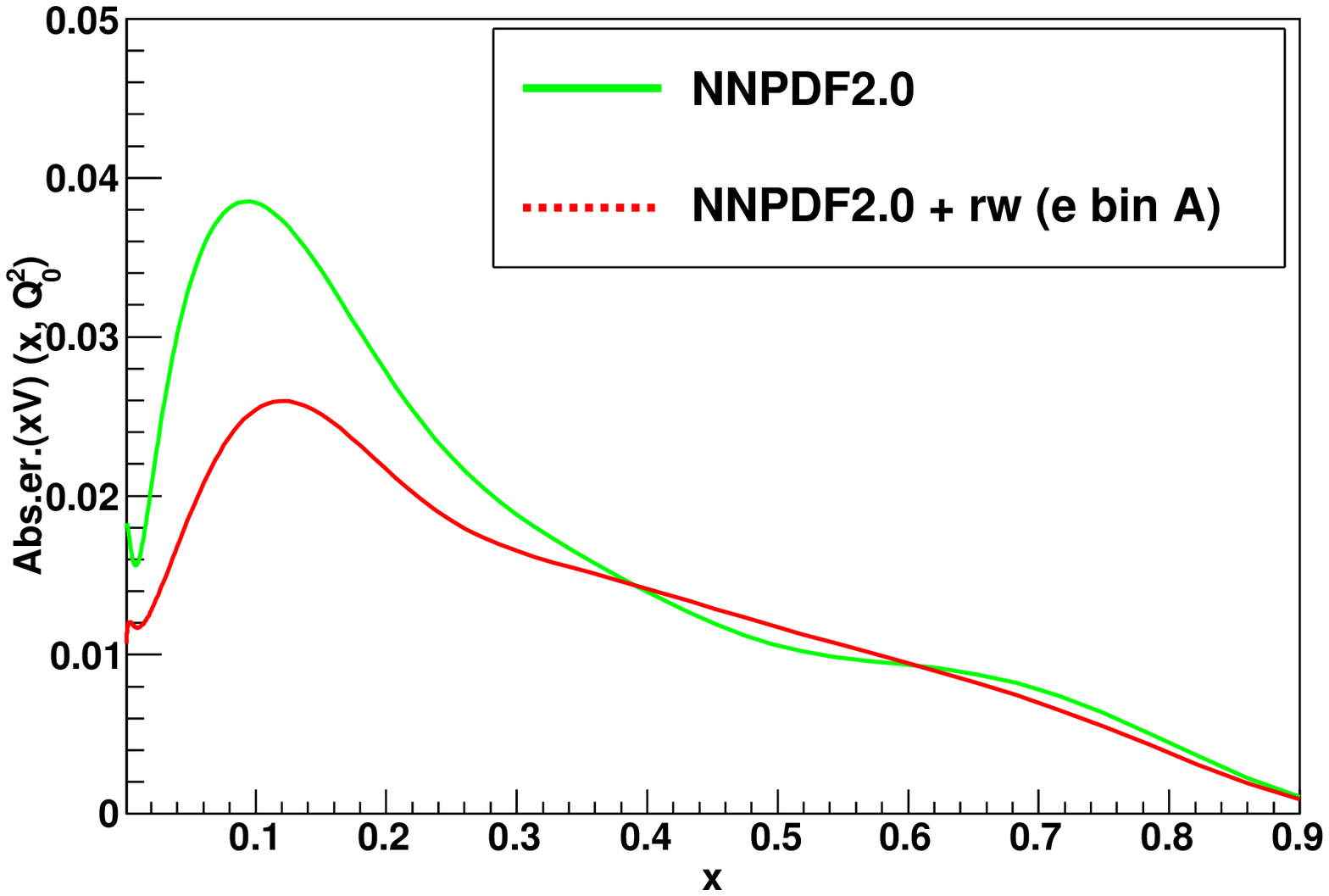}
  \end{center}
  \caption{\small Total valence PDF for NNPDF2.0 and NNPDF2.0 + D0 electron data 
    (bin A).  
    \label{fig:valence-binA}}
\end{figure}
\begin{figure}[t]
  \begin{center}
    \includegraphics[width=0.9\textwidth]{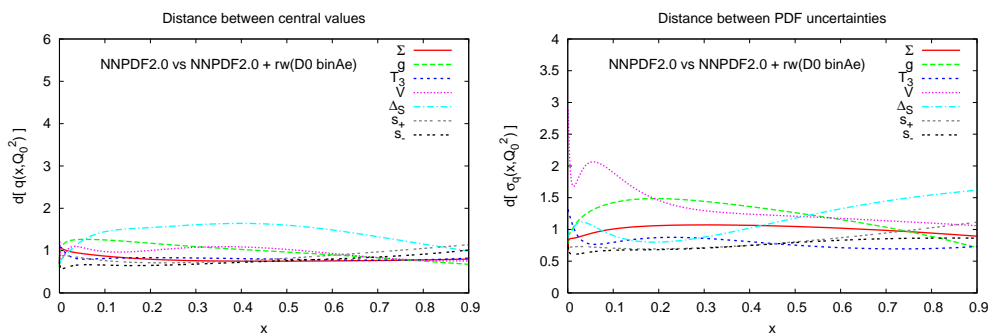}
  \end{center}
  \caption{\small Distances between NNPDF2.0 and NNPDF2.0 + D0 W lepton charge 
    asymmetry  measurements from the electron bin A dataset.
    The NNPDF2.0 set with $N_{\rep}=100$ has been used in the computation of the 
    distances.
\label{fig:distances-binA}}
\end{figure}

The improvement in the description of the electron asymmetry after 
reweighting in Fig.~\ref{fig:obs-incl}, while the 
fit to the other datasets included in the NNPDF2.0 fit shows no significant 
deterioration: if any change has to be noticed, is a slight improvement in 
particular in the fit to the CDF $W$ asymmetry data 
(see Tab.~\ref{tab:chi2-20y}). 
 
In Fig.~\ref{fig:distances-binA} we plot the distances between the prior set 
NNPDF2.0 and the reweighted set: it is clear that the most significant effect
is on the uncertainty in the valence PDF.
Indeed, in Fig.~\ref{fig:valence-binA} we show the error reduction that comes 
from the inclusion of the inclusive D0 electron charge asymmetry data on 
the valence PDF. While the central value remains essentially unchanged, the 
uncertainty is significantly reduced. Small improvements in the precision of 
the singlet and triplet quark distributions can also be observed, while other 
PDFs combination remain unchanged.

Having found that both the inclusive muon and electron (bin A) D0 asymmetry 
data are each consistent with the datasets used in NNPDF2.0, it is 
interesting to ask whether they are also consistent with each other. This 
is not obvious a priori: it is in principle possible for each dataset to 
prefer a different subset of the NNPDF2.0 replicas. 

\begin{figure}[t]
  \begin{center}
    \includegraphics[width=0.45\textwidth]{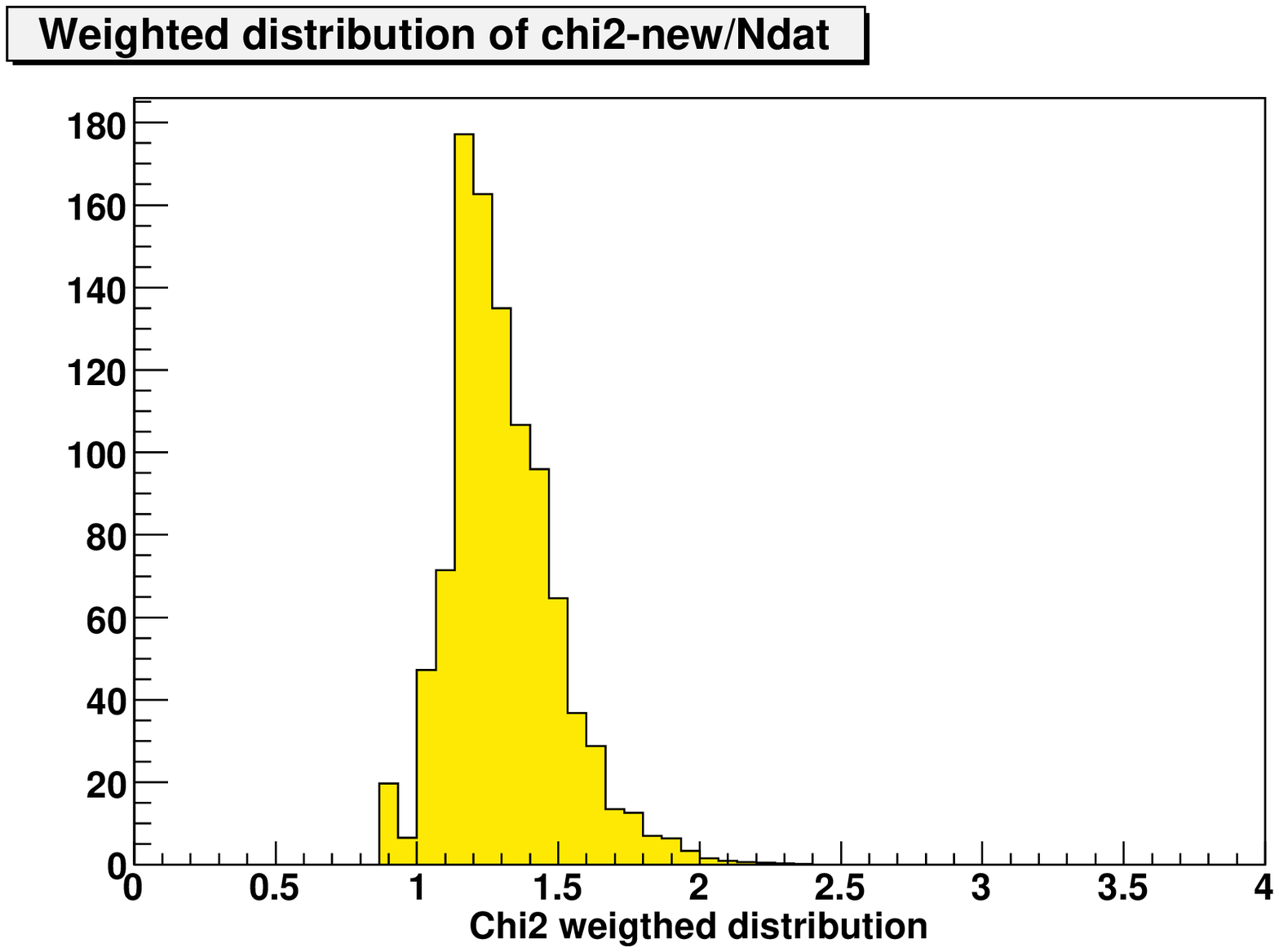}
    \includegraphics[width=0.45\textwidth]{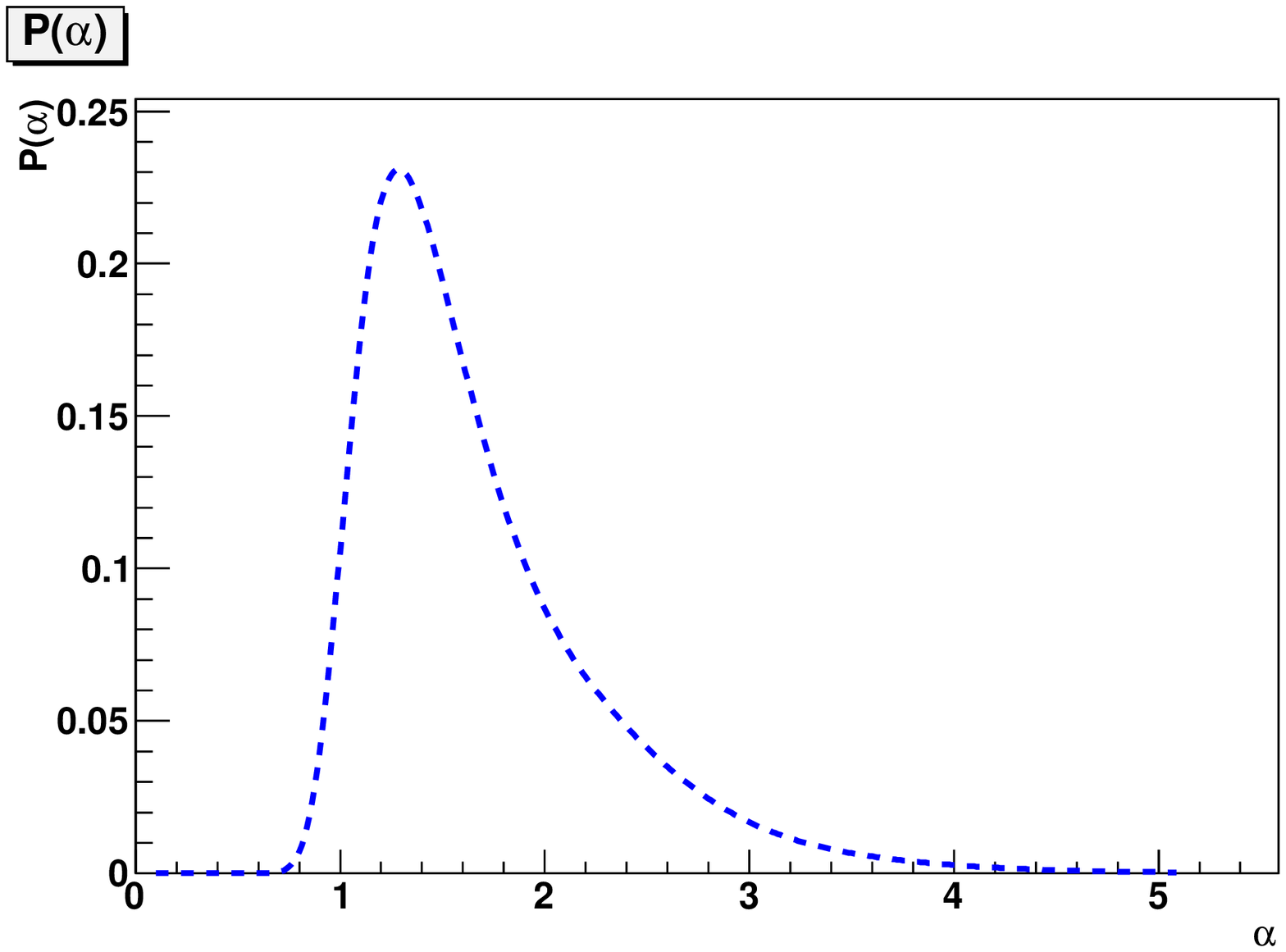}
  \end{center}
  \caption{\small The reweighted $\chi^2$-distribution and the probability 
    distribution ${\mathcal P}(\alpha)$ in the reweighting of the NNPDF2.0 
    PDF set using the combined D0 muon asymmetry data and electron asymmetry 
    data (bin A).
    \label{fig:chi2-binAe-muon} }
\end{figure}

To examine this question we performed a reweighting analysis with the 
combined dataset: the $\chi^2_k$  used to determine the weights are then the sum 
of those from the D0 muon asymmetry data and the D0 electron asymmetry data, i.e. 
$N_{\rm dat}=22$ data points.
The number of effective replicas is then reduced to $N_{\rm eff} = 356$, 
actually a number larger than the case where electron data alone were considered: 
the muon data soften the impact of these data. 
The combined $\chi_k^2$ and $\alpha$ distributions (see 
Fig.~\ref{fig:chi2-binAe-muon}) are now better behaved: while before the 
reweighting only $49\%$ of replicas have a $\half<\chi^2<2$, after reweighting 
this now rises to $99\%$. The peak of the $\alpha$ distribution is now closer 
to one: the overestimated uncertainties of the muon data in part compensate for the 
underestimated uncertainties of the electron data. The quality of the fit 
to the other datasets included in the NNPDF2.0 fit shows no significant 
deterioration, and again there is a slight improvement, in the 
fit to the CDF $W$ asymmetry data (see Tab.~\ref{tab:chi2-20y}).  

\begin{figure}
  \begin{center}
    \includegraphics[width=0.45\textwidth]{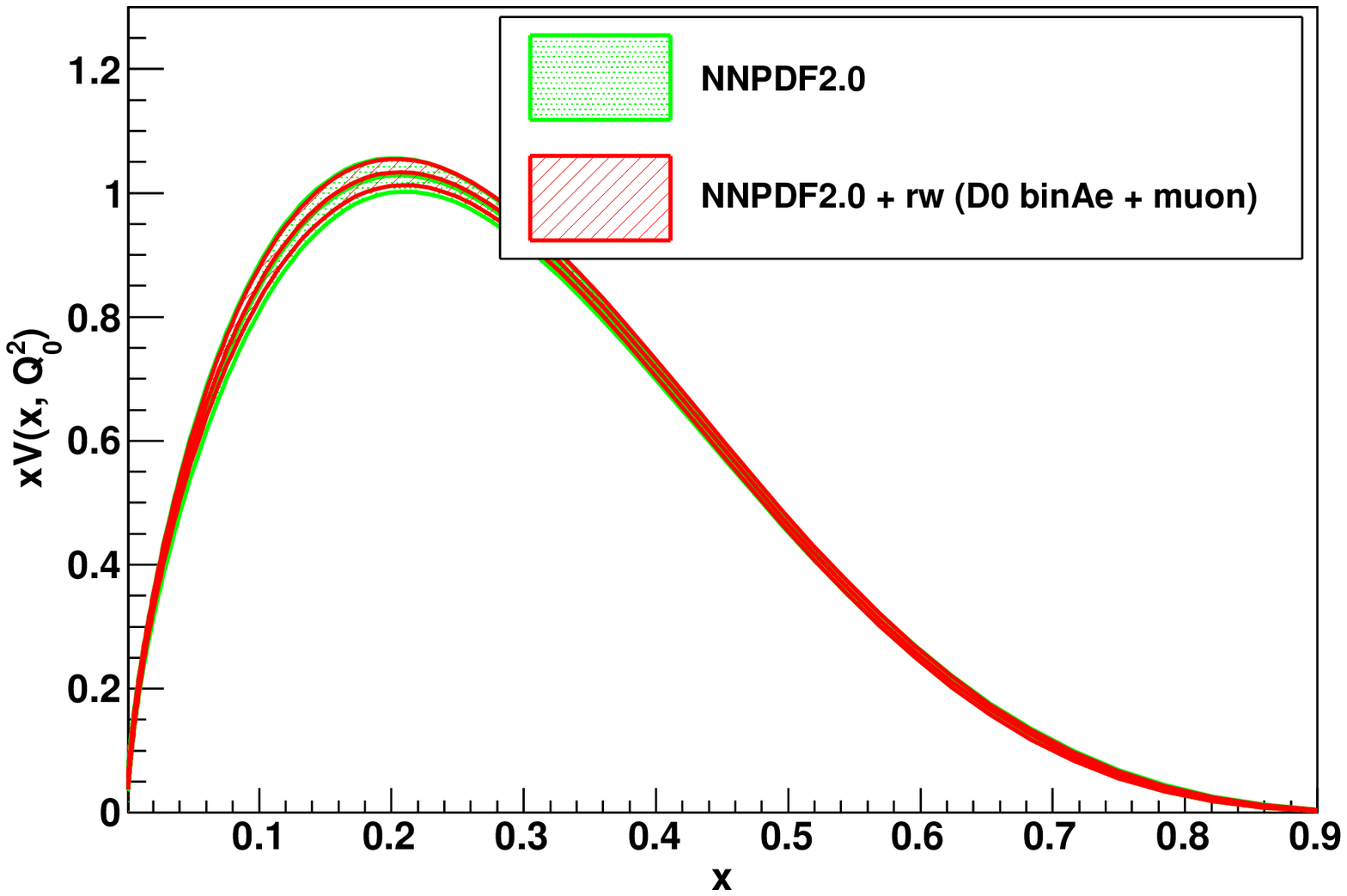}
    \includegraphics[width=0.45\textwidth]{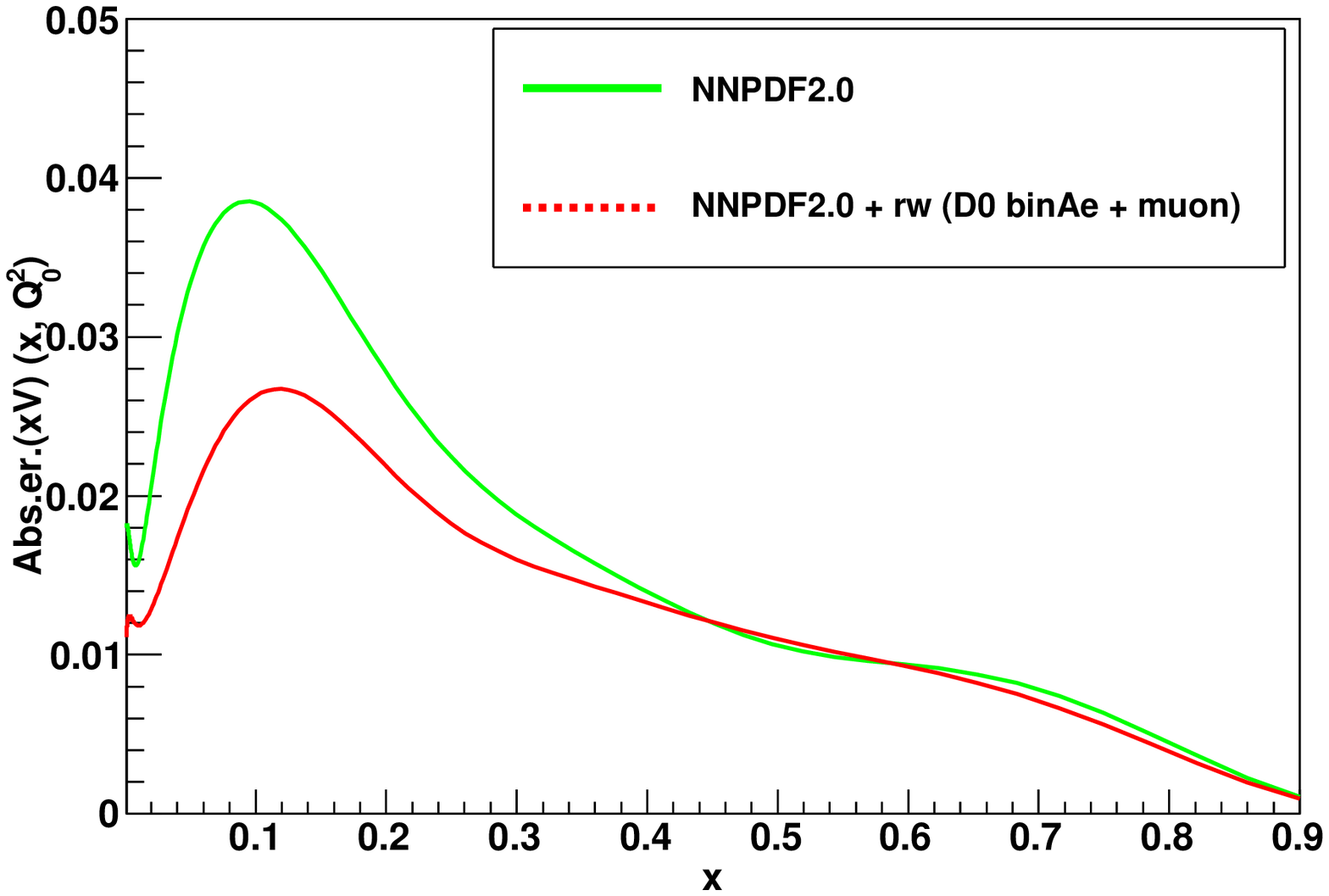}
  \end{center}
  \caption{\small Total valence PDF for the NNPDF2.0 and NNPDF2.0 + D0 muon + D0 
    electron data (bin A) sets.
    \label{fig:valence-binA-muon}}
\end{figure}

In Fig.~\ref{fig:valence-binA-muon} we show the effect of the addition of the D0 
muon and the D0 electron inclusive data on the valence distribution. The precision 
of the valence distribution is significantly improved, though without shifting 
its central value significantly. This implies that the NNPDF2.0 set is quite 
consistent with the inclusive data, so that their addition entails only 
PDF uncertainty reduction without affecting central values. 
It follows that the $d/u$ ratio extracted from the DIS deuterium data and the CDF 
direct $W$ charge asymmetry data will be consistent with the information 
included in the D0 inclusive muon and electron data.

The main statistical estimators for the lepton charge asymmetry data sets
are summarized in Tab.~\ref{tab:chi2}. The two inclusive sets have a significant 
impact on PDFs, and are reasonably consistent with themselves
(though the experimental uncertainties on the inclusive D0 electron charge asymmetry 
data, bin A, may be a little underestimated), with the other data used in 
the NNPDF2.0 fit, in particular the CDF $W$ charge asymmetry data, and with each 
other.

The $\chi^2$ values for the total dataset and for the individual experiments 
in the NNPDF2.0 analysis are shown in Table~\ref{tab:chi2-20y}. The sets that 
differ sizably from the reference results have been highlighted in boldface 
in the different cases. As far as the inclusive muon and electron datasets are
concerned we notice that both are consistent with the NNPDF2.0 datasets,
and their inclusion improves the fit to the $W$ asymmetry data. Furthermore they 
are both consistent with each other.
These conclusions do not support the conclusions of the MSTW08 
analysis~\cite{MSTWasym}, 
which finds that inclusion of 
the D0 electron inclusive bin in
the global fit, without significant deterioration in the fit to the other 
datasets, requires sizable nuclear corrections to deuterium data.

\subsection{More exclusive data } 

\begin{figure}[t!]
  \begin{center}
    \includegraphics[width=0.45\textwidth]{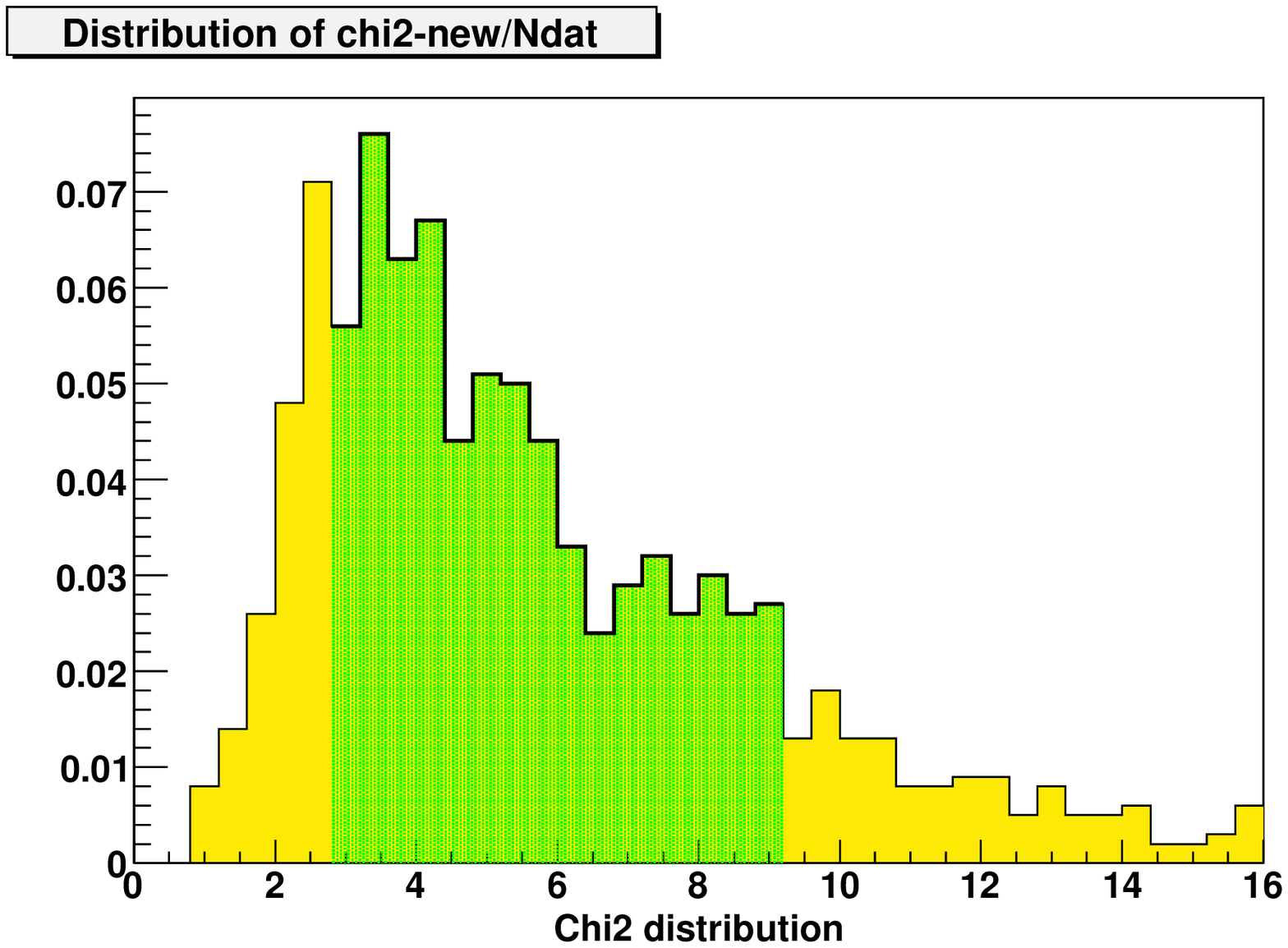}
    \includegraphics[width=0.45\textwidth]{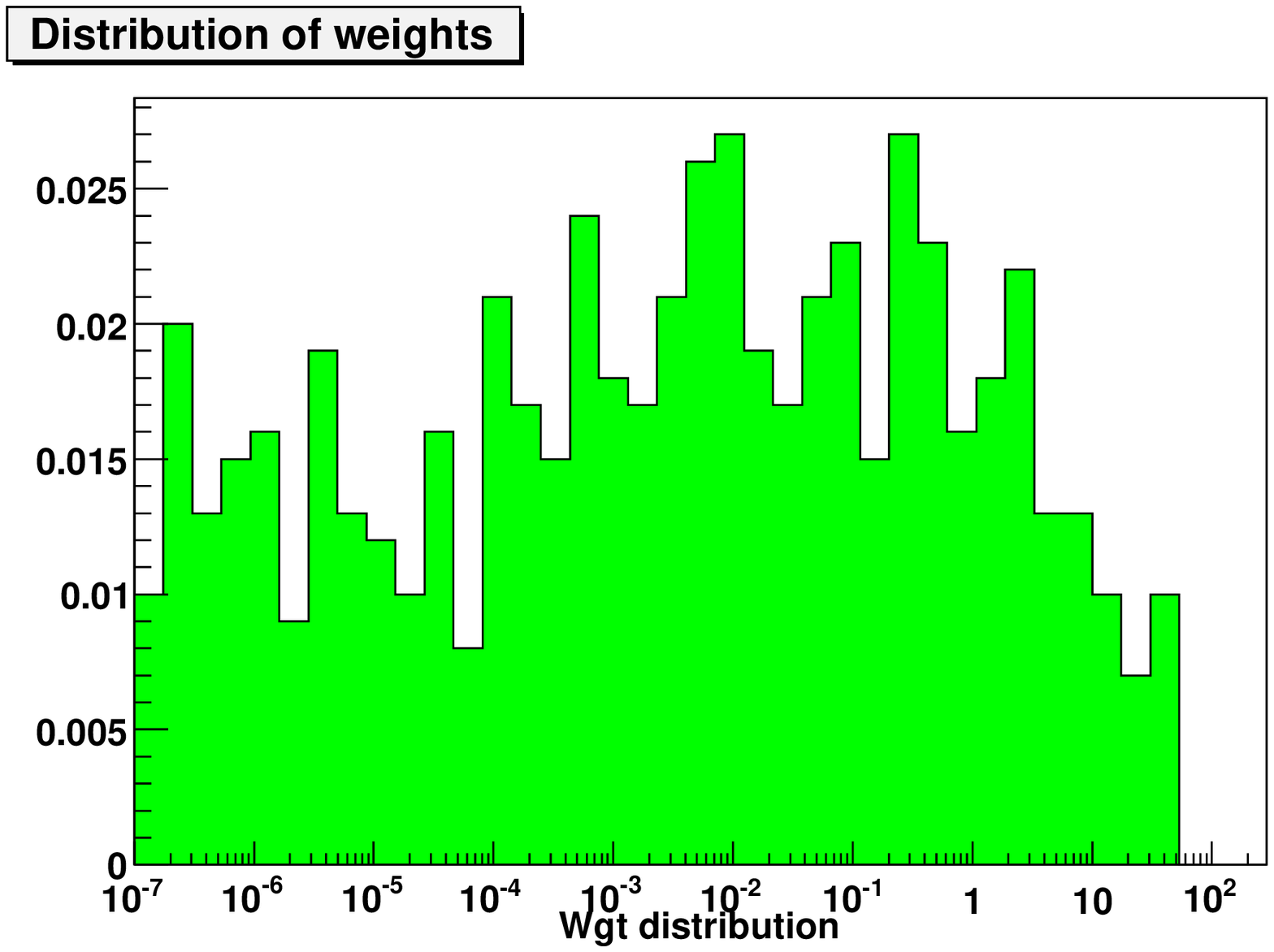}\\
    \includegraphics[width=0.45\textwidth]{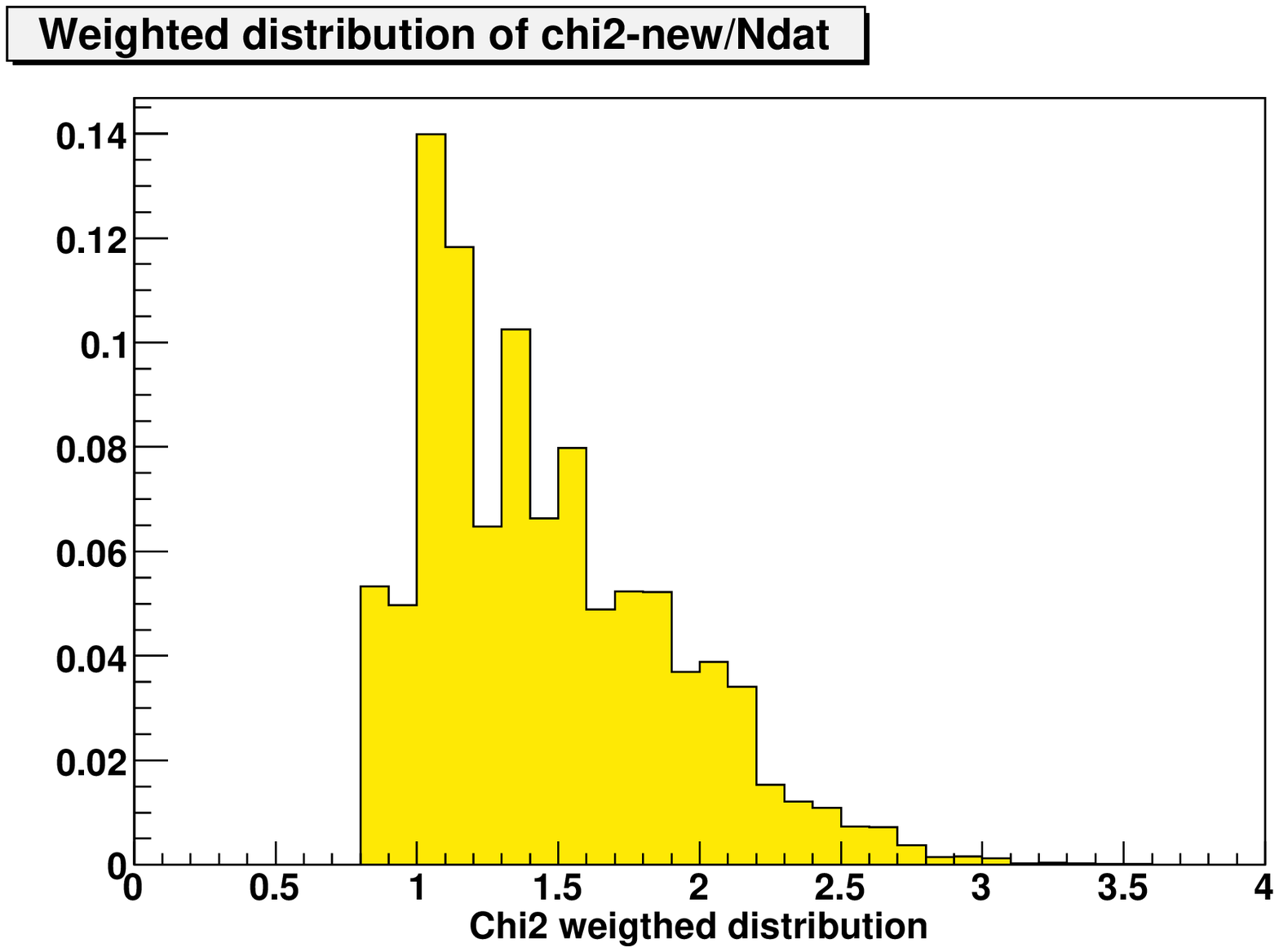}
    \includegraphics[width=0.45\textwidth]{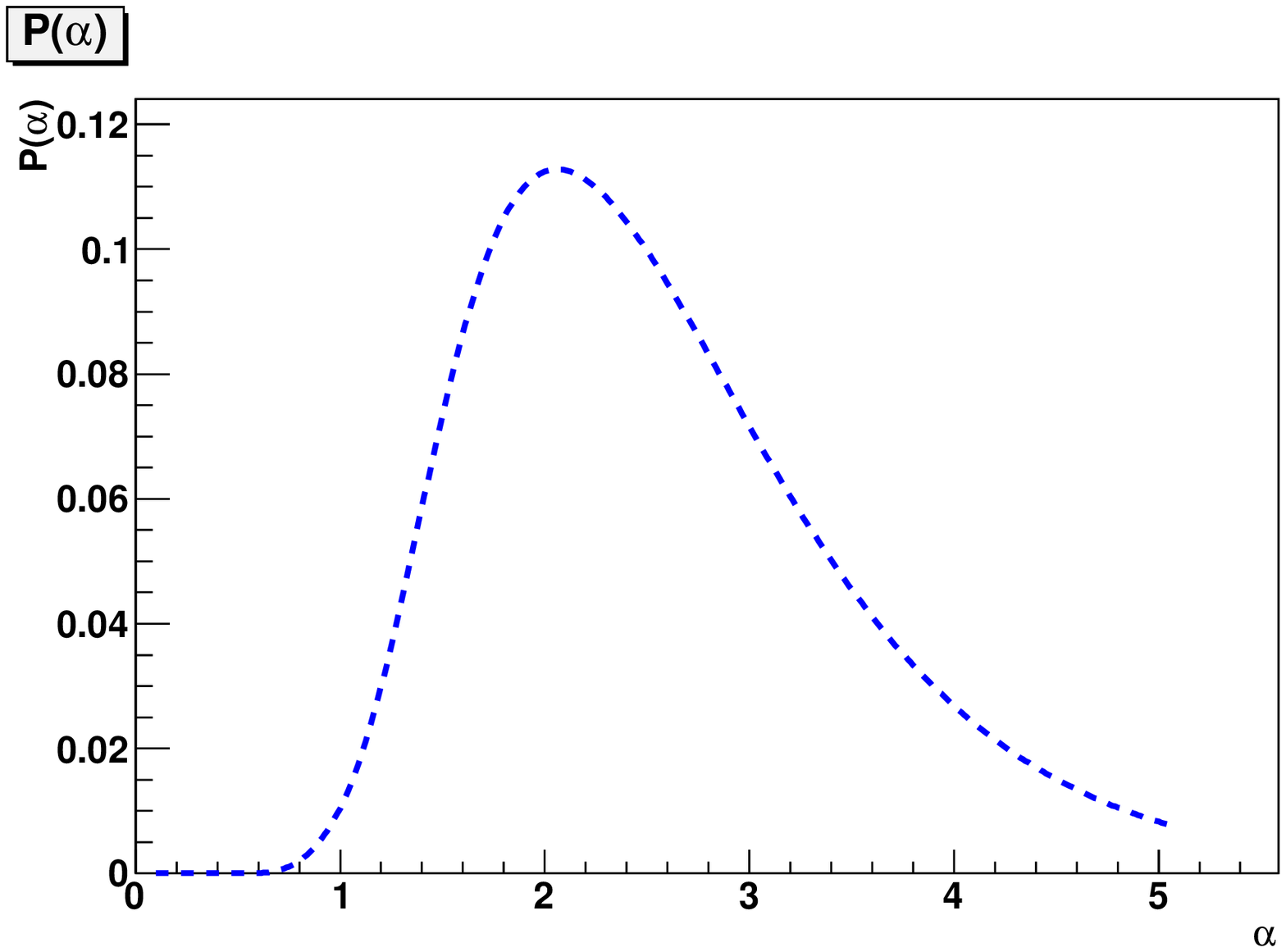}
  \end{center}
  \caption{\small Distribution of the $\chi^2_k$ and the weights $w_k$, the 
    reweighted $\chi^2$-distribution and the probability distribution 
    ${\mathcal P}(\alpha)$ in the reweighting of the NNPDF2.0 PDF set
    using the D0 electron asymmetry data bin B~\cite{d0electron}.
    \label{fig:chi2-binB}}
\end{figure}

We now turn to the less inclusive D0 electron charge asymmetry data (bin B 
and bin C), where the transverse energy of the electron 
is restricted to the range $25$ GeV$<E^e_T<35$ GeV and $E_T^e>35$ GeV 
respectively. 

We first consider each bin separately, and we turn then to their combination. 
Considering first the lower $E_T^e$ bin (bin B), the number of effective replicas 
is now reduced to $N_{\rm eff} = 61$, indicating that, as expected, these data 
are more constraining than those of the inclusive bin. 
This is so because the data binned in $E^e_T$ probe a more localized region in 
$x$ of the PDFs as compared to the inclusive data. 
The $\chi^2$ for this set drops from 4.75 to 1.12 after the data is included. 
From the plots in Fig.~\ref{fig:chi2-binB} we see that indeed there is now a 
significant fraction of very small weights, because many of the replicas 
fit the new data rather badly. After reweighting the $\chi^2$ 
distribution improves very significantly: while before reweighting only 
$4.8\%$ of replicas were in the range $\half < \chi^2 < 2$, after 
reweighting this increases to $86.5 \%$. However the rescaling plot peaks at 
around $\alpha\sim 2$ indicating that the errors on the data are 
significantly underestimated.

\begin{figure}[t]
  \begin{center}
    \includegraphics[width=0.45\textwidth]{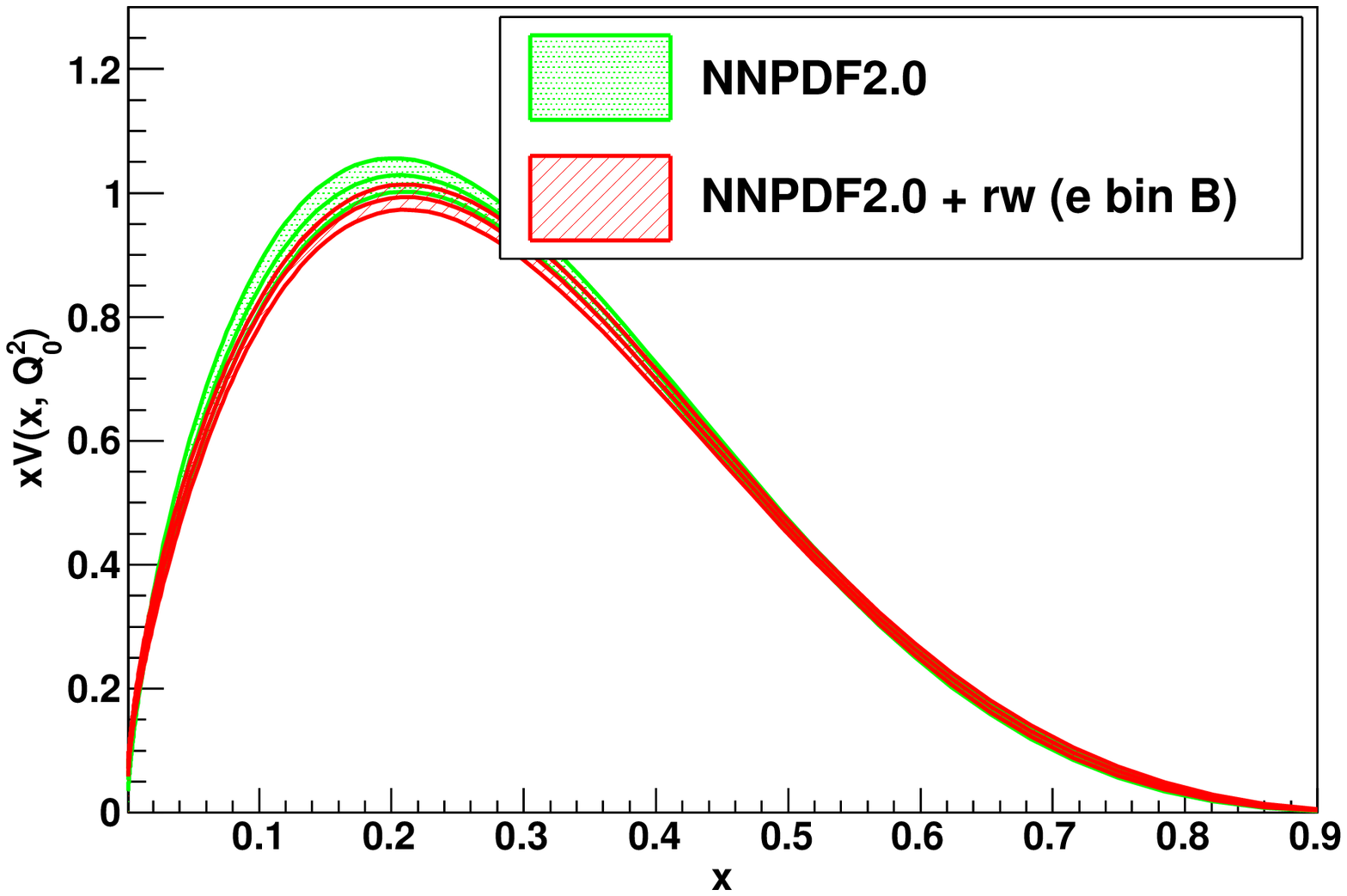}
    \includegraphics[width=0.45\textwidth]{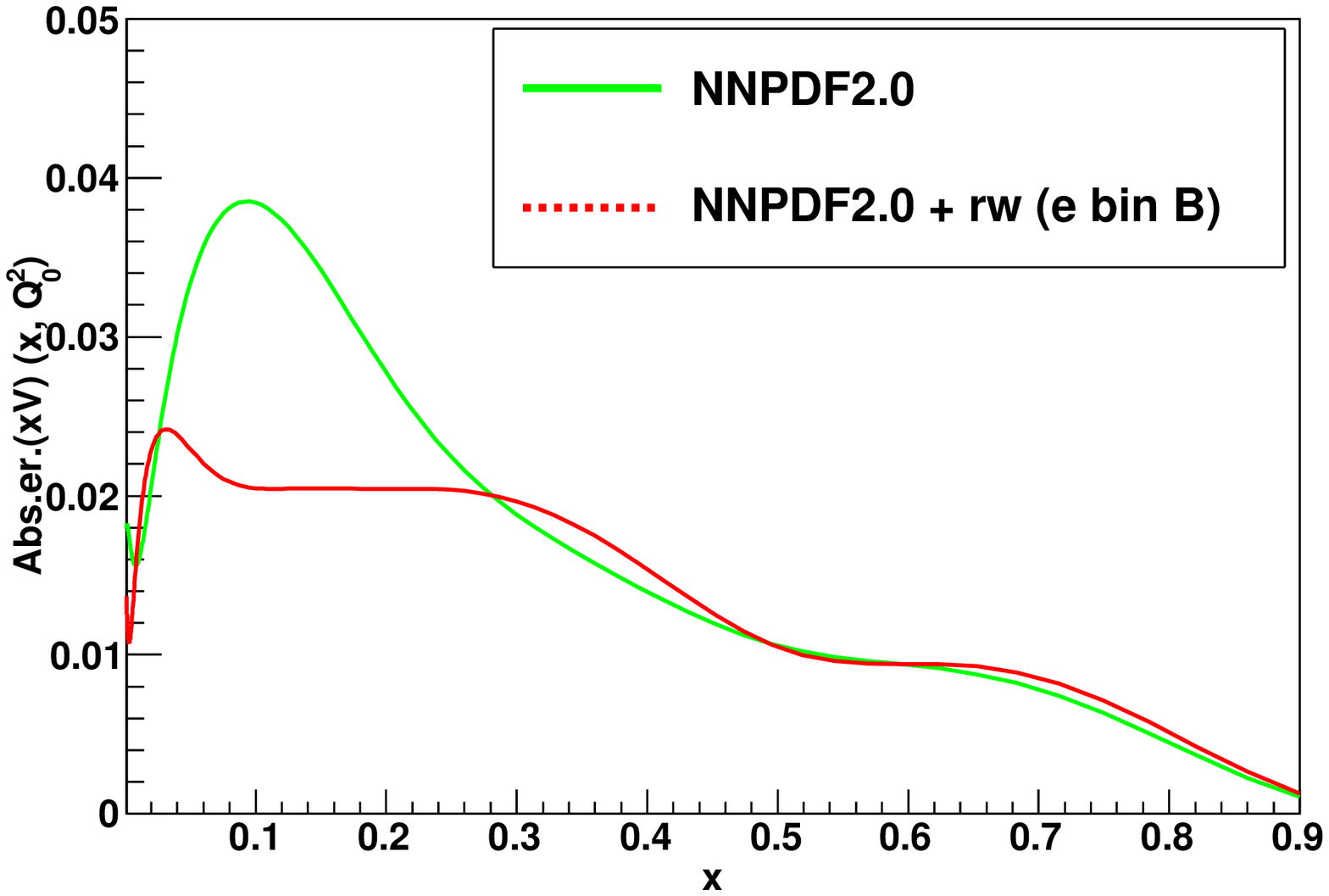}
  \end{center}
  \caption{\small Total valence PDF  for NNPDF2.0 and NNPDF2.0 + D0 electron 
    data (bin B).}
  \label{fig:valence-binB}
\end{figure}
\begin{figure}[t]
  \begin{center}
    \includegraphics[width=0.9\textwidth]{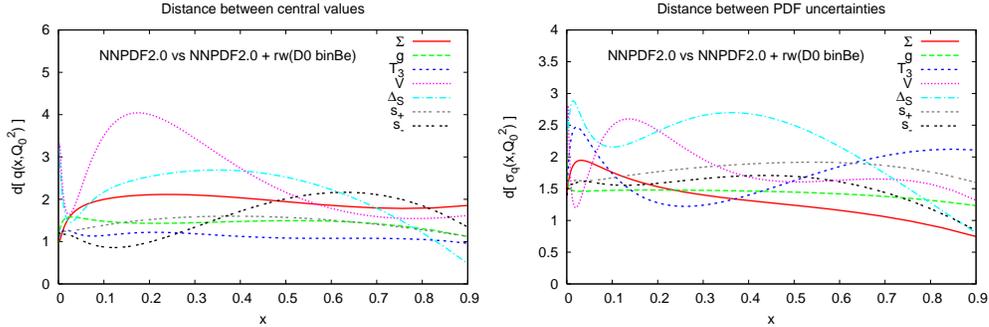}\\
  \end{center}
  \caption{\small Distances between NNPDF2.0 and NNPDF2.0 + D0 W lepton 
    asymmetry measurements from the electron bin B dataset. The NNPDF2.0 
    set with $N_{\rep}=100$ has been used in the computation of the distances.
    \label{fig:distances-binB}}
\end{figure}

\begin{figure}
  \begin{center}
    \includegraphics[width=0.45\textwidth]{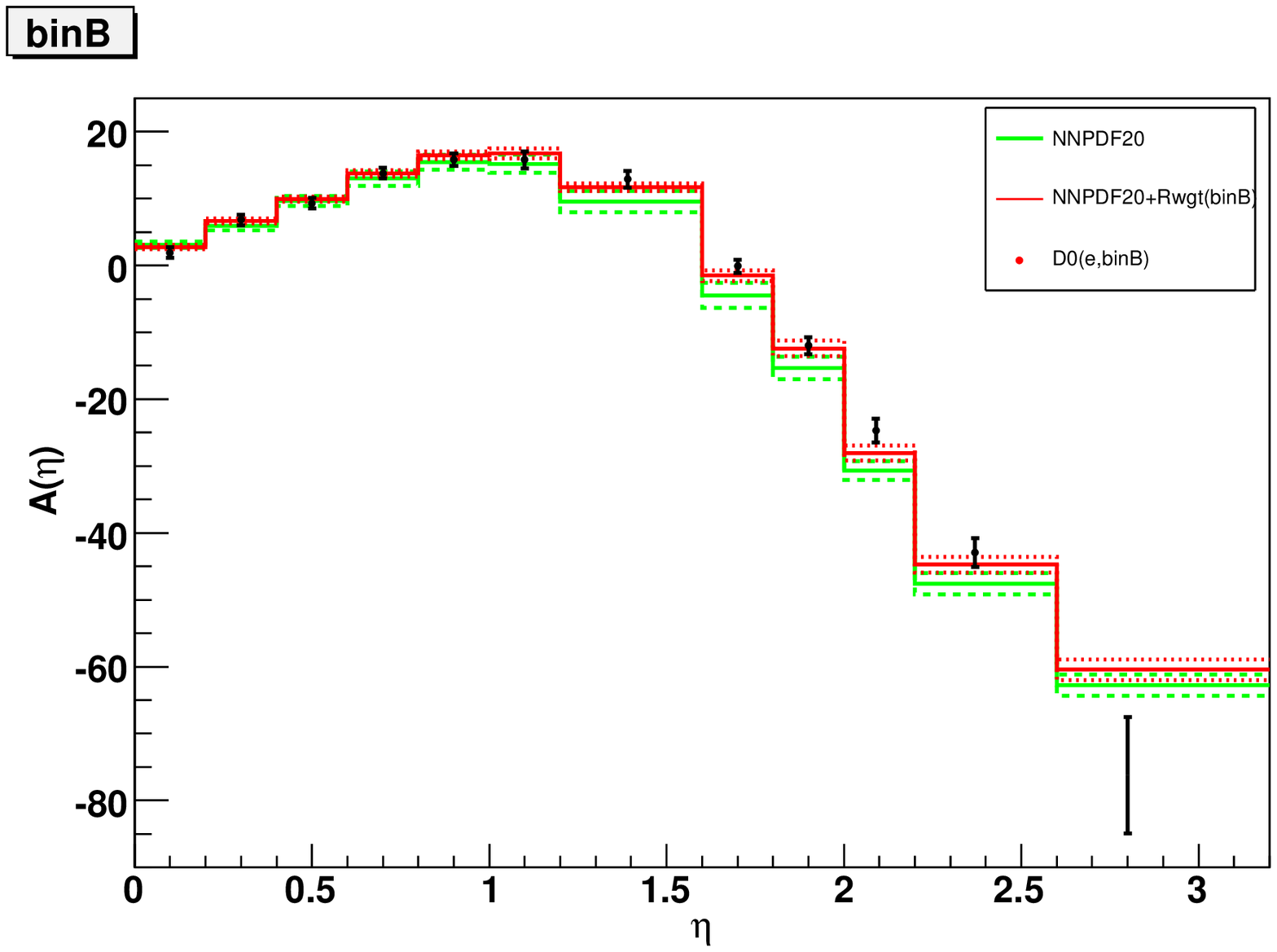}
    \includegraphics[width=0.45\textwidth]{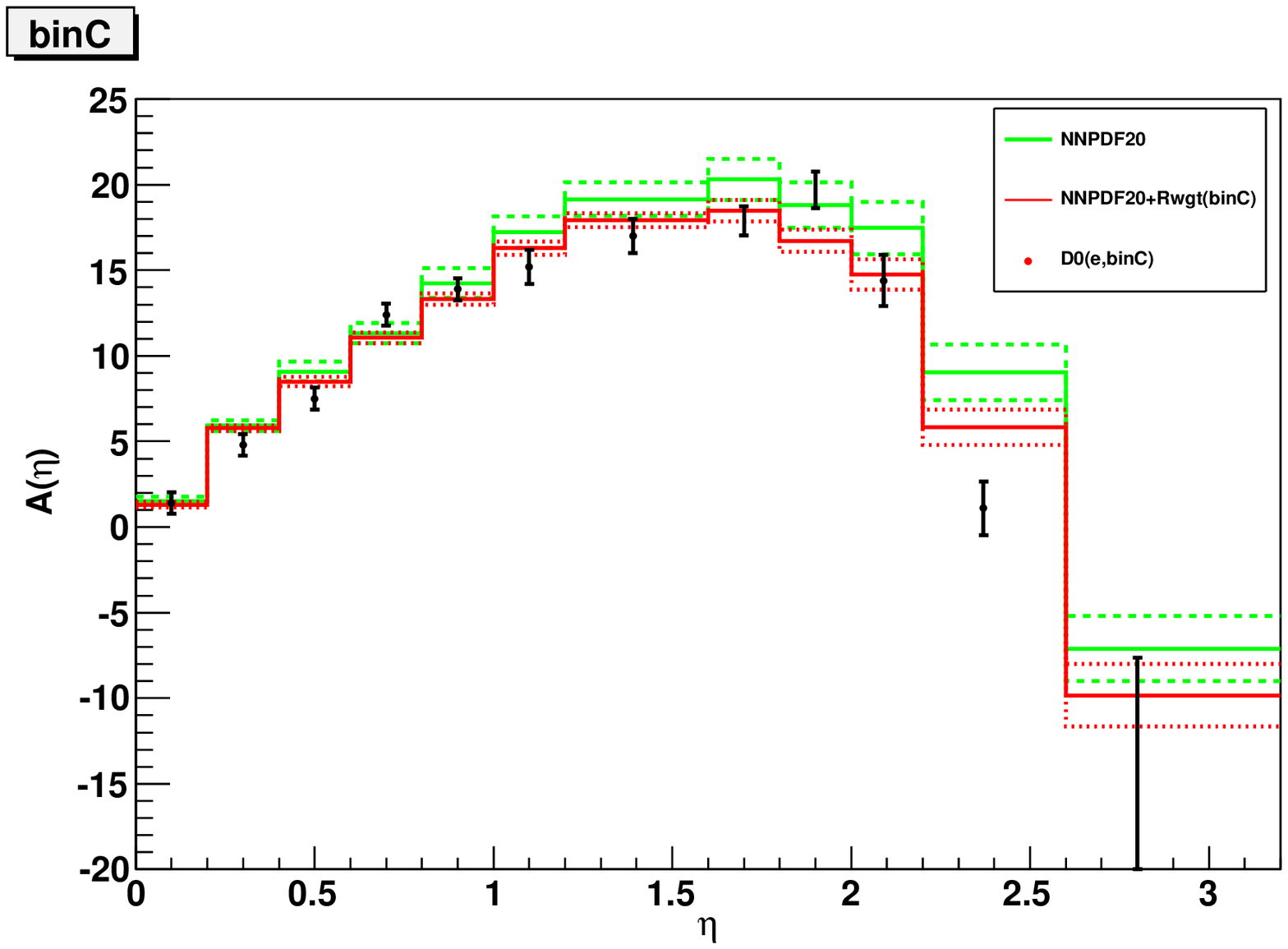}
    \caption{\small W electron asymmetry computed on the NNPDF2.0 set before 
      and after the reweighting of the D0 W electron asymmetry: bin B (on the left) and bin C (on the right).
      \label{fig:obs-excl}}
  \end{center}
\end{figure}

The improvement in the fit to the lowest $E_T^e$ bin charge asymmetry data 
is manifest on the left of Fig.~\ref{fig:obs-excl}. However the fit to some 
of the other datasets in the NNPDF2.0 fit, in particular BCDMSp and BCDMSd, 
becomes significantly worse (see Table~\ref{tab:chi2-20y}). 
The fact that there is as much tension here with the proton data as 
with the deuteron data suggests that it is unlikely that nuclear corrections 
to the deuteron target can help (contrary to the claim 
in Ref.~\cite{MSTWasym}). 
The overall $\chi^2$ per degree of freedom rises from $1.14$ to $1.16$: 
this is rather significant, given that we are only adding $12$ new data points 
to the $3415$ used in NNPDF2.0. The decrease in the fit quality is driven by 
the large weight that BCDMS carry in the global fit. It should be further noted 
that the fit to the inclusive jet data and CDF W asymmetry also worsens.

These problems are also apparent when we look at the effect on individual PDFs: 
in particular while the valence distribution, Fig.~\ref{fig:valence-binB}, 
is now better determined in some ranges of $x$, elsewhere the uncertainty 
increases.
While this may in part be due to the rather limited statistics of the reweighted 
distribution, it is probably also a sign of some inconsistency with the other 
data used in the NNPDF2.0 fit. 
The statistical distances, plotted in Fig.~\ref{fig:distances-binB}, are also 
sizable for some PDFs, especially the valence distribution. 

\begin{figure}[t]
  \begin{center}
    \includegraphics[width=0.45\textwidth]{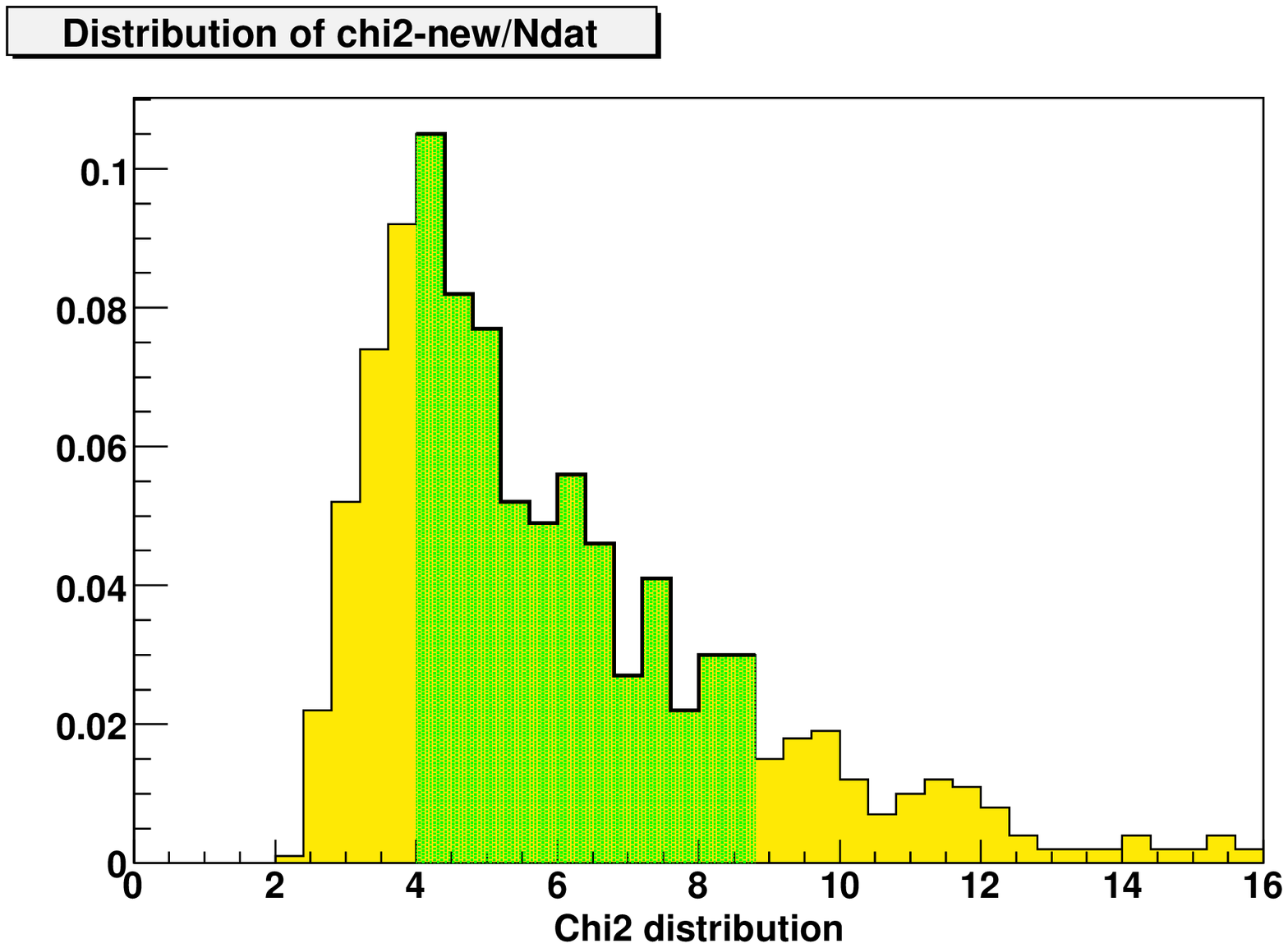}
    \includegraphics[width=0.45\textwidth]{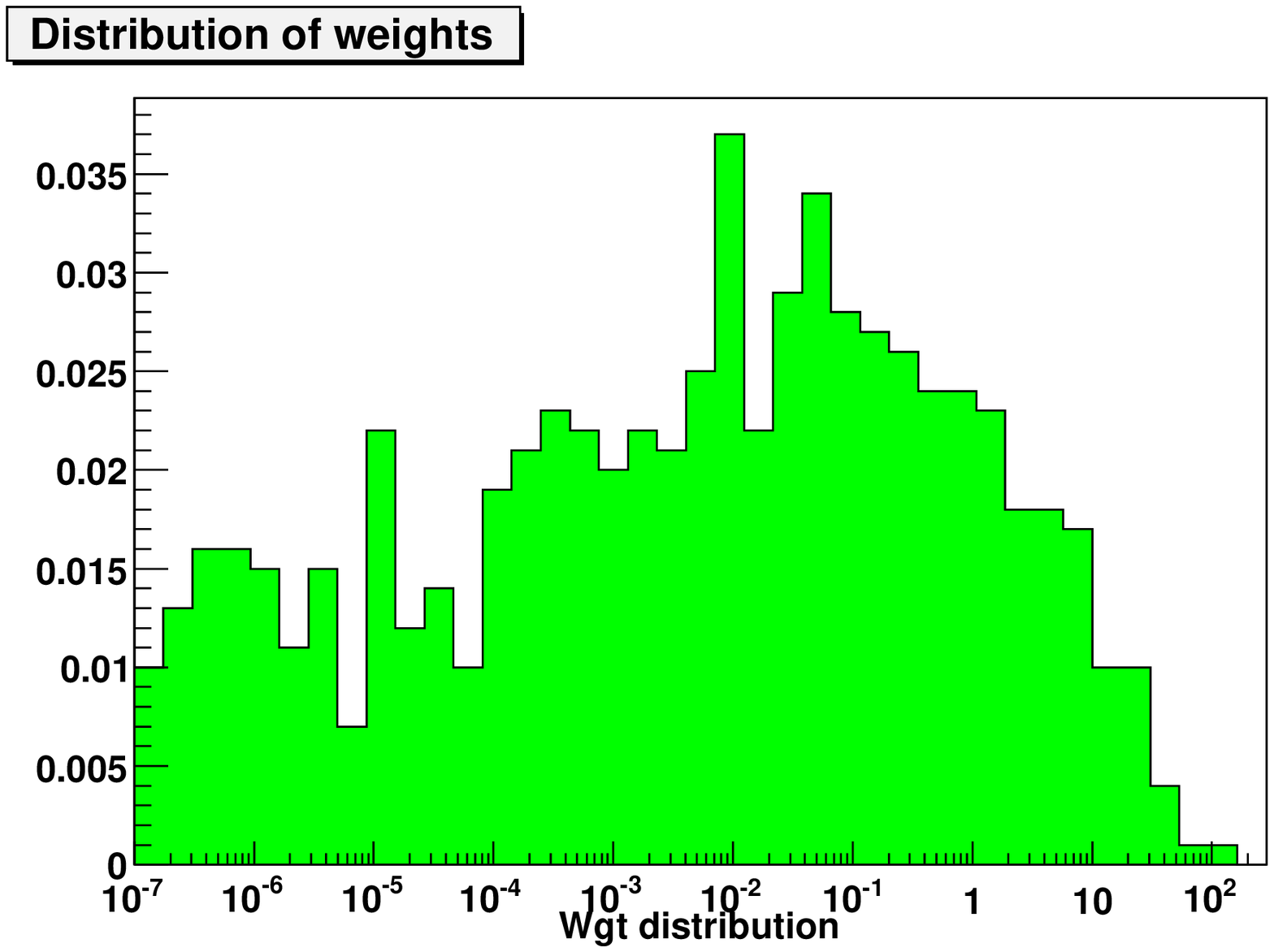}\\
    \includegraphics[width=0.45\textwidth]{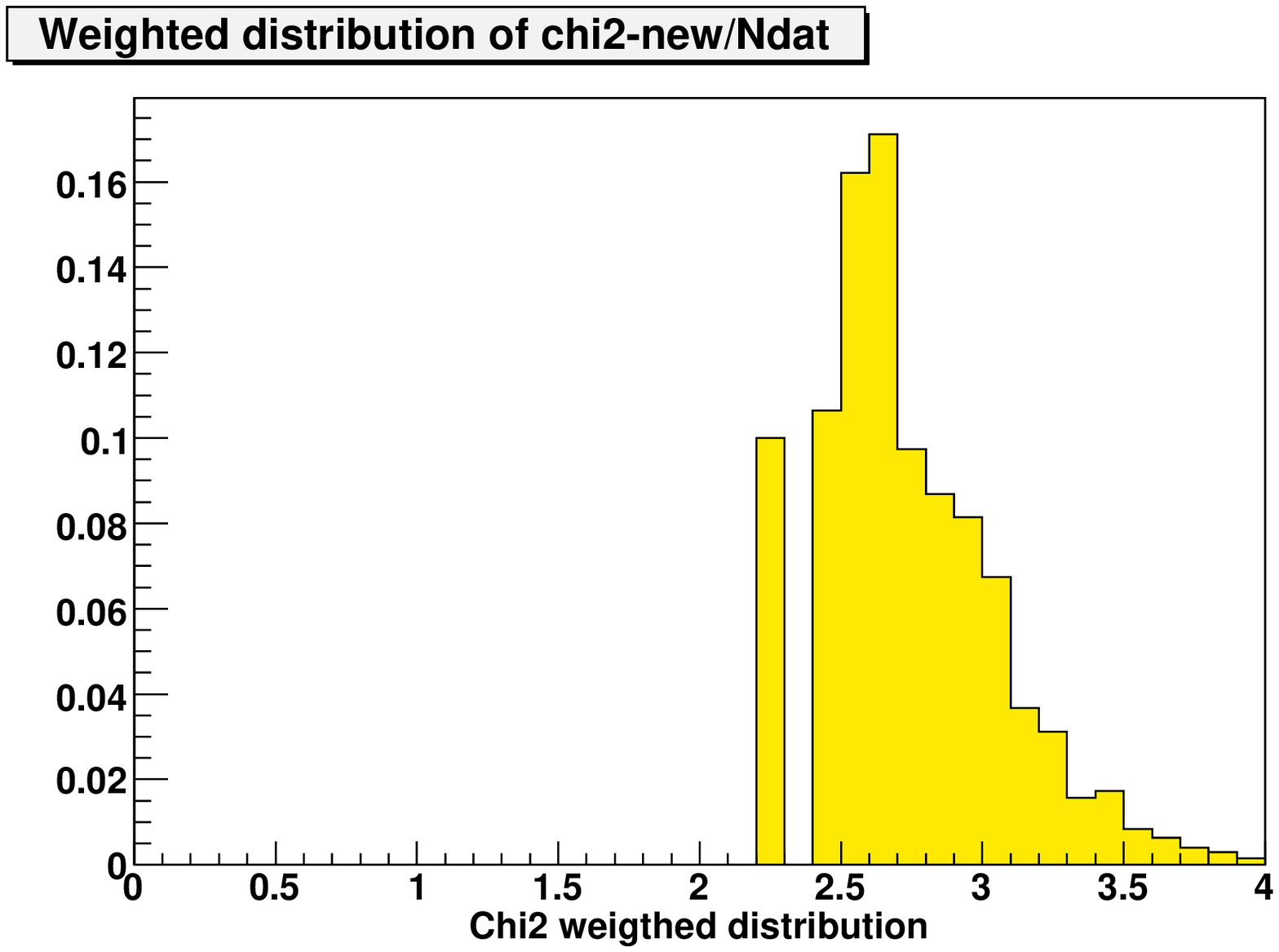}
    \includegraphics[width=0.45\textwidth]{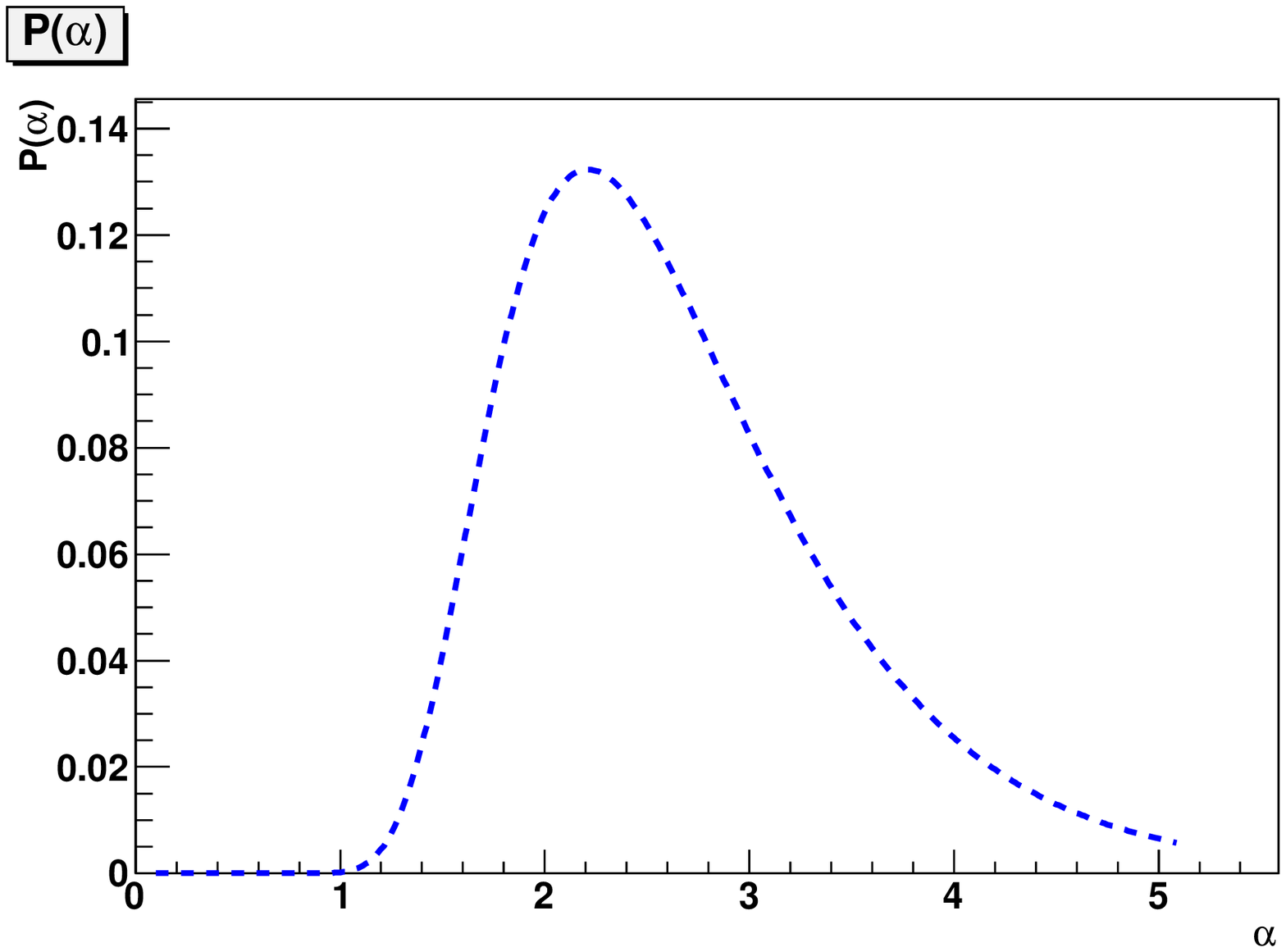}
  \end{center}
  \caption{\small Distribution of the $\chi^2_k$ and the weights $w_k$, the 
    reweighted $\chi^2$-distribution and the probability distribution 
    ${\mathcal P}(\alpha)$ in the reweighting of the NNPDF2.0 PDF set
    using the D0 electron asymmetry data bin C~\cite{d0electron} }
  \label{fig:chi2-binC}
\end{figure}
\begin{figure}[b]
  \begin{center}
    \includegraphics[width=0.45\textwidth]{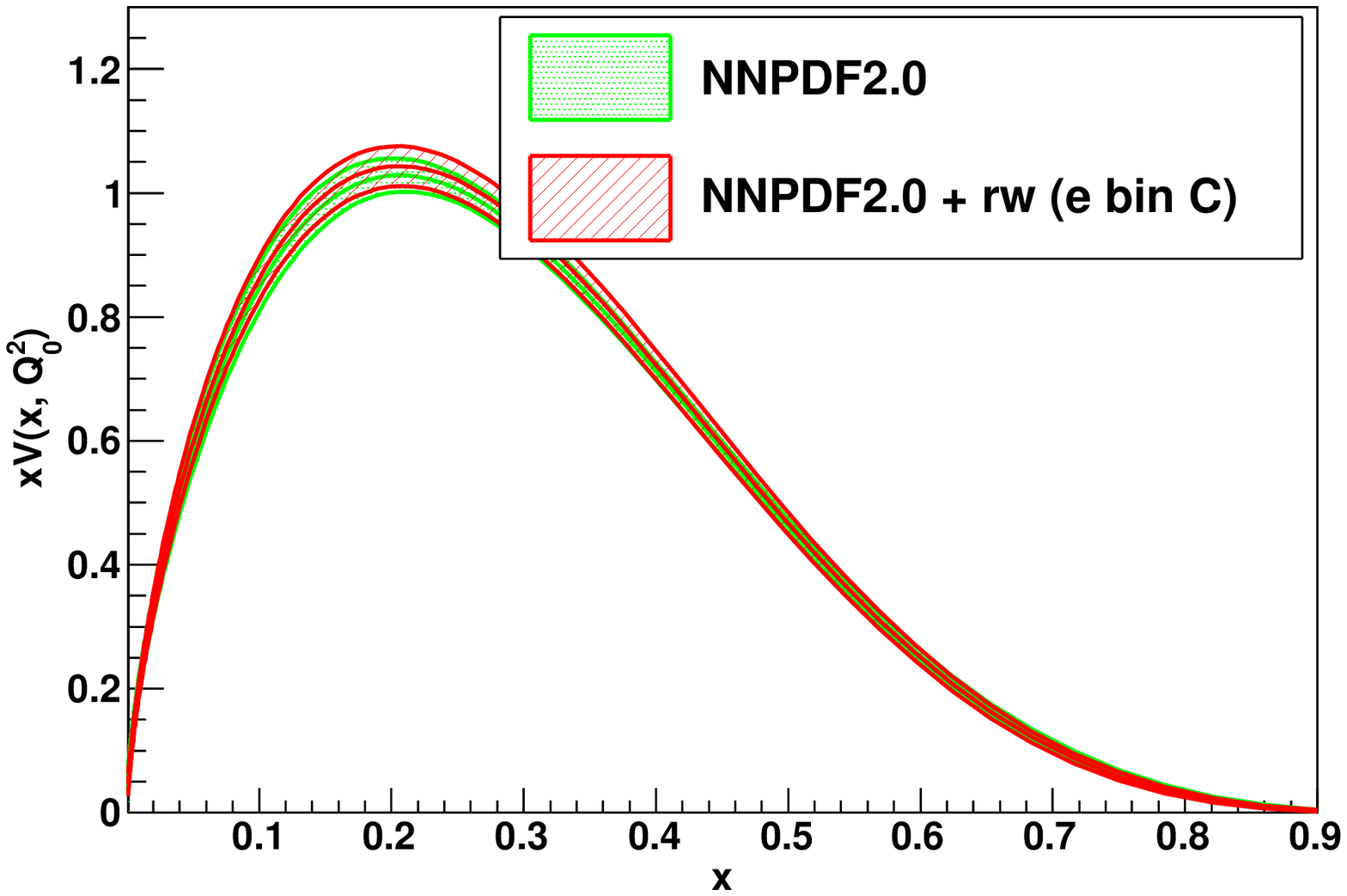}
    \includegraphics[width=0.45\textwidth]{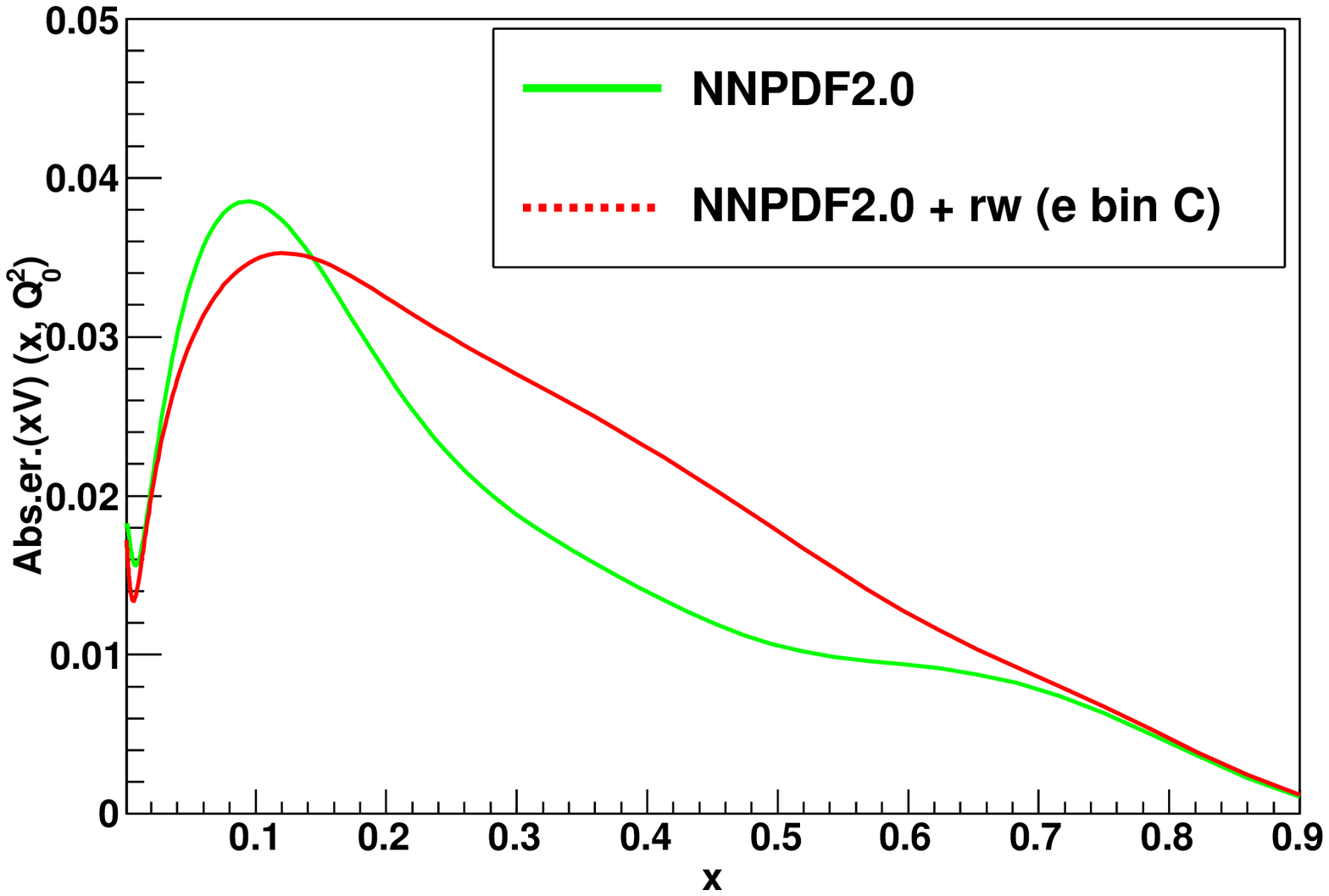}
  \end{center}
  \caption{\small Total valence for NNPDF2.0 and NNPDF2.0 + D0 electron data 
    (bin C).}
  \label{fig:valence-binC}
\end{figure}

We finally consider the remaining D0 electron asymmetry dataset at highest 
$E_T$ (bin C): the results are displayed in Fig.~\ref{fig:chi2-binC}. 
While the impact of these data is similar to that of the lowest $E_T$ bin, 
(bin B), with the effective number of replicas dropping to 68, the quality of 
the fit to the unweighted replicas is so poor (there are no replicas with a 
$\chi^2$ below $2$) that even after reweighting the quality of the fit is 
still not very good, the average $\chi^2$ per data point dropping from 
$5.06$ to $2.51$. 

The rescaling plot shows a preferred value of $\alpha \sim 2.3$, suggesting 
that again the experimental errors in these data are seriously underestimated.
This might be caused by some underestimated systematic uncertainties
in the separation of the data into bins of different $E^e_T$.
The poor quality of the fit, even after reweighting, is again apparent in 
Fig.~\ref{fig:obs-excl}: it is clear that some of the bins simply cannot 
be fitted with a reasonably smooth distribution. There is also tension between
these data and some of the other datasets included in NNPDF2.0 (see 
Table~\ref{tab:chi2-20y}): in particular while BCDMS is now fine, the 
fit to the NMC-pd ratio is spoiled.

When we examine the effect of these data on the PDFs, we 
see (Fig.~\ref{fig:valence-binC}) that rather than making 
the PDFs more precise, in many regions of $x$ the uncertainty increases 
substantially. The enlarging of the uncertainty is of course what one
would expect when inconsistent data are combined, and it was
previously seen to occur in NNPDF parton fits (see e.g. Sect.~3.4.1 of
Ref.~\cite{Dittmar:2009ii}). Here, it is shown to occcur as a consequence of
standard statistical inference.

We also attempted a combined fit of the D0 electron asymmetry data bins B 
and C, but the constraint imposed on PDFs by including these data togehter is 
so severe that the number of effective replicas is reduced to one. 
This shows that not only are these data each inconsistent with other data 
included in the global fit, but they are also inconsistent with each other. 

The main statistical estimators for the exclusive electron charge asymmetry data 
sets are summarized in Tab.~\ref{tab:chi2}. In contrast to what we observe for 
the inclusive sets, these data sets, while having an even greater effect on the 
PDFs, appear to be internally inconsistent (bin C), inconsistent with other data 
used in NNPDF2.0, particularly BCDMS proton and deuteron data (bin B), and also inconsistent with each other.

\begin{table}[b]
\begin{center}
  \begin{tabular}{|c|c|c|c|c|c|}
    \hline
    Set & $N_{\rm eff}/1000$ &$\alpha_{\rm opt}$& $\chi^2$ & $\chi^2_{\rm rw}$ 
    &$\chi^2_{\rm tot-rw}$ \\ 
    \hline
    \hline
    D0 $\mu$ $(E_T>20$ GeV)            & 0.795 & 0.7 & 0.62 & 0.51 & 1.14 \\
    D0 e bin A ($E_T>25$ GeV)          & 0.262 & 1.7 & 2.12 & 1.55 & 1.13 \\ 
    D0 $\mu$ + e bin A                 & 0.356 & 1.3 & 1.44 & 1.11 & 1.13 \\ 
    D0 e bin B ($25$ GeV$<E_T<35$ GeV) & 0.061 & 1.3 & 4.75 & 1.12 & 1.16 \\ 
    D0 e bin C ($E_T>35$ GeV)          & 0.068 & 2.7 & 5.06 & 2.51 & 1.16 \\
    \hline
  \end{tabular}
\end{center}
\caption{\small A summary of the results of reweighting with the D0 $W$ lepton 
  asymmetry data: the fraction $N_{\rm eff}/1000$ of replicas left after reweighting, 
  the most probable value $\alpha_{\rm opt}$ of the error rescaling parameter 
  $\alpha$, the $\chi^2$ per data point to the D0 $W$ lepton data evaluated 
  before and after the reweighting, and the total $\chi^2$ per data point to all 
  the other data in the NNPDF2.0 fit. 
  \label{tab:chi2}}
\end{table}

The results in the present study cannot be directly compared to the ones obtained
in the CT10 analysis, because there the three 
electron $E^e_T$ bins are added simultaneously to the fit. 
On top of the double counting problem, this is problematic because internal 
tensions of experimental origin between the different bins might be
mistaken for a physical effect, such as nuclear corrections. 
Indeed, we have shown that the more exclusive data sets are not only inconsistent 
with other sets in the global analysis but also inconsistent among themselves, 
so that it is probably 
not a good idea to include both in the fit simultaneously.

\begin{figure}[t]
  \begin{center}
    \includegraphics[width=0.45\textwidth]{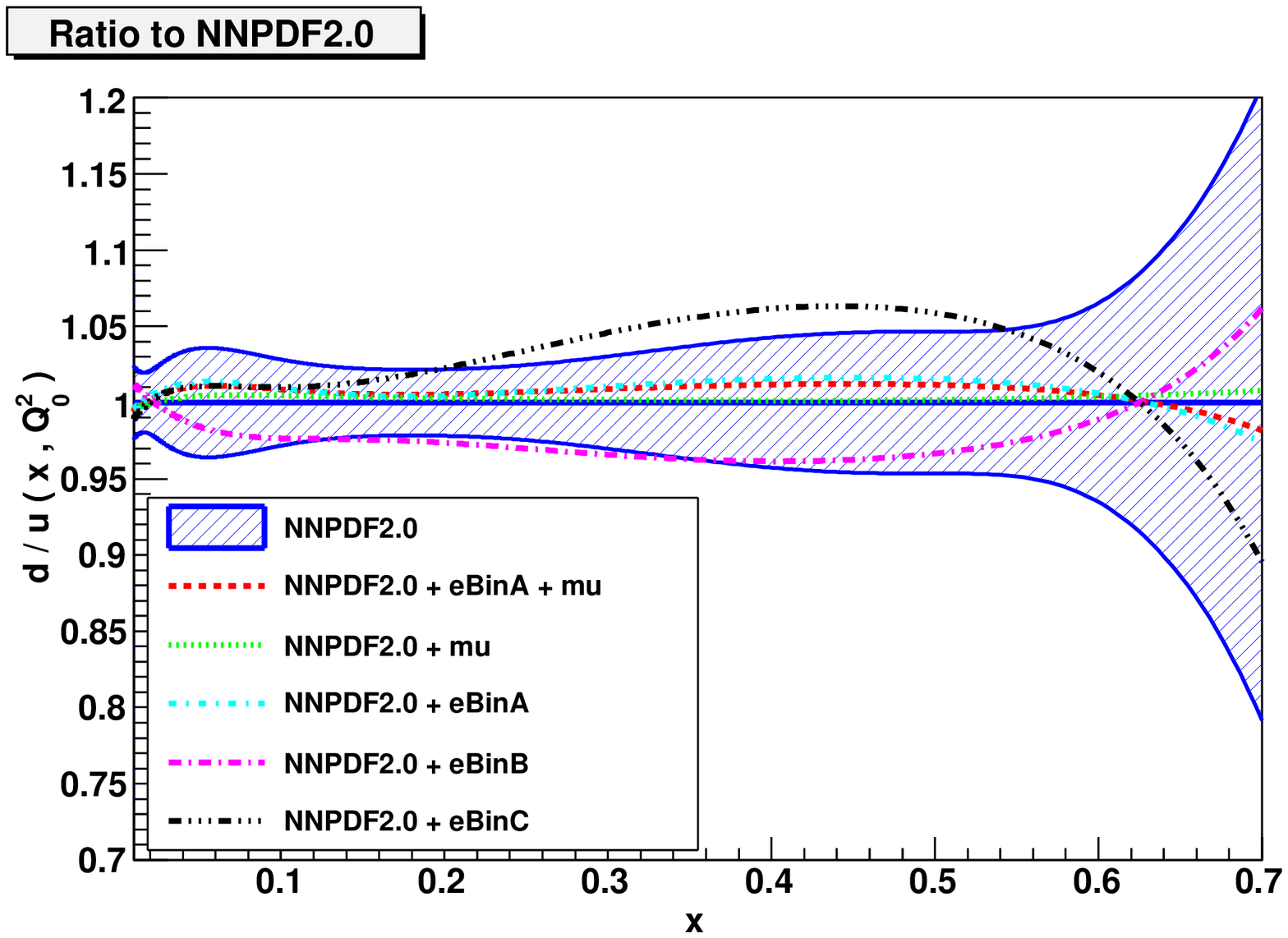}
    \includegraphics[width=0.45\textwidth]{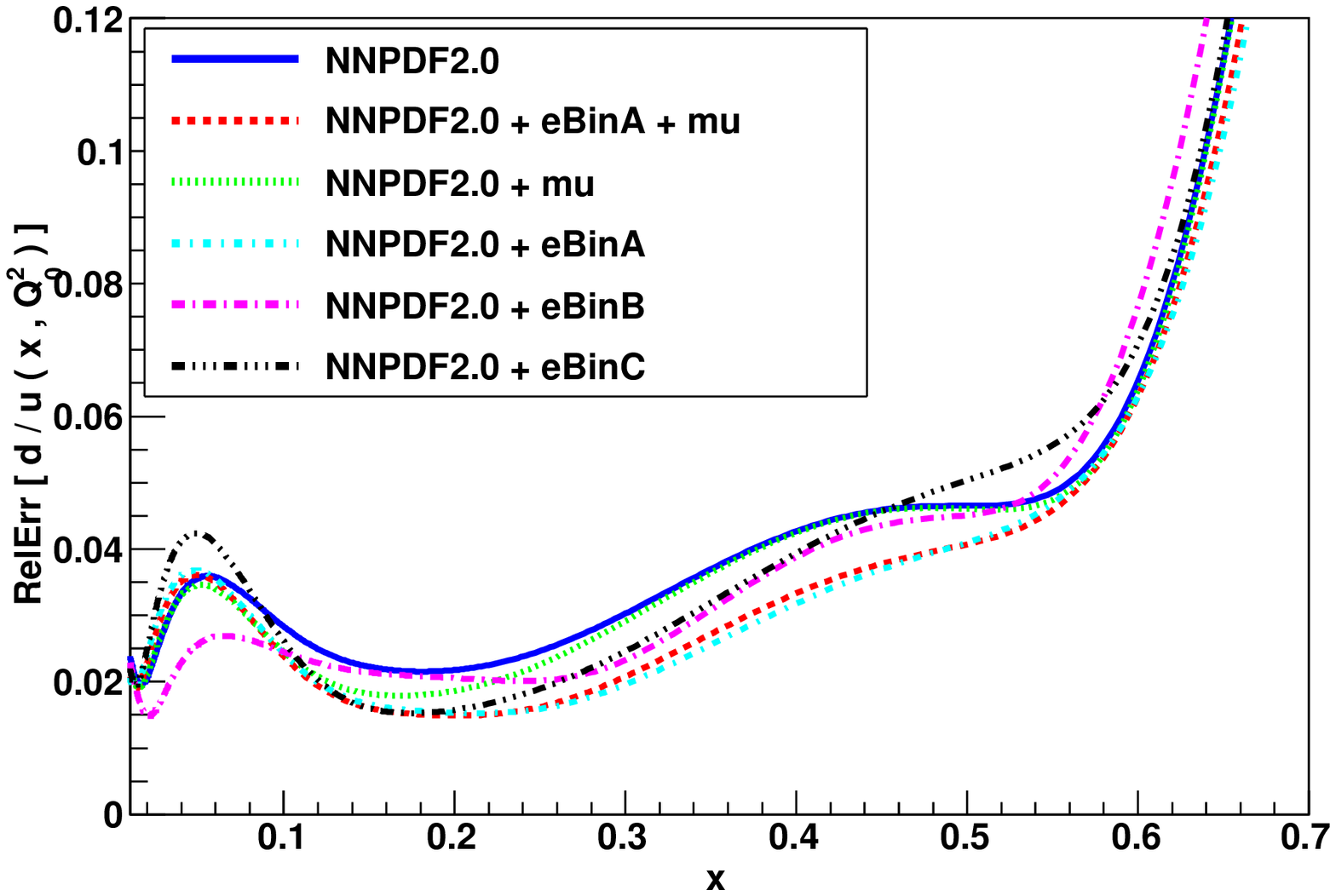}
  \end{center}
  \caption{\small The $d/u$ ratio at large $x$ computed at $Q_0^2=2$ GeV$^2$ from 
    the original NNPDF2.0 sets and the various sets obtained through reweighting 
    of NNPDF2.0. We show the results for the ratio normalized to NNPDF2.0 
    (left plot) and the relative PDF uncertainties in each case (right plot). 
    All uncertainties are 1$\sigma$. 
    \label{fig:d-over-u-final} }
\end{figure}
\begin{figure}[t]
  \begin{center}
    \includegraphics[width=0.45\textwidth]{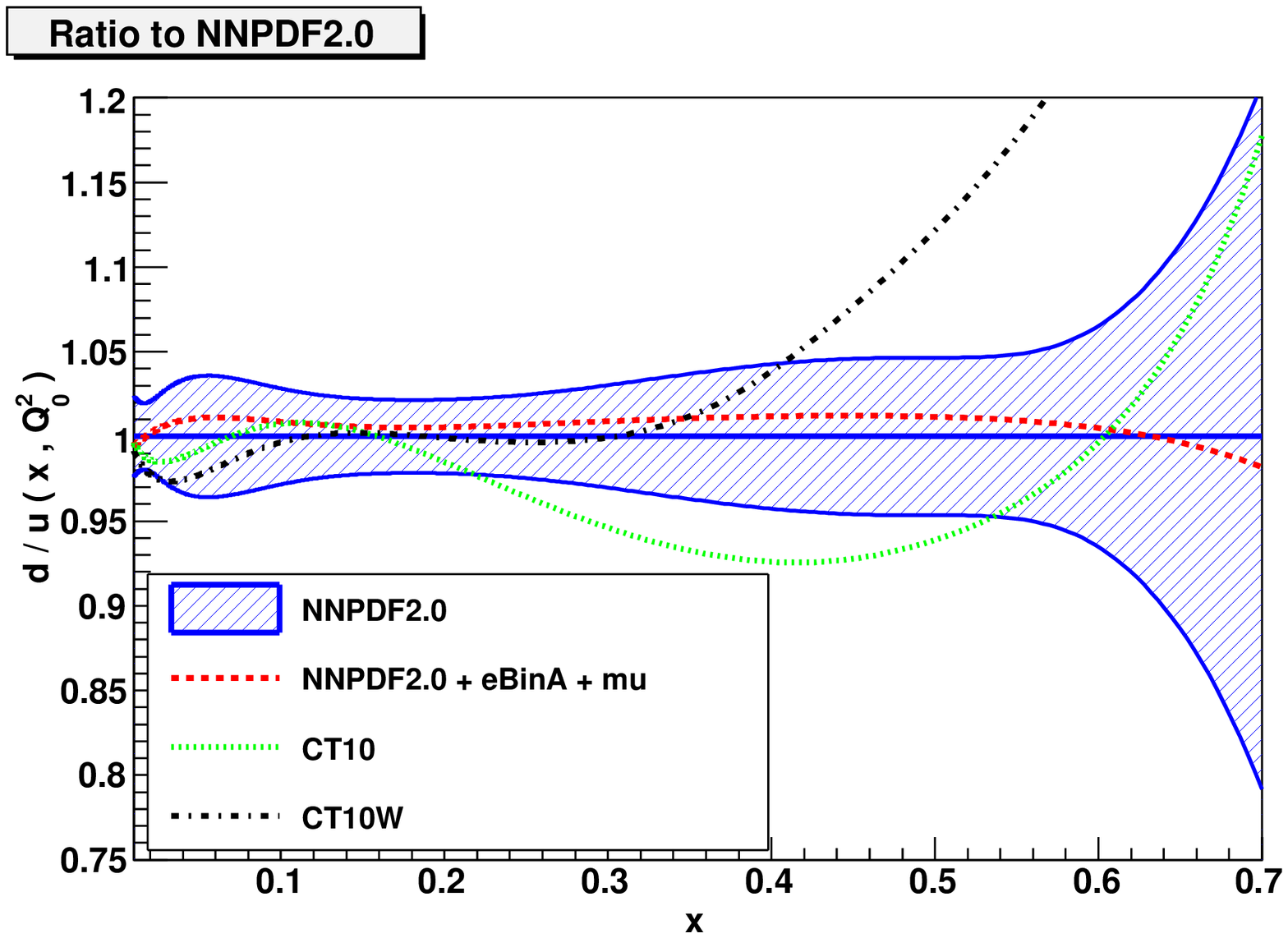}
    \includegraphics[width=0.45\textwidth]{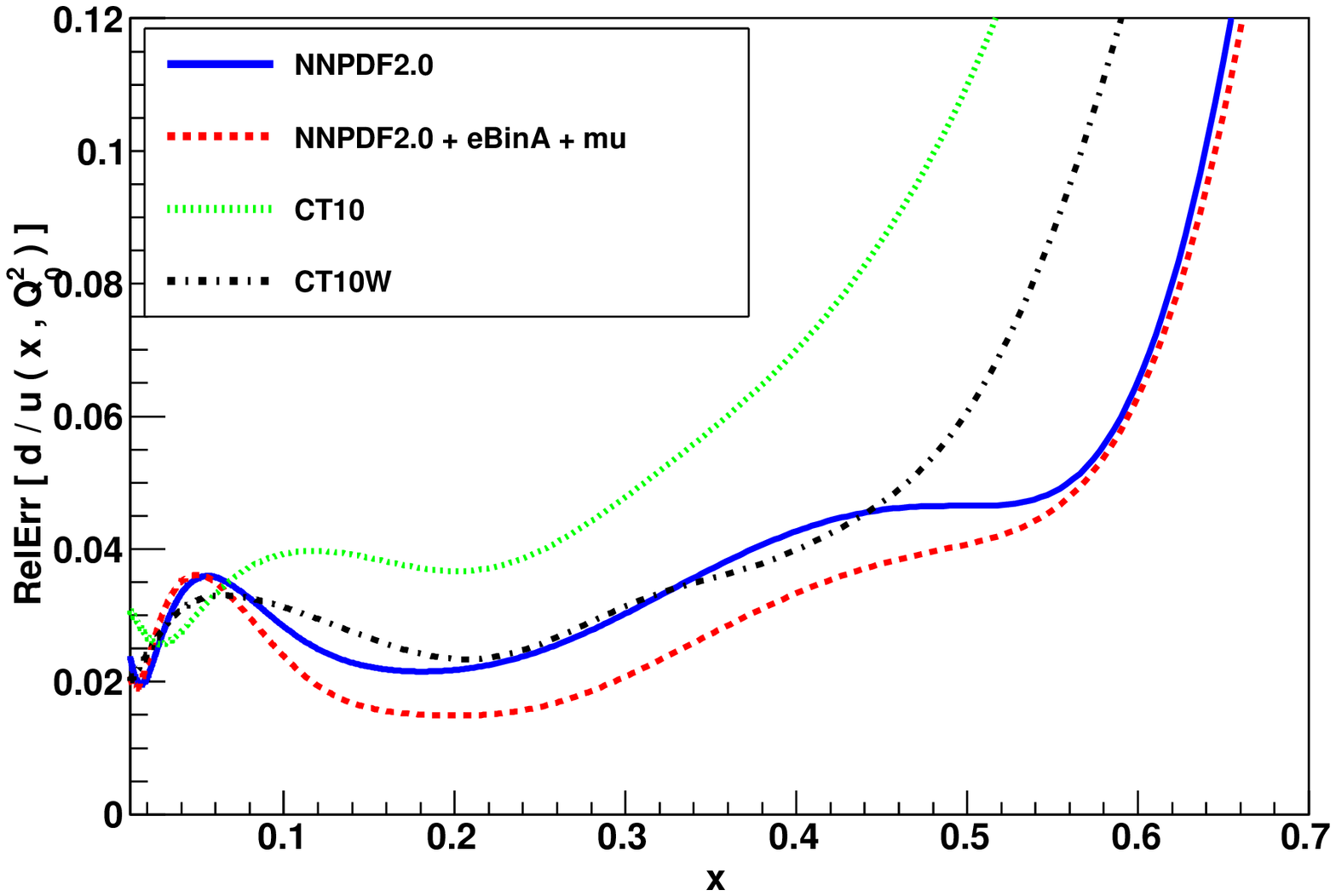}
  \end{center}
  \caption{\small Lower plots: The same $d/u$ ratio from the original NNPDF2.0 sets, 
    the NNPDF2.0 set reweighted by the maximally consistent combination of the D0 
    lepton asymmetry data (the muon data plus the inclusive electron data) and the 
    CT10 and CT10W sets. 
    \label{fig:d-over-u-ct10-incl}}
\end{figure}

\subsection{Implications for the $d/u$ ratio, and LHC benchmarks}

Up to now we have considered the impact of the D0 data on different PDF combinations, 
noticing that the most relevant effect was on the total valence distribution. 
To conclude our analysis we assess the impact of $W$ lepton charge asymmetry
data on the $d/u$ ratio. In Fig.~\ref{fig:d-over-u-final} we display the $d/u$ ratio 
at large $x$ computed at $Q_0^2=2$ GeV$^2$ from the original NNPDF2.0 set and the four sets
obtained through reweighting of NNPDF2.0 with the D0 lepton asymmetry data. 
The effect of the inclusive datasets (muon and electron bin A) is rather small, 
even when they are combined together. The less inclusive sets (bin B and bin C) 
have a rather larger effect, but pull in opposite directions. Even so, the 
effect is only of the same order as the PDF uncertainty.

In Fig.~\ref{fig:d-over-u-ct10-incl} we compare the $d/u$ ratio 
obtained with NNPDF2.0 and with NNPDF2.0 reweighted by the 
maximally consistent combination of the D0 data (muons and electrons bin A) 
with the CT10 and CT10W results, normalized to NNPDF2.0. It can be seen that the combination of D0 
muon and electron bin A data leads
to a substantial error reduction of $\sim 25\%$ in the $d/u$ ratio in the 
$0.1 \lsim x \lsim 0.5$ region, with almost no change in the central value. 
Note also that the $d/u$ ratio obtained from 
the NNPDF2.0 + D0(ebinA+$\mu$) set is rather more precise than that from CT10W, 
despite the fact that they include all the D0 lepton datasets,
with larger weights than the other datasets in the global analysis.

\begin{table}[t]
  \centering
  \footnotesize{
    \begin{tabular}{|l|c|c|c|c|c|}
      \hline 
      & $\sigma(Z\to ll^-)$ & $\sigma(W^-\to l\nu)$ & $\sigma(W^+\to l\nu)$ &
      $\sigma(h^0)$ & $\sigma_{t\bar{t}}$\\
      \hline
      2.0 & 911 $\pm$ 16 pb & 3.98 $\pm$ 0.08 nb 
      & 5.80 $\pm$ 0.12 nb & 11.59 $\pm$ 0.18 pb & 169 $\pm$ 6 pb \\
      \hline
      2.0+D0($\mu$) & 911 $\pm$ 16 pb & 3.98 $\pm$ 0.08 nb 
      & 5.80 $\pm$ 0.12 nb & 11.58 $\pm$ 0.19 pb & 169 $\pm$ 6 pb \\
      2.0+D0(bin A) & 914 $\pm$ 15 pb & 4.00 $\pm$ 0.07 nb 
      & 5.81 $\pm$ 0.11 nb & 11.58 $\pm$ 0.22 pb & 168 $\pm$ 6 pb \\
      2.0+D0($\mu+$ bin A) & 913 $\pm$ 15 pb & 4.00 $\pm$ 0.07 nb 
      & 5.81 $\pm$ 0.11 nb & 11.58 $\pm$ 0.20 pb & 168 $\pm$ 5 pb \\
      \hline
      2.0+D0(bin B) & 904 $\pm$ 15  pb & 3.92 $\pm$ 0.07 nb 
      & 5.78 $\pm$ 0.11 nb & 11.66 $\pm$ 0.15 pb & 172 $\pm$ 5 pb \\
      2.0+D0(bin C) & 913 $\pm$ 22 pb & 4.01 $\pm$ 0.10 nb 
      & 5.78 $\pm$ 0.17 nb & 11.52 $\pm$ 0.28 pb & 168 $\pm$ 8 pb \\
      \hline
    \end{tabular}
  }
  \caption{\small Cross sections for different Standard Candle processes at 
    the LHC  (7 TeV) computed using NNPDF2.0 reweighted PDFs including D0 $W$ 
    lepton asymmetry data. The Higgs cross-section is computed for $m_h = 120$ 
    GeV. 
    \label{tab:lhc-xsects}}
\end{table}
Finally, it is interesting to ask to what extent the inclusion 
of the $W$ lepton charge asymmetry data through reweighting affects the 
determination of some of the LHC standard cross-sections. 
Results for vector boson production, Higgs and $t\bar t$ at $\sqrt{s}=7$ TeV 
are collected in Table~\ref{tab:lhc-xsects}.
They have been computed using MCFM~\cite{Campbell:2000bg,Campbell:2002tg,MCFMurl} 
to determine the cross-section for each replica, and then the weighted 
average of the results evaluated using Eq.~(\ref{eq:avgrw}). 
The uncertainties in each case are purely PDF uncertainties obtained from a 
reweighted evaluation of the variance of the cross-section.
Clearly all these cross-sections are by and large insensitive to the addition 
of the D0 lepton charge asymmetry data, even those data (bins B and C) which show 
inconsistencies with the global dataset and thus have the largest (though least 
reliable) effect.
This is to be expected since the LHC observables we have considered are not 
directly sensitive to large-$x$ quarks, for which the impact of the D0 data is 
the largest.

\clearpage

\section{Conclusions}

In this paper we have developed a method for determining the effect of new data on 
PDFs without the need for a global refitting. 
The method relies on the existence of an ensemble of PDFs, distributed according 
to the uncertainties in a global set of older data, and thus representing the 
prior probability distribution of the PDFs. Such ensembles are provided by the 
NNPDF collaboration. 
The effect of new data is then accounted for by reweighting the PDF replicas 
in the ensemble according to their relative probabilities given the new dataset. 
These probabilities are determined simply and easily by computing the $\chi^2$ 
of the new data to the prediction obtained using a given replica.

We have provided a careful derivation of our formula used to determine the 
weights. This is important because our result differs from that
obtained in a previous attempt to use a reweighting
method~\cite{Giele:1998gw}.
The derivation is subtle because it is necessary to deal with multi-dimensional 
probability densities, where unless one is careful one can fall into inconsistencies 
due to the Borel-Kolmogorov paradox~\cite{jaynes}.

The main advantage of the new method is clear: computing the weights is no more 
difficult or computer intensive than the usual procedure of preparing a plot 
comparing the new dataset with predictions from given PDFs. 
However the information provided is much more substantial - one can 
assess quantitatively 
the impact of the new data on the PDFs, whether the new data are consistent with 
all the older data encoded within the PDF ensemble and the theoretical assumptions
on which it was based, and then whether the new data have any effect on other 
observables of interest such as benchmark cross-sections. 
Only when the impact of the new data is very large does a full refitting of 
the PDF ensemble become necessary, due to the loss of efficiency in the reweighted 
ensemble. 

We thus envisage our method being useful to experimentalists in all sorts of 
situations: testing the reliability of preliminary datasets and their uncertainties, 
assessing the credibility of possible indications of new physics, or in optimizing 
the design of new experiments using pseudodata.

We have shown explicitly that the method works by considering the addition of 
Tevatron inclusive jet data to a prior parton fit using only DIS and DY data. 
We have seen that when reweighted by the inclusive jet data, this fit becomes 
statistically equivalent to a refitting using all the data. 
The statistical equivalence has been quantified using the distance between prior and 
reweighted sets. This confirms that the refitted and the reweighted 
PDF sets can be seen as two samples of the same underlying probability
distribution. This is simultaneously a validation of the reweighting
methodology, and an important a posteriori consistency
check of the fitting procedure: an explicit confirmation that reweighting 
is equivalent to refitting for all data included in the global fit would amount to a
proof that the fitted result is indeed that dictated by the laws of
statistical inference. 
 
Using the reweighting formalism we have determined the impact of recent high 
luminosity D0 Run II lepton asymmetry data on the NNPDF2.0 PDFs. The lepton
asymmetry data has been historically an important constraint on the large-$x$ $d/u$ 
ratio, but recent attempts~\cite{MSTWasym,CT10} to include the new D0 data into global 
fits have been problematic. 
We find  instead that the data which are inclusive in $E_T^l$, the muon asymmetry 
data~\cite{d0muon} and electron asymmetry data~\cite{d0electron} with $E_t^l > 25$~GeV, 
are fully consistent with the NNPDF2.0 predictions and have a have a moderate impact 
on PDFs, showing up as a modest though noticable reduction in the uncertainty of the 
valence quark distribution. 
Moreover they are consistent with each other and with all the other datasets included 
in NNPDF2.0.

The consistency of these data has been recently studied also by the
MSTW and CTEQ collaborations. In particular
MSTW~\cite{MSTWasym} finds
 that it is not possible to fit the inclusive D0 electron dataset without 
affecting the description of the rest of the experiments in the global analysis unless
large nuclear corrections for the DIS deuteron data are applied at the same time. 
The CT10 analysis~\cite{CT10} also suggests a sizable tension 
between the D0 lepton asymmetry
data and the DIS deuteron data. Our results do not support
these conclusions.
Since the predictions for the lepton asymmetry depend strongly on the $d/u$ slope, 
it is possible that the origin of the problems in the CT10 and MSTW analysis 
is that they are based on refitting using a fixed parametrization, and are 
thus subject to the functional biases such a procedure necessarily entails. 

We further find that the less inclusive electron asymmetry data~\cite{d0electron} 
binned in $E_T^e$, the two datasets with $25~{\rm GeV} <E_T^e < 35~{\rm GeV}$ and 
$35~{\rm GeV} <E_T^e$, while having potentially more impact on the PDFs, 
are problematic: the former data set is inconsistent with some of the DIS data 
(specifically BCDMS, both proton and deuteron), while the latter seems to have 
problems of internal consistency.\footnote{Similar difficulties in fitting 
these datasets have been reported by MSTW~\cite{MSTWasym} and 
CTEQ~\cite{CT10}, though it is not easy to make a direct comparison 
since they attempt to fit all 
three D0 electron bins simultaneously.}
Consequently the effect on PDFs of including these datasets is to 
actually increase uncertainties in some regions of $x$. Furthermore, we 
find evidence that these two datasets are also mutually inconsistent. 
We think it likely that 
the experimental errors on these data have been substantially underestimated. 
Until these problems are better understood, we believe that is safer to include 
in the global fit only the inclusive datasets, which even if less constraining
are more robust experimentally. 

The reweighting methodology described here should allow anybody to perform their 
own updates of NNPDF fits, to incorporate whatever new datasets 
they are interested in, by following the same procedure we used 
here for the specific case of the W lepton asymmetry. We very much 
hope that they will exploit this possibility.


\bigskip

{\bf\noindent  Acknowledgments \\}
We would like to thank G. Ferrera and M. Grazzini for useful discussions and 
for instructing us in using the DYNNLO code. We are also grateful to G. Cowan and 
W. Giele for discussions about reweighting and Bayesian statistics, 
P. Nadolsky for discussions about the W asymmetry data, and R. McNulty and 
F. De Lorenzi for discussions about the LHCb study. We would also like to 
thank J.C.~Collins and J. Pumplin for remarks which prompted this 
revised version. M.U. is supported by 
the Bundesministerium f\"ur Bildung and Forschung (BmBF) of the Federal 
Republic of Germany (project code 05H09PAE). This work was 
partly supported by HEPTOOLS under contract MRTN-CT-2006-035505. We would also 
like to acknowledge the use of the computing resources provided by the 
Edinburgh Computer and Data Facility (ECDF) (http://www.ecdf.ed.ac.uk/). The 
ECDF is partially supported by the eDIKT initiative (http://www.edikt.org.uk).

\clearpage
\newpage

\appendix


\section{Distances between reweighted PDFs}
\label{sec:distances}

Given two sets of $N^{(1)}_{\mathrm{rep}}$ and $N^{(2)}_{\mathrm{rep}}$
replicas, in general reweighted, it is possible to use the distance
estimators defined in Appendix A of Ref.~\cite{nnpdf20}
to determine whether they correspond
to different instances of the same underlying probability distribution,
or whether instead they come from different underlying
distributions.

The discussion in Ref.~\cite{nnpdf20} applies also
to reweighted PDF sets with the corresponding modifications
that we list below. 
For example, expectation values have to be computed with the
associated weights. For the first, second and fourth moments
of the PDFs one then has to use
\begin{equation}
  \label{eq:expval}
  \langle
  q^{(k)}\rangle_{(i)}=\frac{1}{N^{(i)}_{\mathrm{rep}}}\sum_{k=1}^{N^{(i)}_{\mathrm{rep}}} w_k^{(i)}
  q^{(i)}_k \ ,
\end{equation}
 \begin{equation}
    \label{eq:std}
    \sigma^2_{(i)}[
    q^{(i)}]=\frac{1}{N^{(i)}_{\mathrm{rep}}-1}
    \sum_{k=1}^{N^{(i)}_{\mathrm{rep}}} w_k^{(i)}\left(q^{(i)}_k-\langle q^{(i)}\rangle\right)^2,
  \end{equation}
 \begin{equation}\label{eq:mfour}
    m_4[q^{(i)}]=\frac{1}{N^{(i)}_{\mathrm{rep}}}
    \sum_{k=1}^{N^{(i)}_{\mathrm{rep}}} w_k^{(i)}\left(q^{(i)}_k-\langle q^{(i)}\rangle\right)^4 \ .
  \end{equation}
Note that in the above equations 
 the unweighted expressions are trivially reproduced setting
$w_k^{(i)}=1$.

Another difference arises when computing the variance of the mean
and the variance of the variance with weighted PDF sets. In this case,
this estimators scale not with the total number of
replicas but with an effective number of replicas after reweighting
\begin{equation}
N_{\rm rep,eff}^{(i)} = \frac{\lp \sum_{k=1}^{N_{\rm rep}^{(i)}}w_k^{(i)}\rp^2}{
\sum_{k=1}^{N_{\rm rep}^{(i)}}w_k^{(i),2}} = 
N_{\rm rep}^2\Bigg/ \lp
\sum_{k=1}^{N_{\rm rep}^{(i)}}w_k^{(i),2}\rp
\end{equation}
that reduces to $N_{\rm rep}$ in the unweighted case (note that this is not the same as the $N_{\rm eff}$ given by the Shannon entropy \eq{eq:effective}.

The variance of the mean for reweighted sets is then 
given by
\begin{equation}
  \label{eq:stdmean}
  \sigma^2_{(i)}[\langle q^{(i)}\rangle]=
  \frac{1}{N^{(i)}_{\mathrm{rep},{\rm eff}}} \sigma^2_{(i)}[q^{(i)}]
\end{equation}
while the variance of the sample variance is
 \begin{equation}
    \label{eq:ases}
    \sigma^2_{(i)}[\bar\sigma^2_{(i)}]= 
    \frac{1}{N_{\mathrm{rep,eff}}^{(i)}}\left[m_4[q^{(i)}]
      -\frac{N_{\mathrm{rep}}^{(i)}-3}{N_{\mathrm{rep}}^{(i)}-1}
      \left(\bar\sigma^2_{(i)}\right)^2\right] \ . 
  \end{equation}
Again in the unweighted case everything reduces to the
expression in Ref.~\cite{nnpdf20}.





\end{document}